\documentclass[notitlepage,a4paper,aps,prd,onecolumn,superscriptaddress,nofootinbib,groupedaddress]{revtex4}

\usepackage{amsmath}
\usepackage{amsfonts,color}
\usepackage{amsmath}
\usepackage{amsthm}
\usepackage{pdfpages}
\usepackage{amsmath}
\usepackage{empheq}
\usepackage{amsfonts,color}
\usepackage{amssymb,float}
\usepackage{physics}
\usepackage{booktabs,multirow}
\usepackage{titlesec}

\usepackage{hyphenat}
\usepackage{ragged2e}

\def\be{\begin{equation}}
\def\ee{\end{equation}}
\def\bea{\begin{eqnarray}}
\def\eea{\end{eqnarray}}
\def\bean{\begin{eqnarray*}}
\def\eean{\end{eqnarray*}}

\def\Z{{}^{(1)}\tilde Z}
\usepackage{titlesec}

\titleformat{\paragraph}
{\normalfont\normalsize\bfseries}{\theparagraph}{1em}{}
\titlespacing*{\paragraph}
{0pt}{3.25ex plus 1ex minus .2ex}{1.5ex plus .2ex}

\usepackage{amsmath}
\usepackage[utf8]{inputenc}
\allowdisplaybreaks
\usepackage{color}
\usepackage{hyperref}
\hypersetup{
    colorlinks=true,
    linkcolor=blue,
    filecolor=magenta,      
   citecolor=blue
}

\usepackage{enumitem}

\usepackage{tikz}
\usetikzlibrary{shapes.geometric}
\usetikzlibrary{arrows.meta,arrows,positioning}
\usetikzlibrary{tikzmark}

\begin{document}
\title{Algebraic classification of the gravitational field in general metric-affine geometries}

\author{Sebastian Bahamonde}
\email{sbahamondebeltran@gmail.com, sebastian.bahamonde@ipmu.jp}
\affiliation{Kavli Institute for the Physics and Mathematics of the Universe (WPI), The University of Tokyo Institutes
for Advanced Study (UTIAS), The University of Tokyo, Kashiwa, Chiba 277-8583, Japan.}

\author{Jorge Gigante Valcarcel}
\email{jorgevalcarcel@ibs.re.kr, gigante.j.aa@m.titech.ac.jp}
\affiliation{Center for Geometry and Physics, Institute for Basic Science (IBS), Pohang 37673, Korea.}
\affiliation{Department of Physics, Tokyo Institute of Technology, 1-12-1 Ookayama, Meguro-ku, Tokyo 152-8551, Japan.}

\author{José M.M. Senovilla}
\affiliation{Departamento de Física, Facultad de Ciencia y Tecnología, Universidad del País Vasco UPV/EHU, Apartado 644, 48080 Bilbao, Spain.}
\affiliation{EHU Quantum Center, Universidad del País Vasco UPV/EHU, Bilbao, Spain.}

\begin{abstract}

We present the algebraic classification of the gravitational field in four-dimensional general metric-affine geometries, thus extending the current results of the literature in the particular framework of Weyl-Cartan geometry by the presence of the traceless nonmetricity tensor. This quantity switches on four of the eleven fundamental parts of the irreducible representation of the curvature tensor under the pseudo-orthogonal group, in such a way that three of them present similar algebraic types as the ones obtained in Weyl-Cartan geometry, whereas the remaining one includes thirty independent components and gives rise to a new algebraic classification. The latter is derived by means of its principal null directions and their levels of alignment, obtaining a total number of sixteen main algebraic types, which can be split into many subtypes. As an immediate application, we determine the algebraic types of the broadest family of static and spherically symmetric black hole solutions with spin, dilation and shear charges in Metric-Affine Gravity.

\end{abstract}

\maketitle

%\tableofcontents

\section{Introduction}

Algebraic classification has certainly played a significant role in the development of General Relativity (GR). Indeed, as featured in the Einstein’s field equations, our current understanding of the gravitational interaction is based on the physical correspondence between the space-time curvature and the energy-momentum tensor of matter, which naturally leads to the study of the algebraic properties of these quantities, in order to find out, analyse and interpret different classes of solutions~\cite{Stephani:2003tm,Griffiths:2009dfa}.

From a mathematical point of view, some tensor quantities on a Lorentzian manifold can be recast as linear maps acting on a vector space, which lays the foundations of algebraic classification via the resolution of an eigenvalue problem~\cite{petrov2016einstein}. In the framework of GR, the gravitational field is fully ascribed to the Riemann curvature tensor, whose irreducible decomposition under the pseudo-orthogonal group expresses it as a linear combination of the Ricci scalar and the completely traceless Weyl and Ricci tensors; the latter presenting nontrivial eigenvalue problems that lead to the so called Petrov and Segre classifications, respectively~\cite{Petrov:2000bs,bona1992intrinsic}.

These classifications have numerous applications in the study of black holes, cosmology and gravitational waves. Of particular interest is the formulation of the Goldberg-Sachs theorem, which states that any vacuum solution of the Einstein's field equations admits a shear-free null geodesic congruence if and only if the Weyl tensor is algebraically special~\cite{goldberg1962theorem}. Such a congruence defines a null vector field that is, at each point, multiply aligned with the algebraic structure of the Weyl tensor, which is a manifestation of its ``speciality''.  In fact, the consideration of an algebraically special Type D Weyl tensor ---in which case there are two such doubly aligned null vector fields--- turned out to be crucial to find the first known rotating black hole solution in GR, namely the stationary and axially symmetric Kerr solution~\cite{Kerr:1963ud}. Thereby, despite of the cumbersome form of the field equations for a stationary and axially symmetric configuration, the Kerr solution possesses a significant degree of symmetry, which is actually realised by the existence of a closed nondegenerate conformal Killing-Yano tensor~\cite{Houri:2007xz} ---actually, all type D vacuum solutions in GR admit a conformal Killing tensor as proven by Walker and Penrose~\cite{Stephani:2003tm}. These objects provide a separability structure for the wave and geodesic equations defined on the space-time, which in turn implies the complete integrability of causal geodesics and the algebraic Type D of the Weyl tensor~\cite{Frolov:2006pe,Oota:2007vx,Frolov:2017kze,Lunin:2017drx,Krtous:2018bvk,Frolov:2018ezx,Stephani:2003tm}. Likewise, a simple example of gravitational radiation is described by the plane-fronted waves with parallel rays (``pp waves"), which include exact vacuum solutions of the Einstein's field equations and correspond to an algebraically special Type N Weyl tensor~\cite{Sippel:1986if,Stephani:2003tm}, that is, with a unique multiply aligned null vector field. On the other hand, different matter sources of physical interest, such as the electromagnetic field, pure radiation matter field and the perfect fluid, also lead in the framework of GR to algebraically special types of the Ricci tensor that are included in the Segre classification~\cite{Stephani:2003tm}.

Given the relationship between the algebraic classification of the Weyl tensor and the existence of aligned null directions such as those discovered by Goldberg and Sachs, alternative classifications have been investigated relying exclusively on the existence of aligned null vectors. Here the concept of alignment is somewhat complicated and depends on the particular properties of the target tensor, but they can be rigorously defined in general by the vanishing of well-defined contractions and exterior products of the target tensor with the null vector. These are referred to as the principal null directions (PNDs) of the tensor. Such alternative classifications are then based on the number of different PNDs and their multiplicities, or their level of alignment. In four dimensions, this approach provides an algebraic classification for the Weyl tensor that is fully equivalent to the Petrov classification~\cite{debever1959tenseur,debever1959super} (see~\cite{Coley:2004jv,Ortaggio:2012jd} for further generalisations in higher dimensions), whereas a richer algebraic classification is obtained for the traceless Ricci tensor, in comparison with the Segre classification~\cite{Milson:2004jx}.

The PNDs of any target tensor can also be characterised, in a much direct and simpler manner, as the null vectors whose contraction on all indices with the {\em superenergy tensor}~\cite{Senovilla:1999xz} of the target tensor vanishes, see~\cite{Senovilla:2010ej,Senovilla:2006gf,Ortaggio:2012jd} and references therein. This generally leads to a refined classification~\cite{Senovilla:2010ej} depending on the number of contractions of the null vector with the superenergy tensor needed to get a vanishing result. The superenergy tensor is the (basically) unique tensor quadratic on the target tensor that satisfies a generalised dominant energy condition, and can be seen as a mathematical generalisation of the traditional energy-momentum tensor. The paradigmatic example is the famous Bel-Robinson tensor, which is just the superenergy tensor of the Weyl tensor~\cite{Senovilla:1999xz}. 

Therefore, the problem of the algebraic classification of the gravitational field in GR and other theories of gravity based on Riemannian geometry is settled, but the presence of additional degrees of freedom in the geometry requires an extension of these results. In particular, a post-Riemannian description of the space-time in the presence of torsion and nonmetricity leads to the formulation of Metric-Affine Gravity (MAG), which constitutes a viable extension of GR and provides a diverse phenomenology at astrophysical and cosmological scales~\cite{Hehl:1994ue,adamowicz1980plane,Baekler:1981lkh,Gonner:1984rw,Bakler:1988nq,Gladchenko:1994wu,Tresguerres:1995js,tresguerres1995exact,Hehl:1999sb,Garcia:2000yi,Puetzfeld:2001hk,Chen:2009at,Lu:2016bcx,Cembranos:2016gdt,Cembranos:2017pcs,Blagojevic:2017wzf,Blagojevic:2017ssv,Obukhov:2019fti,Olmo:2019flu,Guerrero:2020azx,Bahamonde:2020fnq,Obukhov:2020hlp,Iosifidis:2020gth,Aoki:2020zqm,Iosifidis:2020upr,Bahamonde:2021akc,Bahamonde:2021qjk,delaCruzDombriz:2021nrg,Iosifidis:2021xdx,Harko:2021tav,Jimenez-Cano:2022arz,Bahamonde:2022meb,Boudet:2022wmb,Pantig:2022ely,Boudet:2022nub,Bahamonde:2022kwg,Yang:2022icz,Bulgakova:2022bsk,Iosifidis:2022xvp,Iosifidis:2023pvz,Ditta:2023wye,Yasir:2023qho,Javed:2023lmv,Aoki:2023sum,DeMartino:2023qkl,Iosifidis:2023kyf,Mustafa:2023ngp,Yasir:2024nfn,McNutt:2024uks,Barker:2024dhb,Jing:2024thh,Yasir:2024wir,Andrei:2024vvy}. Thereby, a gauge invariant Lagrangian can be constructed from the generalised field strength tensors of this framework, in order to introduce the dynamics of the gravitational field enhanced by torsion and nonmetricity. A complete algebraic classification requires then to classify all the field strength tensors of torsion and nonmetricity, which naturally appear in the irreducible decomposition of the curvature tensor under the pseudo-orthogonal group~\cite{McCrea:1992wa}. Indeed, this problem has been recently addressed in the particular case of Weyl-Cartan geometry~\cite{Bahamonde:2023piz}, while the case of general metric-affine geometries remains unsolved.

In this work we perform a complete algebraic classification of general metric-affine geometries by means of PNDs. Such a case is modeled by an affinely connected metric space-time that is characterised by completely general curvature, torsion and nonmetricity tensors. In particular, in contrast with a Weyl-Cartan space-time, it includes a traceless nonmetricity tensor, whose dynamics is described in the gravitational action of MAG by four field strength tensors; all of them obeying their own algebraic classifications. Indeed, even though the methods of algebraic classification are well-known in the literature, only a small number of tensors, including the Weyl, Ricci and Faraday tensors, have been formally classified. Therefore, the purpose of this work is twofold: we aim to obtain a full algebraic classification of the gravitational field in general metric-affine geometries, which on the other hand demands to obtain a new algebraic classification for a completely traceless and cyclic tensor that constitutes one of the field strengths of the traceless nonmetricity tensor. In comparison with the Weyl, Ricci and Faraday tensors, which carry ten, nine and six independent components in four dimensions, this field strength tensor carries thirty independent components, giving rise to a more complicated problem and in fact to a much richer algebraic classification. Apart from that, it is worthwhile to stress that, although our study refers to the framework of metric-affine geometry, the results are valid for any tensor quantities presenting the same algebraic properties as the ones considered in this work.

This paper is organised as follows. In Sec.~\ref{sec:IrrRep}, we introduce the irreducible decomposition of the curvature tensor in metric-affine geometry, which is determined by eleven building blocks that provide the dynamics of the gravitational field with curvature, torsion and nonmetricity. Taking into account the algebraic symmetries of the mentioned building blocks, they can be sorted into four different categories, each one characterised by its own type of algebraic classification. In fact, it was recently shown that three of these types appear in the framework of Weyl-Cartan geometry~\cite{Bahamonde:2023piz}, hence we briefly revisit them in Sec.~\ref{sec:AlgClas1},~\ref{sec:AlgClas2} and~\ref{sec:AlgClas3}. The main study is then addressed in Sec.~\ref{sec:AlgClas4}, where we obtain the last type of algebraic classification that can take place in general metric-affine geometries. This requires a thorough analysis on the algebraic structure of one of the field strength tensors of the traceless nonmetricity tensor, for which we find its PNDs and their respective levels of alignment in Sec.~\ref{ap:classi}. We then apply, in Sec.~\ref{sec:supere}, the refinements derived by using the more elaborated classification using the superenergy tensor of this field strength. Once the algebraic classification is settled, in Sec.~\ref{sec:application} we determine the algebraic types of all of the field strength tensors of torsion and nonmetricity for the broadest family of static and spherically symmetric black hole solutions with spin, dilation and shear charges in MAG, finding that the gravitational field of the solution is indeed algebraically special. Finally, we present the conclusions in Sec.~\ref{sec:conclusions}, while some technical details are relegated to the appendices.

We work in natural units $c=G=1$ and consider the metric signature $(+,-,-,-)$. On the other hand, we use a tilde accent to denote those quantities that are defined from the general affine connection, in contrast to their unaccented counterparts constructed from the Levi-Civita connection. In addition, we denote with a diagonal arrow the traceless and pseudotraceless pieces of tensors (e.g. ${\nearrow\!\!\!\!\!\!\!Q}\,^{\lambda}{}_{\mu\nu}$ and ${\nearrow\!\!\!\!\!\!\!\tilde{R}}\,^{\lambda}{}_{[\rho\mu\nu]}$). Latin and Greek indices run from $0$ to $3$, referring to anholonomic and coordinate bases, respectively.

\section{Irreducible decomposition of the curvature tensor in metric-affine geometry}\label{sec:IrrRep}

An independent affine connection includes the torsion and nonmetricity tensors
\begin{equation}
    T^{\lambda}\,_{\mu \nu}=2\tilde{\Gamma}^{\lambda}\,_{[\mu \nu]}\,, \quad Q_{\lambda \mu \nu}=\tilde{\nabla}_{\lambda}g_{\mu \nu}\,,
\end{equation}
as its antisymmetric part and as the covariant derivative of the metric tensor, which gives rise to a general curvature tensor that can be expressed as the sum of the Riemann tensor and further post-Riemannian corrections
\begin{equation}\label{totalcurvature}
\tilde{R}^{\lambda}\,_{\rho\mu\nu}=R^{\lambda}\,_{\rho\mu\nu}+\nabla_{\mu}N^{\lambda}\,_{\rho \nu}-\nabla_{\nu}N^{\lambda}\,_{\rho \mu}+N^{\lambda}\,_{\sigma \mu}N^{\sigma}\,_{\rho \nu}-N^{\lambda}\,_{\sigma \nu}N^{\sigma}\,_{\rho \mu}\,,
\end{equation}
with
\begin{equation}
    N^{\lambda}\,_{\mu\nu}=\frac{1}{2}\left(T^{\lambda}\,_{\mu \nu}-T_{\mu}\,^{\lambda}\,_{\nu}-T_{\nu}\,^{\lambda}\,_{\mu}\right)+\frac{1}{2}\left(Q^{\lambda}\,_{\mu \nu}-Q_{\mu}\,^{\lambda}\,_{\nu}-Q_{\nu}\,^{\lambda}\,_{\mu}\right)\,.
\end{equation}

Thereby, whereas in Riemannian geometry the irreducible decomposition of the curvature tensor into irreducible pieces under the pseudo-orthogonal group simply expresses this tensor as a linear combination of the Ricci scalar and the completely traceless Weyl and Ricci tensors, its general form in metric-affine geometry turns out to present a much richer structure~\cite{McCrea:1992wa}. Specifically, it includes eleven irreducible pieces, which can be grouped into antisymmetric and symmetric components
\begin{equation}\label{irrdec}
    \tilde{R}_{\lambda\rho\mu\nu}= \tilde{W}_{\lambda\rho\mu\nu}+\tilde{Z}_{\lambda\rho\mu\nu}\,,
\end{equation}
with
\begin{equation}
\tilde{W}_{\lambda\rho\mu\nu}:=\tilde{R}_{[\lambda\rho]\mu\nu}\,,\quad \tilde{Z}_{\lambda\rho\mu\nu}:=\tilde{R}_{(\lambda\rho)\mu\nu}\,.
\end{equation}

In general, the antisymmetric component includes both Riemannian and post-Riemannian contributions, whereas the symmetric one is switched on only in the presence of nonmetricity. In fact, the nonmetricity tensor can also be separated as the sum of two trace and traceless parts
\begin{equation}
Q_{\lambda\mu\nu}=\frac{1}{4}g_{\mu\nu}Q_{\lambda\rho}\,^{\rho}+{\nearrow\!\!\!\!\!\!\!Q}_{\lambda\mu\nu}\,,
\end{equation}
in such a way that each of these parts provides its own contribution in the aforementioned components.

The decomposition can then be expressed in a convenient way by the definition of the following building blocks~\cite{Bahamonde:2023piz}:
\begin{eqnarray}
      {\nearrow\!\!\!\!\!\!\!\tilde{R}}_{(\mu\nu)}&=&{\nearrow\!\!\!\!\!\!\!{R}}_{\mu\nu}+\nabla_{\lambda}T_{(\mu\nu)}{}^{\lambda} -\nabla_{(\mu}T^{\lambda}{}_{\nu)\lambda}+\frac{1}{2}g_{\mu\nu}\nabla_{\lambda}T^{\rho\lambda}\,_{\rho}-\nabla_{\lambda}Q_{(\mu\nu)}{}^{\lambda}+\frac{1}{2}\nabla_{(\mu}Q_{\nu)}{}^{\lambda}{}_{\lambda}+\frac{1}{2}\nabla_{\lambda}Q^{\lambda}{}_{\mu\nu}\nonumber\\
    &&+\,\frac{1}{4}g_{\mu\nu}\bigl(\nabla_{\lambda}Q^{\rho}\,_{\rho}\,^{\lambda}-\nabla_{\lambda}Q^{\lambda}\,_{\rho}\,^{\rho}\bigr)+ \frac{1}{2}T^{\rho}{}_{\lambda(\mu}T_{\nu)}{}^{\lambda}{}_{\rho}+T^{\rho\lambda}{}_{\rho}T_{(\mu\nu)\lambda}+\frac{1}{4}T_{\mu\lambda\rho} T_{\nu}{}^{\lambda\rho}\nonumber\\
    &&+\,\frac{1}{4}g_{\mu\nu}\Bigl(T^{\lambda}\,_{\lambda\sigma}T^{\rho}\,_{\rho}\,^{\sigma}-\frac{1}{2}T_{\lambda\rho\sigma}T^{\rho\lambda\sigma}-\frac{1}{4}T_{\lambda\rho\sigma}T^{\lambda\rho\sigma}\Bigr)+Q_{\lambda\mu\rho}Q^{[\lambda}{}_{\nu}{}^{\rho]}+\frac{1}{2}Q^{\lambda \rho}{}_{\rho}Q_{(\mu\nu)\lambda}\nonumber\\
    &&-\,\frac{1}{4}\bigl(Q_{\lambda\mu\nu}Q^{\lambda\rho}{}_{\rho}+Q_{\mu\lambda\rho}Q_{\nu}{}^{\lambda\rho}\bigr)+\frac{1}{16}g_{\mu\nu}\bigl(2Q_{\lambda\rho\sigma}Q^{\rho\lambda\sigma}+Q^{\rho\lambda}\,_{\lambda}Q_{\rho}\,^{\sigma}\,_{\sigma}-Q_{\lambda\rho\sigma}Q^{\lambda\rho\sigma}-2Q^{\sigma\lambda}\,_{\lambda}Q^{\rho}\,_{\rho\sigma}\bigr)\nonumber\\
    &&+\,\frac{1}{2}\bigl(T^{\lambda}{}_{(\mu}{}^{\rho}Q_{\nu)\lambda\rho}-Q_{\lambda\rho(\mu}T^{\lambda}{}_{\nu)}{}^{\rho}-2Q_{(\mu\nu)\lambda}T^{\rho\lambda}{}_{\rho}+Q_{\lambda\rho(\mu}T^{\rho}{}_{\nu)}{}^{\lambda}-2Q_{\lambda\rho(\mu}T_{\nu)}{}^{\lambda\rho}-Q_{\lambda}{}^{\rho}{}_{\rho}T_{(\mu\nu)}{}^{\lambda}+Q_{\lambda\mu\nu}T^{\rho\lambda}{}_{\rho}\bigr)\nonumber\\
    &&+\,\frac{1}{4}g_{\mu\nu}\bigl(T^{\lambda}\,_{\lambda\rho}Q^{\rho\sigma}\,_{\sigma}-T_{\lambda\rho\sigma}Q^{\sigma\lambda\rho}-T^{\lambda}\,_{\lambda\rho}Q^{\sigma\rho}\,_{\sigma}\bigr)\,,\label{BB1}\\
    \tilde{R}^{(T)}_{[\mu\nu]}&=&\tilde{\nabla}_{[\mu}T^{\lambda}\,_{\nu]\lambda}+\frac{1}{2}\tilde{\nabla}_{\lambda}T^{\lambda}\,_{\mu\nu}-\frac{1}{2}T^{\lambda}\,_{\rho\lambda}T^{\rho}\,_{\mu\nu}\,, \quad \tilde{R}^{\lambda}\,_{\lambda\mu\nu}=\nabla_{[\nu}Q_{\mu]\lambda}{}^{\lambda}\,,\label{BB2}\\
    {\nearrow\!\!\!\!\!\!\!\hat{R}}^{(Q)}_{(\mu\nu)}&=&\tilde\nabla_{\lambda}{\nearrow\!\!\!\!\!\!\!Q}_{(\mu \nu )}{}^{\lambda}-\tilde\nabla_{(\mu}{\nearrow\!\!\!\!\!\!\!Q}^\lambda{}_{\nu )\lambda}+{\nearrow\!\!\!\!\!\!\!Q}^{\lambda\rho}\,_{\lambda}{\nearrow\!\!\!\!\!\!\!Q}_{(\mu\nu)\rho}-{\nearrow\!\!\!\!\!\!\!Q}_{\lambda\rho(\mu}{\nearrow\!\!\!\!\!\!\!Q}_{\nu)}{}^{\lambda\rho}-T_{\lambda\rho(\mu}{\nearrow\!\!\!\!\!\!\!Q}^{\lambda\rho}{}_{\nu)}\,,\label{BB3}\\
     \hat{R}^{(Q)}_{[\mu\nu]}&=&\tilde{\nabla}_{[\mu}{\nearrow\!\!\!\!\!\!\!Q}^{\lambda}{}_{\nu]\lambda}-\tilde{\nabla}_{\lambda}{\nearrow\!\!\!\!\!\!\!Q}_{[\mu\nu]}{}^{\lambda}-\frac{1}{2}\tilde{\nabla}_{[\mu}{\nearrow\!\!\!\!\!\!\!Q}_{\nu]\lambda}{}^{\lambda}+{\nearrow\!\!\!\!\!\!\!Q}_{[\nu\mu]\lambda}{\nearrow\!\!\!\!\!\!\!Q}_{\rho}{}^{\lambda\rho}-{\nearrow\!\!\!\!\!\!\!Q}_{\rho\lambda [\mu}{\nearrow\!\!\!\!\!\!\!Q}_{\nu] }{}^{\rho\lambda} + {\nearrow\!\!\!\!\!\!\!Q}_{\lambda\rho[\mu}T^{\lambda}{}_{\nu]}{}^{\rho}+ \frac{1}{4}{\nearrow\!\!\!\!\!\!\!Q}^{\lambda\rho}{}_{\rho} T_{\lambda\mu\nu} \,,\label{BB4}\\
    {\nearrow\!\!\!\!\!\!\!\tilde{R}}^{(T)}_{\lambda[\rho\mu\nu]}&=& \frac{1}{2} g_{\lambda[\rho|}\tilde{\nabla}_{\sigma}T^{\sigma}{}_{|\mu\nu]}+g_{\lambda[\rho}\tilde{\nabla}_{\mu}T^{\sigma}{}_{\nu]\sigma}-g_{\lambda\sigma}\tilde{\nabla}_{[\rho}T^{\sigma}{}_{\mu\nu]}+\frac{1}{24}\varepsilon_{\lambda\rho\mu\nu}\varepsilon_{\sigma}{}^{\alpha\beta\gamma}\bigl(\tilde{\nabla}_{\gamma}T^{\sigma}{}_{\beta\alpha}+T_{\beta\omega\gamma}T^{\omega\sigma}{}_{\alpha}\bigr)\nonumber\\
    &&+\,T_{\lambda\sigma[\rho}T^{\sigma}{}_{\mu\nu]}-\frac{1}{2}g_{\lambda[\rho}T^{\sigma}{}_{\mu\nu]} T^{\omega}{}_{\sigma\omega}\,,\label{BB5}\\
    {\nearrow\!\!\!\!\!\!\!\tilde{R}}^{(Q)}_{\lambda[\rho\mu\nu]} &=&\frac{3}{2}\Big( g_{\lambda[\rho|}\tilde{\nabla}_{\sigma}{\nearrow\!\!\!\!\!\!\!Q}_{|\mu\nu]}{}^{\sigma}-g_{\lambda[\rho}\tilde{\nabla}_{\mu}{\nearrow\!\!\!\!\!\!\!Q}^{\sigma}{}_{\nu]\sigma}-2\tilde{\nabla}_{[\rho}{\nearrow\!\!\!\!\!\!\!Q}_{\mu\nu]\lambda}+g_{\lambda[\rho}{\nearrow\!\!\!\!\!\!\!Q}_{\mu\nu]\sigma}{\nearrow\!\!\!\!\!\!\!Q}_{\omega}{}^{\sigma\omega}+g_{\lambda[\rho}{\nearrow\!\!\!\!\!\!\!Q}^{\sigma }{}_{\mu}{}^{\omega}{\nearrow\!\!\!\!\!\!\!Q}_{\nu]\sigma\omega}+{\nearrow\!\!\!\!\!\!\!Q}_{\sigma\lambda[\rho}T^{\sigma}{}_{\mu\nu]}\nonumber\\
    &&+\,g_{\lambda[\rho|}{\nearrow\!\!\!\!\!\!\!Q}_{\sigma|\mu|}{}^{\omega}T^{\sigma}{}_{\omega|\nu]}+\frac{1}{2}Q_{[\rho|\sigma}{}^{\sigma}{\nearrow\!\!\!\!\!\!\!Q}_{|\mu\nu]\lambda}\Big)\,,\label{BB6}\\  {}^{(1)}\tilde{W}_{\lambda\rho\mu\nu}&=&\tilde{R}_{[\lambda\rho]\mu\nu}-\frac{3}{4}\Big({\nearrow\!\!\!\!\!\!\!\tilde{R}}^{(T)}_{\lambda[\rho\mu\nu]}+{\nearrow\!\!\!\!\!\!\!\tilde{R}}^{(T)}_{\nu[\lambda\rho\mu]}-{\nearrow\!\!\!\!\!\!\!\tilde{R}}^{(T)}_{\rho[\lambda\mu\nu]}-{\nearrow\!\!\!\!\!\!\!\tilde{R}}^{(T)}_{\mu[\lambda\rho\nu]}\Bigr)-\frac{1}{2}\Bigl({\nearrow\!\!\!\!\!\!\!\tilde{R}}^{(Q)}_{\mu[\lambda\rho\nu]}-{\nearrow\!\!\!\!\!\!\!\tilde{R}}^{(Q)}_{\nu[\lambda\rho\mu]}\Big)+\,\frac{1}{24}\left.\ast\tilde{R}\right. \varepsilon_{\lambda\rho\mu\nu}\nonumber\\
&&-\,\frac{1}{4}\Big[g_{\lambda\mu}\left(2{\nearrow\!\!\!\!\!\!\!\tilde{R}}_{(\rho\nu)}+{\nearrow\!\!\!\!\!\!\!\hat{R}}^{(Q)}_{(\rho \nu) }\right)+g_{\rho\nu}\left(2{\nearrow\!\!\!\!\!\!\!\tilde{R}}_{(\lambda\mu)} +{\nearrow\!\!\!\!\!\!\!\hat{R}}^{(Q)}_{(\lambda\mu)}\right)-g_{\lambda\nu}\left(2 {\nearrow\!\!\!\!\!\!\!\tilde{R}}_{(\rho\mu)}+{\nearrow\!\!\!\!\!\!\!\hat{R}}^{(Q)}_{(\rho\mu)}\right)-g_{\rho\mu}\left(2{\nearrow\!\!\!\!\!\!\!\tilde{R}}_{(\lambda\nu)}+{\nearrow\!\!\!\!\!\!\!\hat{R}}^{(Q)}_{(\lambda\nu)}\right)\Big]\nonumber\\
&&-\,\frac{1}{4}\Big[g_{\lambda\mu}\Bigl(2\tilde{R}^{(T)}_{[\rho\nu]}+\hat{R}^{(Q)}_{[\rho\nu]}\Bigr)+g_{\rho\nu}\Bigl(2\tilde{R}^{(T)}_{[\lambda\mu]}+\hat{R}^{(Q)}_{[\lambda\mu]}\Bigr)-g_{\lambda\nu}\Bigl(2\tilde{R}^{(T)}_{[\rho\mu]}+\hat{R}^{(Q)}_{[\rho\mu]}\Bigr)-g_{\rho\mu}\Bigl(2\tilde{R}^{(T)}_{[\lambda\nu]}+\hat{R}^{(Q)}_{[\lambda\nu]}\Bigr)\nonumber\\
 &&+\,\tilde{R}^{\sigma}{}_{\sigma\lambda[\mu}g_{\nu]\rho}-\tilde{R}^{\sigma}{}_{\sigma\rho[\mu}g_{\nu]\lambda}\Big]-\frac{1}{6}\tilde{R}\,g_{\lambda[\mu}g_{\nu]\rho}\,,\label{BB7}\\     
   {}^{(1)}\tilde{Z}_{\lambda\rho\mu\nu} &=&\tilde{R}_{(\lambda\rho)\mu\nu}-\frac{1}{4}\Big({\nearrow\!\!\!\!\!\!\!\tilde{R}}^{(Q)}_{\lambda[\rho\mu\nu]}+{\nearrow\!\!\!\!\!\!\!\tilde{R}}^{(Q)}_{\rho[\lambda\mu\nu]}\Big)-\frac{1}{6}\Big(g_{\lambda\nu}\hat{R}^{(Q)}_{[\rho\mu]}+g_{\rho\nu}\hat{R}^{(Q)}_{[\lambda\mu]}-g_{\lambda\mu}\hat{R}^{(Q)}_{[\rho\nu]}-g_{\rho\mu}\hat{R}^{(Q)}_{[\lambda\nu]}+g_{\lambda\rho}\hat{R}^{(Q)}_{[\mu\nu]}\Big)\nonumber\\
 &&-\,\frac{1}{4}g_{\lambda\rho}\tilde{R}^{\sigma}{}_{\sigma\mu\nu}-\frac{1}{8}\Big(g_{\lambda\nu}{\nearrow\!\!\!\!\!\!\!\hat{R}}^{(Q)}_{(\rho\mu)}+g_{\rho\nu} {\nearrow\!\!\!\!\!\!\!\hat{R}}^{(Q)}_{(\lambda\mu)}-g_{\lambda\mu} {\nearrow\!\!\!\!\!\!\!\hat{R}}^{(Q)}_{(\rho\nu)}-g_{\rho\mu} {\nearrow\!\!\!\!\!\!\!\hat{R}}^{(Q)}_{(\lambda\nu)}\Big)\,,\label{BB8}\\
    \tilde{R}&=& R-2\nabla_{\mu}T^{\nu \mu}\,_{\nu}+\nabla_{\mu}Q^{\mu}\,_{\nu}\,^{\nu}-\nabla_{\mu}Q^{\nu}\,_{\nu}\,^{\mu}+\frac{1}{4}T_{\lambda \mu \nu}T^{\lambda \mu \nu}+\frac{1}{2}T_{\lambda \mu \nu}T^{\mu \lambda \nu}-T^{\lambda}\,_{\lambda\nu}T^{\mu}\,_{\mu}\,^{\nu}+T_{\lambda\mu\nu}Q^{\nu\lambda\mu}\nonumber\\
    &&+\,T^{\lambda}\,_{\lambda\nu}Q^{\mu\nu}\,_{\mu}-T^{\lambda}\,_{\lambda\nu}Q^{\nu\mu}\,_{\mu}+\frac{1}{4}Q_{\lambda\mu\nu}Q^{\lambda\mu\nu}-\frac{1}{2}Q_{\lambda\mu\nu}Q^{\mu\lambda\nu}+\frac{1}{2}Q^{\nu\lambda}\,_{\lambda}Q^{\mu}\,_{\mu\nu}-\frac{1}{4}Q^{\nu\lambda}\,_{\lambda}Q_{\nu}\,^{\mu}\,_{\mu}\,,\label{BB9}\\
\ast\tilde{R}&=&\varepsilon^{\lambda\rho\mu\nu}\Bigl(\nabla_{\lambda}T_{\rho\mu\nu}+\frac{1}{2}T^{\sigma}{}_{\lambda\rho}T_{\sigma\mu\nu}-Q_{\lambda\sigma\rho}T^{\sigma}{}_{\mu\nu}\Bigr)\,,\label{BB10}
\end{eqnarray}
which gives rise to six irreducible parts $\tilde{W}_{\lambda\rho\mu\nu}=\displaystyle\sum_{i=1}^{6}{}^{(i)}\tilde{W}_{\lambda\rho\mu\nu}$ in the antisymmetric component:
\begin{eqnarray}
{}^{(1)}\tilde{W}_{\lambda\rho\mu\nu}&=&\tilde{W}_{\lambda\rho\mu\nu}-\sum_{i=2}^{6}{}^{(i)}\tilde{W}_{\lambda\rho\mu\nu}\,,\label{first_irreducible_antisymmetric_piece}\\
{}^{(2)}\tilde{W}_{\lambda\rho\mu\nu}&=&\frac{3}{4}\Big({\nearrow\!\!\!\!\!\!\!\tilde{R}}^{(T)}_{\lambda[\rho\mu\nu]}+{\nearrow\!\!\!\!\!\!\!\tilde{R}}^{(T)}_{\nu[\lambda\rho\mu]}-{\nearrow\!\!\!\!\!\!\!\tilde{R}}^{(T)}_{\rho[\lambda\mu\nu]}-{\nearrow\!\!\!\!\!\!\!\tilde{R}}^{(T)}_{\mu[\lambda\rho\nu]}\Bigr)+\frac{1}{2}\Bigl({\nearrow\!\!\!\!\!\!\!\tilde{R}}^{(Q)}_{\mu[\lambda\rho\nu]}-{\nearrow\!\!\!\!\!\!\!\tilde{R}}^{(Q)}_{\nu[\lambda\rho\mu]}\Big)\,,\\
    {}^{(3)}\tilde{W}_{\lambda\rho\mu\nu}&=&-\,\frac{1}{24}\left.\ast\tilde{R}\right. \varepsilon_{\lambda\rho\mu\nu}\,,\\
     {}^{(4)}\tilde{W}_{\lambda\rho\mu\nu}&=&\frac{1}{4}\Big[g_{\lambda\mu}\left(2{\nearrow\!\!\!\!\!\!\!\tilde{R}}_{(\rho\nu)}+{\nearrow\!\!\!\!\!\!\!\hat{R}}^{(Q)}_{(\rho \nu) }\right)+g_{\rho\nu}\left(2{\nearrow\!\!\!\!\!\!\!\tilde{R}}_{(\lambda\mu)} +{\nearrow\!\!\!\!\!\!\!\hat{R}}^{(Q)}_{(\lambda\mu)}\right)-g_{\lambda\nu}\left(2 {\nearrow\!\!\!\!\!\!\!\tilde{R}}_{(\rho\mu)}+{\nearrow\!\!\!\!\!\!\!\hat{R}}^{(Q)}_{(\rho\mu)}\right)-g_{\rho\mu}\left(2{\nearrow\!\!\!\!\!\!\!\tilde{R}}_{(\lambda\nu)}+{\nearrow\!\!\!\!\!\!\!\hat{R}}^{(Q)}_{(\lambda\nu)}\right)\Big]\,,\nonumber \\ \\
{}^{(5)}\tilde{W}_{\lambda\rho\mu\nu}&=&\frac{1}{4}\Big[g_{\lambda\mu}\Bigl(2\tilde{R}^{(T)}_{[\rho\nu]}+\hat{R}^{(Q)}_{[\rho\nu]}\Bigr)+g_{\rho\nu}\Bigl(2\tilde{R}^{(T)}_{[\lambda\mu]}+\hat{R}^{(Q)}_{[\lambda\mu]}\Bigr)-g_{\lambda\nu}\Bigl(2\tilde{R}^{(T)}_{[\rho\mu]}+\hat{R}^{(Q)}_{[\rho\mu]}\Bigr)-g_{\rho\mu}\Bigl(2\tilde{R}^{(T)}_{[\lambda\nu]}+\hat{R}^{(Q)}_{[\lambda\nu]}\Bigr)\nonumber\\
 &&+\,\tilde{R}^{\sigma}{}_{\sigma\lambda[\mu}g_{\nu]\rho}-\tilde{R}^{\sigma}{}_{\sigma\rho[\mu}g_{\nu]\lambda}\Big]\,,\\
   {}^{(6)}\tilde{W}_{\lambda\rho\mu\nu}&=&\frac{1}{6}\,\tilde{R}\,g_{\lambda[\mu}g_{\nu]\rho}\,,\label{irreducible_pieces_antisymmetric_component}
\end{eqnarray}
as well as to five $\tilde{Z}_{\lambda\rho\mu\nu}=\displaystyle\sum_{i=1}^{5}{}^{(i)}\tilde{Z}_{\lambda\rho\mu\nu}$ in the symmetric one:
\begin{eqnarray}
 {}^{(1)}\tilde{Z}_{\lambda\rho\mu\nu}&=& \tilde{Z}_{\lambda\rho\mu\nu}-\sum_{i=2}^{5}{}^{(i)}\tilde{Z}_{\lambda\rho\mu\nu}\,,\label{first_irreducible_symmetric_piece}\\
       {}^{(2)}\tilde{Z}_{\lambda\rho\mu\nu}&=&\frac{1}{4}\Big({\nearrow\!\!\!\!\!\!\!\tilde{R}}^{(Q)}_{\lambda[\rho\mu\nu]}+{\nearrow\!\!\!\!\!\!\!\tilde{R}}^{(Q)}_{\rho[\lambda\mu\nu]}\Big)\,,\\
       {}^{(3)}\tilde{Z}_{\lambda\rho\mu\nu}&=&\frac{1}{6}\Big(g_{\lambda\nu}\hat{R}^{(Q)}_{[\rho\mu]}+g_{\rho\nu}\hat{R}^{(Q)}_{[\lambda\mu]}-g_{\lambda\mu}\hat{R}^{(Q)}_{[\rho\nu]}-g_{\rho\mu}\hat{R}^{(Q)}_{[\lambda\nu]}+g_{\lambda\rho}\hat{R}^{(Q)}_{[\mu\nu]}\Big)\,,\\
       {}^{(4)}\tilde{Z}_{\lambda\rho\mu\nu}&=&\frac{1}{4}g_{\lambda\rho}\tilde{R}^{\sigma}{}_{\sigma\mu\nu}\,,\\
       {}^{(5)}\tilde{Z}_{\lambda\rho\mu\nu}&=&\frac{1}{8}\Big(g_{\lambda\nu}{\nearrow\!\!\!\!\!\!\!\hat{R}}^{(Q)}_{(\rho\mu)}+g_{\rho\nu} {\nearrow\!\!\!\!\!\!\!\hat{R}}^{(Q)}_{(\lambda\mu)}-g_{\lambda\mu} {\nearrow\!\!\!\!\!\!\!\hat{R}}^{(Q)}_{(\rho\nu)}-g_{\rho\mu} {\nearrow\!\!\!\!\!\!\!\hat{R}}^{(Q)}_{(\lambda\nu)}\Big)\,.\label{irreducible_pieces_symmetric_component}
\end{eqnarray}

The resulting eleven irreducible parts of the curvature tensor can then be included in the general action of MAG to provide the dynamics of the gravitational field enhanced by torsion and nonmetricity~\cite{Hehl:1994ue}. Thereby, it is clear that these parts play a crucial role in MAG, which merits the study of their algebraic structure, in line with the analyses carried out for the Weyl and Ricci tensors in GR. In any case, further studies can also be focused on the torsion and nonmetricity tensors per se, which has already found applications in the particular framework of teleparallelism~\cite{Coley:2019zld,McNutt:2021tun}.

Thus, in order to perform the algebraic classification of the building blocks involved in the irreducible decomposition of the curvature tensor, it is first essential to take into account their algebraic symmetries. Specifically, the building block ${}^{(1)}\tilde{W}_{\lambda\rho\mu\nu}$ represents the Weyl tensor in the presence of torsion and nonmetricity, fulfilling the following algebraic symmetries:
\begin{align}
    {}^{(1)}\tilde{W}_{\lambda\rho\mu\nu}&=-\,{}^{(1)}\tilde{W}_{\rho\lambda\mu\nu}=-\,{}^{(1)}\tilde{W}_{\lambda\rho\nu\mu}\,,\label{Wsym1}\\
    {}^{(1)}\tilde{W}_{\lambda[\rho\mu\nu]}&={}^{(1)}\tilde{W}^{\lambda}{}_{\mu\lambda\nu}=0\,.\label{Wsym2}
\end{align}
On the other hand, the antisymmetrised building blocks ${\nearrow\!\!\!\!\!\!\!\tilde{R}}^{(T)}_{\lambda[\rho\mu\nu]}$ and ${\nearrow\!\!\!\!\!\!\!\tilde{R}}^{(Q)}_{\lambda[\rho\mu\nu]}$ are both completely traceless and pseudotraceless tensors:
\begin{align}
    g^{\lambda\rho}{\nearrow\!\!\!\!\!\!\!\tilde{R}}^{(T)}_{\lambda[\rho\mu\nu]}&=g^{\lambda\rho}{\nearrow\!\!\!\!\!\!\!\tilde{R}}^{(Q)}_{\lambda[\rho\mu\nu]}=0\,,\\
    \varepsilon^{\lambda\rho\mu\nu}{\nearrow\!\!\!\!\!\!\!\tilde{R}}^{(T)}_{\lambda[\rho\mu\nu]}&=\varepsilon^{\lambda\rho\mu\nu}{\nearrow\!\!\!\!\!\!\!\tilde{R}}^{(Q)}_{\lambda[\rho\mu\nu]}=0\,,
\end{align}
the symmetric building blocks ${\nearrow\!\!\!\!\!\!\!\tilde{R}}_{(\mu\nu)}$ and ${\nearrow\!\!\!\!\!\!\!\hat{R}}_{(\mu\nu)}^{(Q)}$ are also traceless:
\begin{equation}
    g^{\mu\nu}{\nearrow\!\!\!\!\!\!\!\tilde{R}}_{(\mu\nu)}=g^{\mu\nu}{\nearrow\!\!\!\!\!\!\!\hat{R}}_{(\mu\nu)}^{(Q)}=0\,,
\end{equation}
whereas $\tilde{R}^{(T)}_{[\mu\nu]}$, $\hat{R}^{(Q)}_{[\mu\nu]}$ and  $\tilde{R}^\lambda\,_{\lambda\mu\nu}$ are simply antisymmetric. Finally, the building block $^{(1)}\tilde{Z}_{\lambda\rho\mu\nu}$ also constitutes a traceless tensor, which additionally satisfies a cyclic property:
\begin{align}
    {}^{(1)}\tilde{Z}^{\lambda}{}_{\lambda\mu\nu}&={}^{(1)}\tilde{Z}^{\lambda}{}_{\mu\lambda\nu}=0\,,\label{AlgSym1Z}\\
    {}^{(1)}\tilde{Z}_{\lambda[\rho\mu\nu]}&=0\,.\label{AlgSym2Z}
\end{align}

As is clear, the aforementioned algebraic symmetries constrain the number of independent components of the building blocks, which for the case of a four-dimensional affinely connected metric space-time can be collected in Table~\ref{tab:buildingblocks}. Indeed, it turns out that the sets $\{{\nearrow\!\!\!\!\!\!\!\tilde{R}}^{(T)}_{\lambda[\rho\mu\nu]}, {\nearrow\!\!\!\!\!\!\!\tilde{R}}^{(Q)}_{\lambda[\rho\mu\nu]}, {\nearrow\!\!\!\!\!\!\!\tilde{R}}_{(\mu\nu)}, {\nearrow\!\!\!\!\!\!\!\hat{R}}_{(\mu\nu)}^{(Q)}\}$ and $\{\tilde{R}^{(T)}_{[\mu\nu]}, \hat{R}^{(Q)}_{[\mu\nu]}, \tilde{R}^\lambda\,_{\lambda\mu\nu}\}$ contain building blocks with $9$ and $6$ independent components, which already suggests their respective building blocks may obey the same type of algebraic classification.

Following these lines, in the next sections we shall see there exist in general four different types of algebraic classification in metric-affine geometry.

\begin{table*}
\begin{center}
    \begin{tabular}{| c | c |}
    \hline
    Building block & Number of independent components \\ \hline
    
    $^{(1)}\tilde{Z}_{\lambda\rho\mu\nu}$ & 30  \\ \hline
    
    $^{(1)}\tilde{W}_{\lambda\rho\mu\nu}$ & 10  \\ \hline
    ${\nearrow\!\!\!\!\!\!\!\tilde{R}}^{(T)}_{\lambda[\rho\mu\nu]}$ & 9  \\ \hline
    ${\nearrow\!\!\!\!\!\!\!\tilde{R}}^{(Q)}_{\lambda[\rho\mu\nu]}$ & 9  \\ \hline
    
    ${\nearrow\!\!\!\!\!\!\!\tilde{R}}_{(\mu\nu)}$ & 9  \\ \hline
    
    ${\nearrow\!\!\!\!\!\!\!\hat{R}}_{(\mu\nu)}^{(Q)}$ & 9  \\ \hline
    
    $\tilde{R}^{(T)}_{[\mu\nu]}$ & 6  \\ \hline
    
    $\hat{R}^{(Q)}_{[\mu\nu]}$ & 6  \\ \hline
    
    $\tilde{R}^\lambda\,_{\lambda\mu\nu}$ & 6  \\ \hline

    $\tilde{R}$ & 1  \\ \hline
    
    $\ast\tilde{R}$ & 1  \\ \hline
    \end{tabular}
\end{center}
\caption{Number of independent components of the building blocks.}
\label{tab:buildingblocks}
\end{table*}

\section{Algebraic classification of $^{(1)}\tilde{W}_{\lambda\rho\mu\nu}$}\label{sec:AlgClas1}

The fact that the tensor $^{(1)}\tilde{W}_{\lambda\rho\mu\nu}$ represents the Weyl part of the curvature tensor in the presence of torsion and nonmetricity, fulfilling the algebraic symmetries~\eqref{Wsym1} and~\eqref{Wsym2}, clearly points out that this tensor must obey the Petrov classification. Indeed, this classification can be derived by means of its PNDs, which requires to express the tensor in terms of a set of null vectors $l_{\mu}$, $k_{\mu}$, $m_{\mu}$ and $\bar{m}_{\mu}$ that satisfy the following pseudo-orthogonality and normalisation conditions\footnote{Note that our conventions for the null vectors and for other relevant quantities in algebraic classification, such as the complex scalars of a given tensor, differ throughout the paper from the ones considered in~\cite{Stephani:2003tm}.}:
\begin{eqnarray}
    k^\mu l_\mu&=&-\,m^\mu \bar{m}_\mu=1\,,\label{ort1}\\
    k^\mu m_\mu&=& k^\mu \bar{m}_\mu=l^\mu m_\mu= l^\mu \bar{m}_\mu=0\,,\\
    k^\mu k_\mu&=& l^\mu l_\mu= m^\mu m_\mu= \bar{m}^\mu \bar{m}_\mu=0\,.\label{ort3}
\end{eqnarray}
Thereby, the $10$ independent components of the tensor $^{(1)}\tilde{W}_{\lambda\rho\mu\nu}$ can be described by five complex scalars $\{\Sigma_{i}\}_{i=0}^{4}$ as
\begin{eqnarray}
\label{W1id}
    ^{(1)}\tilde{W}_{\lambda\rho\mu\nu}&=&-\,\frac{1}{2}\left(\Sigma_2+\bar{\Sigma}_2\right)\left(\{l_\lambda k_{\rho}l_{\mu}k_\nu\}+\{m_\lambda \bar{m}_{\rho}m_{\mu}\bar{m}_\nu\}\right)+\left(\Sigma_2-\bar{\Sigma}_2\right)\{l_\lambda k_{\rho}m_{\mu}\bar{m}_\nu\}\nonumber\\
    &&-\,\frac{1}{2}\left(\bar{\Sigma}_0\{k_{\lambda}m_{\rho}k_{\mu}m_\nu\}+\Sigma_0\{k_\lambda \bar{m}_{\rho}k_{\mu}\bar{m}_\nu\}\right)-\frac{1}{2}\left(\Sigma_4\{l_\lambda m_{\rho}l_{\mu}m_\nu\}+\bar{\Sigma}_4\{l_\lambda \bar{m}_{\rho}l_{\mu}\bar{m}_\nu\}\right)\nonumber\\
    &&+\,\left(\Sigma_2\{l_\lambda m_{\rho}k_{\mu}\bar{m}_\nu\}+\bar{\Sigma}_2\{l_\lambda \bar{m}_{\rho}k_{\mu}m_\nu\}\right)\nonumber\\
    &&-\,\bar{\Sigma}_1\left(\{l_\lambda k_{\rho}k_{\mu}m_\nu\}+\{k_\lambda m_{\rho}m_{\mu}\bar{m}_\nu\}\right)-\Sigma_1\left(\{l_\lambda k_{\rho}k_{\mu}\bar{m}_\nu\}+\{k_\lambda \bar{m}_{\rho}\bar{m}_{\mu}m_\nu\}\right)\nonumber\\
    &&+\,\Sigma_3\left(\{l_\lambda k_{\rho}l_{\mu}m_\nu\}-\{l_\lambda m_{\rho}m_{\mu}\bar{m}_\nu\}\right)+\bar{\Sigma}_3\left(\{l_\lambda k_{\rho}l_{\mu}\bar{m}_\nu\}-\{l_\lambda \bar{m}_{\rho}\bar{m}_{\mu}m_\nu\}\right)\,,
    \end{eqnarray}
where
\begin{align}\label{Sigma0}
  \Sigma_0=&-{}^{(1)}\tilde{W}_{\lambda\rho\mu\nu}l^\lambda m^\rho l^\mu m^\nu\,,\quad\Sigma_1=-\,{}^{(1)}\tilde{W}_{\lambda\rho\mu\nu}l^\lambda k^\rho l^\mu m^\nu\,,\quad\Sigma_2=-\,{}^{(1)}\tilde{W}_{\lambda\rho\mu\nu}l^\lambda m^\rho \bar{m}^\mu k^\nu\,,\\
   \Sigma_3=&-{}^{(1)}\tilde{W}_{\lambda\rho\mu\nu}l^\lambda k^\rho \bar{m}^\mu k^\nu\,,\quad\Sigma_4=-\,{}^{(1)}\tilde{W}_{\lambda\rho\mu\nu}k^\lambda \bar{m}^\rho k^\mu \bar{m}^\nu\,,\label{Sigma4}
\end{align}
and
\begin{equation}
  \{w_\lambda x_{\rho}y_{\mu}z_\nu\}=w_\lambda x_\rho y_\mu z_\nu-w_\lambda x_\rho z_\mu y_\nu-x_\lambda w_\rho y_\mu z_\nu+x_\lambda w_\rho z_\mu y_\nu+y_\lambda z_\rho w_\mu x_\nu-y_\lambda z_\rho x_\mu w_\nu-z_\lambda y_\rho w_\mu x_\nu+z_\lambda y_\rho x_\mu w_\nu\,.
\end{equation}

The PNDs can then be found by performing a rotation along the null vector $k^{\mu}$, given by a complex function $\epsilon$:
\begin{eqnarray}\label{complex_rot}
    k'_\mu=k_\mu\,,\quad m'_\mu=m_\mu+\epsilon \,k_\mu\,,\quad \bar{m}'_\mu=\bar{m}_\mu+\bar{\epsilon}\,k_\mu\,,\quad l'_\mu=l_\mu+\bar{\epsilon}\,m_\mu+\epsilon\,\bar{m}_\mu+\epsilon\bar{\epsilon} \,k_\mu\,,
\end{eqnarray}
which transforms the complex scalars as
\begin{eqnarray}
\Sigma'_4&=&\Sigma_4\,,\quad \Sigma'_3=\Sigma_3+\epsilon\,\Sigma_4\,,\quad \Sigma'_2=\Sigma_2+2\epsilon\,\Sigma_3+\epsilon^{2}\Sigma_4\,,\\
\Sigma'_1&=&\Sigma_1+3\epsilon\,\Sigma_2+3\epsilon^{2}\Sigma_3+\epsilon^{3}\Sigma_4\,,\\
\Sigma'_0&=&\Sigma_0+4\epsilon\,\Sigma_1+6\epsilon^{2}\Sigma_2+4\epsilon^{3}\Sigma_3+\epsilon^{4}\Sigma_4\,.
\end{eqnarray}
Thus, the different roots of the quartic polynomial equation $\Sigma'_{0}=0$ and their multiplicities provide the PNDs and their levels of alignment, respectively, which determines the algebraic types of the classification. In particular, it is possible to find a rotated null tetrad where they satisfy the following constraints\footnote{For simplicity, we omit the prime in the notation for each equivalence.}:
\begin{eqnarray}
    l_{[\sigma}{}^{(1)}\tilde{W}_{\lambda] \rho \mu [\nu}l_{\omega]}l^{\rho}l^{\mu}=0 &\iff& \Sigma_0=0\,,\label{equiv1W1}\\
    ^{(1)}\tilde{W}_{\lambda \rho \mu [\nu}l_{\omega]}l^{\rho}l^{\mu}=0 &\iff& \Sigma_0=\Sigma_1=0\,,\\
    ^{(1)}\tilde{W}_{\lambda \rho \mu [\nu}k_{\omega]}k^{\rho}k^{\mu}={}^{(1)}\tilde{W}_{\lambda \rho \mu [\nu}l_{\omega]}l^{\rho}l^{\mu}=0 &\iff& \Sigma_0=\Sigma_1=\Sigma_3=\Sigma_4=0\,,\\
    ^{(1)}\tilde{W}_{\lambda \rho \mu [\nu}l_{\omega]}l^{\mu}=0 &\iff& \Sigma_0=\Sigma_1=\Sigma_2=0\,,\\
    ^{(1)}\tilde{W}_{\lambda \rho \mu \nu}l^{\mu}=0 &\iff& \Sigma_0=\Sigma_1=\Sigma_2=\Sigma_3=0\,.\label{equiv5W1}
\end{eqnarray}

The algebraic classification of the tensor $^{(1)}\tilde{W}_{\lambda \rho \mu \nu}$ can then be outlined in Table~\ref{tab:Algebraictypes1}.

\begin{table*}
\begin{center}
    \begin{tabular}{| c | c | c | c|}
    \hline
    Algebraic type & Segre characteristic & Intrinsic characterisation \\ \hline
    
    I & $[1\,1\,1]$  & $l_{[\sigma}{}^{(1)}\tilde{W}_{\lambda] \rho \mu [\nu}l_{\omega]}l^{\rho}l^{\mu}=0$ \\ \hline
    II & $[2\,1]$  & $^{(1)}\tilde{W}_{\lambda \rho \mu [\nu}l_{\omega]}l^{\rho}l^{\mu}=0$ \\ \hline
    D & $[(1\,1)\,1]$ & $^{(1)}\tilde{W}_{\lambda \rho \mu [\nu}k_{\omega]}k^{\rho}k^{\mu}={}^{(1)}\tilde{W}_{\lambda \rho \mu [\nu}l_{\omega]}l^{\rho}l^{\mu}=0$ \\ \hline
    III & $[3]$ & $^{(1)}\tilde{W}_{\lambda \rho \mu [\nu}l_{\omega]}l^{\mu}=0$ \\ \hline
    N & $[(2\,1)]$ & $^{(1)}\tilde{W}_{\lambda \rho \mu \nu}l^{\mu}=0$ \\ \hline
    O & $[-]$ & $^{(1)}\tilde{W}_{\lambda \rho \mu \nu}=0$ \\ \hline
    \end{tabular}
\end{center}
\caption{Algebraic types for the tensor $^{(1)}\tilde{W}_{\lambda\rho\mu\nu}$.}
\label{tab:Algebraictypes1}
\end{table*}

\section{Algebraic classification of ${\nearrow\!\!\!\!\!\!\!\tilde{R}}^{(T)}_{\lambda[\rho\mu\nu]}, {\nearrow\!\!\!\!\!\!\!\tilde{R}}^{(Q)}_{\lambda[\rho\mu\nu]}, {\nearrow\!\!\!\!\!\!\!\tilde{R}}_{(\mu\nu)}$ and ${\nearrow\!\!\!\!\!\!\!\hat{R}}_{(\mu\nu)}^{(Q)}$}\label{sec:AlgClas2}

In order to classify the tensors ${\nearrow\!\!\!\!\!\!\!\tilde{R}}^{(T)}_{\lambda[\rho\mu\nu]}, {\nearrow\!\!\!\!\!\!\!\tilde{R}}^{(Q)}_{\lambda[\rho\mu\nu]}, {\nearrow\!\!\!\!\!\!\!\tilde{R}}_{(\mu\nu)}$ and ${\nearrow\!\!\!\!\!\!\!\hat{R}}_{(\mu\nu)}^{(Q)}$, it is first worthwhile to stress that ${\nearrow\!\!\!\!\!\!\!\tilde{R}}^{(T)}_{\lambda[\rho\mu\nu]}$ and ${\nearrow\!\!\!\!\!\!\!\tilde{R}}^{(Q)}_{\lambda[\rho\mu\nu]}$ can be expressed as second order symmetric and traceless tensors as follows:
\begin{align}
    {\nearrow\!\!\!\!\!\!\!\tilde{M}}_{\mu\nu}&=\frac{1}{6}\varepsilon_{(\mu}{}^{\lambda\rho\sigma}{\nearrow\!\!\!\!\!\!\!\tilde{R}}^{(T)}_{\nu)[\lambda\rho\sigma]}\,,\\
    {\nearrow\!\!\!\!\!\!\!\tilde{K}}_{\mu\nu}&=\frac{1}{6}\varepsilon_{(\mu}{}^{\lambda\rho\sigma}{\nearrow\!\!\!\!\!\!\!\tilde{R}}^{(Q)}_{\nu)[\lambda\rho\sigma]}\,.
\end{align}
Accordingly, all of the tensors ${\nearrow\!\!\!\!\!\!\!\tilde{R}}^{(T)}_{\lambda[\rho\mu\nu]}, {\nearrow\!\!\!\!\!\!\!\tilde{R}}^{(Q)}_{\lambda[\rho\mu\nu]}, {\nearrow\!\!\!\!\!\!\!\tilde{R}}_{(\mu\nu)}$ and ${\nearrow\!\!\!\!\!\!\!\hat{R}}_{(\mu\nu)}^{(Q)}$ can be ascribed to a set of second order symmetric and traceless tensors $\{{\nearrow\!\!\!\!\!\!\!\tilde{B}}^{(i)}_{\mu\nu}\}_{i=1}^{4}$. The algebraic classification can then be directly derived from the eigenvalue equations
\begin{equation}
    {\nearrow\!\!\!\!\!\!\!\tilde{B}}^{(i)a}{}_{b}v^{b}=\lambda v^{a}\,,
\end{equation}
in such a way that the corresponding characteristic polynomials are determined by the invariants
\begin{equation}
    \tilde{U}^{(i)}={\nearrow\!\!\!\!\!\!\!\tilde{B}}^{(i)a}{}_{b}{\nearrow\!\!\!\!\!\!\!\tilde{B}}^{(i)b}{}_{a}\,, \quad \tilde{V}^{(i)}={\nearrow\!\!\!\!\!\!\!\tilde{B}}^{(i)a}{}_{b}{\nearrow\!\!\!\!\!\!\!\tilde{B}}^{(i)b}{}_{c}{\nearrow\!\!\!\!\!\!\!\tilde{B}}^{(i)c}{}_{a}\,, \quad \tilde{W}^{(i)}={\nearrow\!\!\!\!\!\!\!\tilde{B}}^{(i)a}{}_{b}{\nearrow\!\!\!\!\!\!\!\tilde{B}}^{(i)b}{}_{c}{\nearrow\!\!\!\!\!\!\!\tilde{B}}^{(i)c}{}_{d}{\nearrow\!\!\!\!\!\!\!\tilde{B}}^{(i)d}{}_{a}\,,
\end{equation}
yielding
\begin{equation}\label{CharEqAlg2}
    \lambda^4-\frac{\tilde{U}^{(i)}}{2}\lambda^2-\frac{\tilde{V}^{(i)}}{3}\lambda+\frac{1}{8}\bigl[\bigl(\tilde{U}^{(i)}\bigr)^{2}-2\tilde{W}^{(i)}\bigr]=0\,.
\end{equation}

The multiplicities of the roots of the characteristic equation~\eqref{CharEqAlg2} turn out to be determined by different combinations of signs for the subsequent invariants~\cite{bona1992intrinsic,Senovilla:1999fa}:
\begin{equation}
    \tilde{U}^{(i)}_{*}=\big(\tilde{W}^{(i)}_{*}\big)^{3}-\bigl\{3\tilde{U}^{(i)}\tilde{W}^{(i)}_{*}+4\bigl[3\bigl(\tilde{V}^{(i)}\bigr)^{2}-\bigl(\tilde{U}^{(i)}\bigr)^{3}\bigr]\bigr\}^2\,, \quad \tilde{V}^{(i)}_{*}=2\tilde{U}^{(i)}-|\tilde{W}^{(i)}_{*}|^{1/2} \,, \quad \tilde{W}^{(i)}_{*}=7\bigl(\tilde{U}^{(i)}\bigr)^{2}-12\tilde{W}^{(i)}\,,
\end{equation}
which provides the well-known Segre classification described in Table~\ref{tab:Algebraictypes2}.

\begin{table*}
\begin{center}
    \begin{tabular}{| c | c |}
    \hline
    Segre characteristic & Invariants \\ \hline
    $[1,1\,1\,1]$  & $\tilde{U}^{(i)}_{*}, \tilde{V}^{(i)}_{*} > 0$  \\ \hline
    $[Z\,\bar{Z}\,1\,1]$  & $\tilde{U}^{(i)}_{*} < 0$  \\ \hline
    $[Z\,\bar{Z}\,(1\,1)]$  & $\tilde{U}^{(i)}_{*} = 0\,, \,\, \tilde{V}^{(i)}_{*} < 0\,, \,\,  \tilde{W}^{(i)}_{*} > 0$  \\ \hline
    $[2\,1\,1]$  & $\tilde{U}^{(i)}_{*} = 0\,, \,\, \tilde{V}^{(i)}_{*} > 0\,, \,\,  \tilde{W}^{(i)}_{*} > 0$  \\ \hline
    $[1,1\,(1\,1)]$  & $\tilde{U}^{(i)}_{*} = 0\,, \,\, \tilde{V}^{(i)}_{*} > 0\,, \,\,  \tilde{W}^{(i)}_{*} > 0$  \\ \hline
    $[(1,1)1\,1]$  & $\tilde{U}^{(i)}_{*} = 0\,, \,\, \tilde{V}^{(i)}_{*} > 0\,, \,\,  \tilde{W}^{(i)}_{*} > 0$  \\ \hline
    $[3\,1]$  & $\tilde{U}^{(i)}_{*}=\tilde{W}^{(i)}_{*}=0\,, \,\, \tilde{V}^{(i)}_{*} > 0$  \\ \hline
    $[(2\,1)\,1]$  & $\tilde{U}^{(i)}_{*}=\tilde{W}^{(i)}_{*}=0\,, \,\, \tilde{V}^{(i)}_{*} > 0$  \\ \hline
    $[(1,1\,1)\,1]$  & $\tilde{U}^{(i)}_{*}=\tilde{W}^{(i)}_{*}=0\,, \,\, \tilde{V}^{(i)}_{*} > 0$  \\ \hline
    $[1,(1\,1\,1)]$  & $\tilde{U}^{(i)}_{*}=\tilde{W}^{(i)}_{*}=0\,, \,\, \tilde{V}^{(i)}_{*} > 0$  \\ \hline
    $[2\,(1\,1)]$  & $\tilde{U}^{(i)}_{*}=\tilde{V}^{(i)}_{*}=0\,, \,\,  \tilde{W}^{(i)}_{*} > 0$  \\ \hline
    $[(1,1)\,(1\,1)]$  & $\tilde{U}^{(i)}_{*}=\tilde{V}^{(i)}_{*}=0\,, \,\,  \tilde{W}^{(i)}_{*} > 0$  \\ \hline
    $[(3\,1)]$  & $\tilde{U}^{(i)}_{*}=\tilde{V}^{(i)}_{*}=\tilde{W}^{(i)}_{*}=0$  \\ \hline
    $[(2\,1\,1)]$  & $\tilde{U}^{(i)}_{*}=\tilde{V}^{(i)}_{*}=\tilde{W}^{(i)}_{*}=0$  \\ \hline
    $[(1,1\,1\,1)]$  & $\tilde{U}^{(i)}_{*}=\tilde{V}^{(i)}_{*}=\tilde{W}^{(i)}_{*}=0$  \\ \hline
    \end{tabular}
\end{center}
\caption{Algebraic types for the tensor ${\nearrow\!\!\!\!\!\!\!\tilde{B}}^{(i)}_{\mu\nu}$.}
\label{tab:Algebraictypes2}
\end{table*}

\section{Algebraic classification of $\tilde{R}^{(T)}_{[\mu\nu]}, \hat{R}^{(Q)}_{[\mu\nu]}$ and $\tilde{R}^\lambda\,_{\lambda\mu\nu}$}\label{sec:AlgClas3}

The tensors $\tilde{R}^{(T)}_{[\mu\nu]}, \hat{R}^{(Q)}_{[\mu\nu]}$ and $\tilde{R}^\lambda\,_{\lambda\mu\nu}$ can be ascribed to a set of antisymmetric tensors $\{\tilde{X}^{(i)}_{[\mu\nu]}\}_{i=1}^{3}$, which in turn can be expressed in terms of the null vectors as
\begin{equation}
    \tilde{X}^{(i)}_{[\mu\nu]}=2\left[\Omega_{2}k_{[\mu}m_{\nu]}+\bar{\Omega}_{2}k_{[\mu}\bar{m}_{\nu]}-\Omega_{0}l_{[\mu}\bar{m}_{\nu]}-\bar{\Omega}_{0}l_{[\mu}m_{\nu]}-\left(\Omega_{1}+\bar{\Omega}_{1}\right)k_{[\mu}l_{\nu]}+\left(\Omega_{1}-\bar{\Omega}_{1}\right)m_{[\mu}\bar{m}_{\nu]}\right]\,,
\end{equation}
where
\begin{align}
    \Omega_{0}^{(i)}&=k^{[\mu}m^{\nu]}\tilde{X}^{(i)}_{[\mu\nu]}\,, \quad \Omega_{1}^{(i)}=\frac{1}{2}\Bigl(k^{[\mu}l^{\nu]}-m^{[\mu}\bar{m}^{\nu]}\Bigr)\tilde{X}^{(i)}_{[\mu\nu]}\,, \quad \Omega_{2}^{(i)}=-\,l^{[\mu}\bar{m}^{\nu]}\tilde{X}^{(i)}_{[\mu\nu]}\label{Omega}\,,
\end{align}
constitute three sets of complex scalars, each set $\{\Omega^{(i)}_{1},\Omega^{(i)}_{2},\Omega^{(i)}_{3}\}_{i=1}^{3}$ encoding the six independent components of the associated tensor.

Thereby, a rotation of the form~\eqref{complex_rot} transforms the complex scalars as
\begin{equation}
    \Omega^{(i)'}_0=\Omega^{(i)}_0\,,\quad \Omega^{(i)'}_1=\Omega^{(i)}_1+\bar{\epsilon}\,\Omega^{(i)}_0\,,\quad \Omega^{(i)'}_2=\Omega^{(i)}_2+2\bar{\epsilon}\,\Omega^{(i)}_1+\bar{\epsilon}^2\Omega^{(i)}_0\,,
\end{equation}
which allows the PNDs of the respective tensors to be found by obtaining the roots of the quadratic polynomial equations $\Omega^{(i)'}_2 = 0$, namely
\begin{equation}
    \Omega^{(i)}_2+2\bar{\epsilon}\,\Omega^{(i)}_1+\bar{\epsilon}^2\Omega^{(i)}_0=0\,.\label{quadraticpolOmega}
\end{equation}
The different multiplicities of the PNDs give rise to the algebraic types of the classification, which in this case can be characterised by the following constraints:
\begin{eqnarray}
    \bigl(\tilde{X}^{(i)}_{[\mu\nu]}l_{\lambda}-\tilde{X}^{(i)}_{[\mu\lambda]}l_{\nu}\bigr)l^{\mu}=0 &\iff& \Omega^{(i)}_2 = 0\,,\\
    \tilde{X}^{(i)}_{[\mu\nu]}l^\mu=0 &\iff& \Omega^{(i)}_1 = \Omega^{(i)}_2 = 0\,.
\end{eqnarray}

The algebraic classification of the tensor $\tilde{X}^{(i)}_{[\mu\nu]}$ can then be summarised in Table~\ref{tab:Algebraictypes3}.

\begin{table*}
\begin{center}
    \begin{tabular}{| c | c | c | c|}
    \hline
    Algebraic type & Segre characteristic &  Intrinsic characterisation \\ \hline
    I & $[1\,1]$ & $\bigl(\tilde{X}^{(i)}_{[\mu\nu]}l_{\lambda}-\tilde{X}^{(i)}_{[\mu\lambda]}l_{\nu}\bigr)l^{\mu}=0$ \\ \hline
    N & $[2]$ & $\tilde{X}^{(i)}_{[\mu\nu]}l^{\mu}=0$ \\ \hline
    O & $[-]$ & $\tilde{X}^{(i)}_{[\mu\nu]}=0$ \\ \hline
    \end{tabular}
\end{center}
\caption{Algebraic types for the tensor $\tilde{X}^{(i)}_{[\mu\nu]}$.}
\label{tab:Algebraictypes3}
\end{table*}

\section{Algebraic classification of $^{(1)}\tilde{Z}_{\lambda\rho\mu\nu}$}\label{sec:AlgClas4}

As pointed out in Sec.~\ref{sec:IrrRep}, the tensor $^{(1)}\tilde{Z}_{\lambda\rho\mu\nu}$ constitutes one of the irreducible parts of the symmetric component of the curvature tensor in the presence of torsion and nonmetricity. Therefore, besides being symmetric in the first pair of indices and antisymmetric in the last pair, it fulfils the algebraic symmetries~\eqref{AlgSym1Z} and~\eqref{AlgSym2Z}. In terms of null vectors $l_{\mu}$, $k_{\mu}$, $m_{\mu}$ and $\bar{m}_{\mu}$, the $30$ dof of such a tensor can then be distributed into $15$ complex scalars $\{\Delta_{i}\}_{i=0}^{14}$ as
\begin{align}
    ^{(1)}\tilde{Z}_{\lambda\rho\mu\nu}=&-2 \Delta_{0}[k_{\lambda }k_{\rho }k_{\mu }\bar{m}_{\nu }]-2\bar{\Delta}_{0}[k_{\lambda }k_{\rho }k_{\mu }m_{\nu }]+\displaystyle\frac{2}{3}\bigl(\Delta_{1}+\bar{\Delta}_{1}\bigr) \bigl(3\left[k_\lambda \bar{m}_\rho k_\mu m_\nu \right]-\left[k_\lambda k_\rho l_\mu k_\nu \right]\bigr)\nonumber\\
    &+2\bigl(\Delta_{1}-\bar{\Delta}_{1}\bigr) \bigl(\left[k_\lambda k_\rho m_\mu \bar{m}_\nu \right]+\left[k_\lambda \bar{m}_\rho k_\mu m_\nu \right]\bigr)+2 \Delta _2 [k_{\lambda }\bar{m}_{\rho }k_{\mu }\bar{m}_{\nu }]+2 \bar{\Delta }_2 [k_{\lambda }m_{\rho }k_{\mu }m_{\nu }]\nonumber\\
    &+2 \Delta _3 \bigl([k_{\lambda }m_{\rho }l_{\mu }k_{\nu }]+[k_{\lambda }m_{\rho }\bar{m}_{\mu }m_{\nu }]\bigr)+2 \bar{\Delta }_3 \bigl([k_{\lambda }\bar{m}_{\rho }l_{\mu }k_{\nu }]+[k_{\lambda }\bar{m}_{\rho }m_{\mu }\bar{m}_{\nu }]\bigr)\nonumber\\
     &-2\Delta_{4}\bigl(2 [l_{\lambda }k_{\rho }k_{\mu }\bar{m}_{\nu }]+[k_{\lambda }\bar{m}_{\rho }m_{\mu }\bar{m}_{\nu }]+[\bar{m}_{\lambda }\bar{m}_{\rho }k_{\mu }m_{\nu }]\bigr)-\frac{2}{3}\Delta_{5}[\bar{m}_{\lambda }\bar{m}_{\rho }k_{\mu }\bar{m}_{\nu }]\nonumber\\
    &-2\bar{\Delta}_{4}\bigl(2 [l_{\lambda }k_{\rho }k_{\mu }m_{\nu }]+[k_{\lambda }m_{\rho }\bar{m}_{\mu }m_{\nu }]+[m_{\lambda }m_{\rho }k_{\mu }\bar{m}_{\nu }]\bigr)-\frac{2}{3}\bar{\Delta}_{5}[m_{\lambda }m_{\rho }k_{\mu }m_{\nu }]\nonumber\\
    &-\frac{2}{3} \Delta _6 \bigl(3 [m_{\lambda }m_{\rho }l_{\mu }k_{\nu }]+[m_{\lambda }m_{\rho }\bar{m}_{\mu }m_{\nu }]\bigr)-\frac{2}{3} \bar{\Delta }_6 \bigl(3 [\bar{m}_{\lambda }\bar{m}_{\rho }l_{\mu }k_{\nu }]+[\bar{m}_{\lambda }\bar{m}_{\rho }m_{\mu }\bar{m}_{\nu }]\bigr)\nonumber\\
     &+2\Delta_{7}\bigl(2 [l_{\lambda }k_{\rho }m_{\mu }\bar{m}_{\nu }]+2 [l_{\lambda }\bar{m}_{\rho }k_{\mu }m_{\nu }]+ [m_{\lambda }\bar{m}_{\rho }m_{\mu }\bar{m}_{\nu }]- [l_{\lambda }k_{\rho }l_{\mu }k_{\nu }]\bigr)+\frac{2}{3}  \Delta _9 [m_{\lambda }m_{\rho }l_{\mu }m_{\nu }]\nonumber\\
    &-2\bar{\Delta}_{7}\bigl(2 [l_{\lambda }k_{\rho }m_{\mu }\bar{m}_{\nu }]+2 [m_{\lambda }\bar{m}_{\rho }l_{\mu }k_{\nu }]+[m_{\lambda }\bar{m}_{\rho }m_{\mu }\bar{m}_{\nu }]+[l_{\lambda }k_{\rho }l_{\mu }k_{\nu }]\bigr)+\frac{2}{3} \bar{\Delta }_9 [\bar{m}_{\lambda }\bar{m}_{\rho }l_{\mu }\bar{m}_{\nu }]\nonumber\\
    &+\frac{2}{3}\Delta_{8}\bigl(3[l_{\lambda }\bar{m}_{\rho }k_{\mu }\bar{m}_{\nu }]+[\bar{m}_{\lambda }\bar{m}_{\rho }m_{\mu }\bar{m}_{\nu }]\bigr)+\frac{2}{3} \bar{\Delta}_{8}\bigl(3 [l_{\lambda }m_{\rho }k_{\mu }m_{\nu }]+[m_{\lambda }m_{\rho }\bar{m}_{\mu }m_{\nu }]\bigr)\nonumber\\
    &+2 \Delta _{10} \bigl(2 [l_{\lambda }m_{\rho }l_{\mu }k_{\nu }]+[l_{\lambda }m_{\rho }\bar{m}_{\mu }m_{\nu }]+[m_{\lambda }m_{\rho }l_{\mu }\bar{m}_{\nu }]\bigr)+2 \bar{\Delta }_{10} \bigl(2 [l_{\lambda }\bar{m}_{\rho }l_{\mu }k_{\nu }]+[l_{\lambda }\bar{m}_{\rho }m_{\mu }\bar{m}_{\nu }]+[\bar{m}_{\lambda }\bar{m}_{\rho }l_{\mu }m_{\nu }]\bigr)\nonumber\\
    &-2 \Delta _{11} \bigl([l_{\lambda }l_{\rho }k_{\mu }\bar{m}_{\nu }]+[l_{\lambda }\bar{m}_{\rho }m_{\mu }\bar{m}_{\nu }]\bigr)-2 \bar{\Delta }_{11} \bigl([l_{\lambda }l_{\rho }k_{\mu }m_{\nu }]+[l_{\lambda }m_{\rho }\bar{m}_{\mu }m_{\nu }]\bigr)\nonumber\\
     &-2\Delta_{12}[l_{\lambda }m_{\rho }l_{\mu }m_{\nu }]+2\bigl(\Delta_{13}-\bar{\Delta}_{13}\bigr)\bigl(\left[l_\lambda \bar{m}_\rho l_\mu m_\nu \right]+\left[l_\lambda l_\rho m_\mu \bar{m}_\nu \right]\bigr)+2\Delta_{14}\left[l_\lambda l_\rho l_\mu m_\nu \right]\nonumber\\
    &-2\bar{\Delta}_{12} [l_{\lambda }\bar{m}_{\rho }l_{\mu }\bar{m}_{\nu }]-2\bigl(\Delta_{13}+\bar{\Delta}_{13}\bigr)\bigl(\left[l_\lambda l_\rho l_\mu k_\nu \right]+\left[l_\lambda \bar{m}_\rho l_\mu m_\nu \right]\bigr)+2\bar{\Delta}_{14}\left[l_\lambda l_\rho l_\mu \bar{m}_\nu \right]\,,\label{tensorZ1}
\end{align}
where
\begin{eqnarray}
    \Delta _0&=& {}^{(1)}\tilde{Z}_{\lambda \rho \mu \nu } l^{\lambda }l^{\rho }l^{\mu }m^{\nu }\,,\label{Delta0}\\
    \Delta _1&=& \frac{1}{2} {}^{(1)}\tilde{Z}_{\lambda \rho \mu \nu } \left(l^{\lambda }l^{\rho }l^{\mu }k^{\nu }-l^{\lambda }l^{\rho }m^{\mu }\bar{m}^{\nu }\right)\,,\\
    \Delta _2&=& {}^{(1)}\tilde{Z}_{\lambda \rho \mu \nu } l^{\lambda }m^{\rho }l^{\mu }m^{\nu }\,,\\
    \Delta _3&=& \frac{1}{2} {}^{(1)}\tilde{Z}_{\lambda \rho \mu \nu } \left(l^{\lambda }\bar{m}^{\rho }l^{\mu }k^{\nu }+l^{\lambda }\bar{m}^{\rho }\bar{m}^{\mu }m^{\nu }\right)\,,\\
    \Delta _4&=& \frac{1}{2} {}^{(1)}\tilde{Z}_{\lambda \rho \mu \nu } \left(l^{\lambda }m^{\rho }l^{\mu }k^{\nu }-l^{\lambda }m^{\rho }m^{\mu }\bar{m}^{\nu }\right)\,,\\
    \Delta _5&=& {}^{(1)}\tilde{Z}_{\lambda \rho \mu \nu } m^{\lambda }m^{\rho }l^{\mu }m^{\nu }\,,\\
    \Delta _6&=& \frac{1}{2} {}^{(1)}\tilde{Z}_{\lambda \rho \mu \nu } \left(\bar{m}^{\lambda }\bar{m}^{\rho }\bar{m}^{\mu }m^{\nu }+\bar{m}^{\lambda }\bar{m}^{\rho }l^{\mu }k^{\nu }\right)\,,\\
    \Delta _7&=& \frac{1}{2} {}^{(1)}\tilde{Z}_{\lambda \rho \mu \nu } \left(m^{\lambda }\bar{m}^{\rho }l^{\mu }k^{\nu }-m^{\lambda }\bar{m}^{\rho }m^{\mu }\bar{m}^{\nu }\right)\,,\\
    \Delta _8&=& -\,\frac{1}{2} {}^{(1)}\tilde{Z}_{\lambda \rho \mu \nu } \left(m^{\lambda }m^{\rho }m^{\mu }\bar{m}^{\nu }+m^{\lambda }m^{\rho }k^{\mu }l^{\nu }\right)\,,\\
    \Delta _9&=& -\,{}^{(1)}\tilde{Z}_{\lambda \rho \mu \nu } \bar{m}^{\lambda }\bar{m}^{\rho }k^{\mu }\bar{m}^{\nu }\,,\\
    \Delta _{10}&=& -\,\frac{1}{2} {}^{(1)}\tilde{Z}_{\lambda \rho \mu \nu } \left(k^{\lambda }\bar{m}^{\rho }k^{\mu }l^{\nu }-k^{\lambda }\bar{m}^{\rho }\bar{m}^{\mu }m^{\nu }\right)\,,\\
    \Delta _{11}&=& -\,\frac{1}{2}{}^{(1)}\tilde{Z}_{\lambda \rho \mu \nu } \left(k^{\lambda }m^{\rho }k^{\mu }l^{\nu }+k^{\lambda }m^{\rho }m^{\mu }\bar{m}^{\nu }\right)\,,\\\Delta _{12}&=& -\,{}^{(1)}\tilde{Z}_{\lambda \rho \mu \nu } k^{\lambda }\bar{m}^{\rho }k^{\mu }\bar{m}^{\nu }\,,\\\Delta _{13}&=& -\,\frac{1}{2} {}^{(1)}\tilde{Z}_{\lambda \rho \mu \nu } \left(k^{\lambda }k^{\rho }k^{\mu }l^{\nu }-k^{\lambda }k^{\rho }\bar{m}^{\mu }m^{\nu }\right)\,,\\
    \Delta _{14}&=&-\,{}^{(1)}\tilde{Z}_{\lambda \rho \mu \nu } k^{\lambda }k^{\rho }k^{\mu }\bar{m}^{\nu }\,,\label{Delta14}
\end{eqnarray}
and
\begin{equation}
    \left[x_\lambda y_\rho z_\mu w_\nu\right] =  x_{(\lambda} y_{\rho)} z_{[\mu}w_{\nu]}-y_{(\lambda} w_{\rho)} x_{[\mu} z_{\nu]}-x_{(\lambda} w_{\rho)}y_{[\mu} z_{\nu]}\,.
\end{equation}
Note that, by the interchange $l^\mu \leftrightarrow k^\mu$ (and $m^\mu \leftrightarrow \bar{m}^\mu$), the scalars $\Delta_{i}$ are in direct correspondence with $-\,\Delta_{14-i}$ for all $i\in\{0,1,2,3,4,5,6,7\}$. Thereby, any alignment property referred to $l^\mu$ based on the first set of complex scalars has a replica as an alignment property of $k^\mu$ by replacing with the corresponding ones of the second set.

Under the null rotation~\eqref{complex_rot}, defined by a complex function $\epsilon$ that keeps the null vector $k^{\mu}$ fixed, the scalars transform as
\begin{align}
    \Delta_{0}'=&\;\Delta _0+4 \epsilon  \Delta _1+2 \bar{\epsilon } \Delta _2+6 \epsilon ^2 \Delta _3+8 \epsilon  \bar{\epsilon }
   \Delta _4+\bar{\epsilon }^2 \Delta _5+4 \epsilon ^3 \Delta _6+12 \epsilon ^2 \bar{\epsilon } \Delta _7+4 \epsilon 
   \bar{\epsilon }^2 \Delta _8+\epsilon ^4 \Delta _9\nonumber\\
   &+8 \epsilon ^3 \bar{\epsilon } \Delta _{10}+6 \epsilon ^2
   \bar{\epsilon }^2 \Delta _{11}+2 \epsilon ^4 \bar{\epsilon } \Delta _{12}+4 \epsilon ^3 \bar{\epsilon }^2 \Delta
   _{13}+\epsilon ^4 \bar{\epsilon }^2 \Delta _{14}\,,\label{Delta'0}\\
   \Delta_1'=&\;\Delta _1+3 \epsilon  \Delta _3+2 \bar{\epsilon } \Delta _4+3 \epsilon ^2 \Delta _6+6 \epsilon  \bar{\epsilon }
   \Delta _7+\bar{\epsilon }^2 \Delta _8+\epsilon ^3 \Delta _9+6 \epsilon ^2 \bar{\epsilon } \Delta _{10}+3 \epsilon 
   \bar{\epsilon }^2 \Delta _{11}+2 \epsilon ^3 \bar{\epsilon } \Delta _{12}\nonumber\\
   &+3 \epsilon ^2 \bar{\epsilon }^2 \Delta
   _{13}+\epsilon ^3 \bar{\epsilon }^2 \Delta _{14}\,,\\
    \Delta_{2}'=&\;\Delta _2+4 \epsilon  \Delta _4+\bar{\epsilon } \Delta _5+6 \epsilon ^2 \Delta _7+4 \epsilon  \bar{\epsilon } \Delta _8+4 \epsilon ^3 \Delta
   _{10}+6 \epsilon ^2 \bar{\epsilon } \Delta _{11}+\epsilon ^4 \Delta _{12}+4 \epsilon ^3 \bar{\epsilon } \Delta _{13}+\epsilon ^4
   \bar{\epsilon } \Delta _{14}\,,\\
     \Delta_{3}'=&\;\Delta _3+2 \epsilon  \Delta _6+2 \bar{\epsilon } \Delta _7+\epsilon ^2 \Delta _9+4 \epsilon  \bar{\epsilon } \Delta _{10}+\bar{\epsilon }^2
   \Delta _{11}+2 \epsilon ^2 \bar{\epsilon } \Delta _{12}+2 \epsilon  \bar{\epsilon }^2 \Delta _{13}+\epsilon ^2 \bar{\epsilon }^2 \Delta
   _{14}\,,\\
      \Delta_{4}'=&\;\Delta _4+3 \epsilon  \Delta _7+\bar{\epsilon } \Delta _8+3 \epsilon ^2 \Delta _{10}+3 \epsilon  \bar{\epsilon }
   \Delta _{11}+\epsilon ^3 \Delta _{12}+3 \epsilon ^2 \bar{\epsilon } \Delta _{13}+\epsilon ^3 \bar{\epsilon }
   \Delta _{14}\,,\\
       \Delta_{5}'=&\;\Delta _5+4 \epsilon  \Delta _8+6 \epsilon ^2 \Delta _{11}+4 \epsilon ^3 \Delta _{13}+\epsilon ^4 \Delta _{14}\,,\\
        \Delta_{6}'=&\;\Delta _6+\epsilon  \Delta _9+2 \bar{\epsilon } \Delta _{10}+2 \epsilon  \bar{\epsilon } \Delta _{12}+\bar{\epsilon }^2 \Delta _{13}+\epsilon 
   \bar{\epsilon }^2 \Delta _{14}\,,\\
         \Delta_{7}'=&\;\Delta _7+2 \epsilon  \Delta _{10}+\bar{\epsilon } \Delta _{11}+\epsilon ^2 \Delta _{12}+2 \epsilon  \bar{\epsilon }
   \Delta _{13}+\epsilon ^2 \bar{\epsilon } \Delta _{14}\,,\\
          \Delta_{8}'=&\;\Delta _8+3 \epsilon  \Delta _{11}+3 \epsilon ^2 \Delta _{13}+\epsilon ^3 \Delta _{14}\,,\\
           \Delta_{9}'=&\;\Delta _9+2 \bar{\epsilon } \Delta _{12}+\bar{\epsilon }^2 \Delta _{14}\,,\\
            \Delta_{10}'=&\;\Delta _{10}+\epsilon  \Delta _{12}+\bar{\epsilon } \Delta _{13}+\epsilon  \bar{\epsilon } \Delta _{14}\,,\\
             \Delta_{11}'=&\;\Delta _{11}+2 \epsilon  \Delta _{13}+\epsilon ^2 \Delta _{14}\,,\\
              \Delta_{12}'=&\;\Delta _{12}+\bar{\epsilon } \Delta _{14}\,,\\
               \Delta_{13}'=&\;\Delta _{13}+\epsilon  \Delta _{14}\,,\\
                \Delta_{14}'=&\;\Delta _{14}\,.\label{Delta'14}
    \end{align}

An algebraic classification of the tensor $\Z_{\lambda\rho\mu\nu}$ can then be obtained by defining its PNDs and their levels of alignment. An alternative, which leads to a more refined classification, can also be achieved by establishing the levels of alignment with its superenergy tensor~\cite{Senovilla:1999xz,Senovilla:2010ej}. In this sense, we shall first apply in Sec.~\ref{ap:classi} the method of PNDs to the tensor $\Z_{\lambda\rho\mu\nu}$ alone, in order to derive the basic algebraic classification for this tensor, whereas in Sec.~\ref{sec:supere} we shall show the main refinements that arise when using its superenergy tensor.

\subsection{Algebraic classification by means of the PNDs and their levels of alignment}\label{ap:classi}

For any arbitrary tensor there is a well established definition of PND~\cite{Senovilla:2006gf}, also called {\em aligned null direction} or AND~\cite{Milson:2004jx,Milson:2004wr,Ortaggio:2012jd}, that depends on the index-symmetry properties of the tensor. For the case of the tensor $\Z_{\lambda\rho\mu\nu}$, this definition reads
\be\label{PND1}
l^\mu \mbox{ is a PND } \hspace{1cm} \Longleftrightarrow \hspace{1cm} \Z_{\lambda\rho\mu[\nu}l_{\sigma]}l^\lambda l^\rho l^\mu=0\,, 
\ee
for a (necessarily) null $l^\mu$. This implies that $l^\mu$ is somehow aligned with the structure of the tensor $\Z_{\lambda\rho\mu\nu}$, and it can be seen equivalent to the vanishing of the scalar $\Delta_0$ defined in Expression~\eqref{Delta0}:
\be\label{PND2}
l^\mu \mbox{ is a PND } \hspace{1cm} \Longleftrightarrow \hspace{1cm} \Delta_0=0\,. 
\ee
However, one immediately notices that the above PND condition~\eqref{PND2} is also satisfied for null vectors $l^\mu$ that comply with stricter conditions, such as for instance 
\begin{equation}
    \Z_{\lambda\rho\mu\nu}l^\lambda l^\rho l^\mu = 0\,,
\end{equation}
or even
\begin{equation}
    \Z_{\lambda\rho\mu\nu}l^\lambda = 0\,.
\end{equation}
These stricter conditions entail a {\em higher-order alignment} of $l^\mu$ with the tensor $\Z_{\lambda\rho\mu\nu}$. To provide a measure of the several levels of alignment, one introduces the notions of {\em boost weight} and {\em boost order} associated to any null direction~\cite{Ortaggio:2012jd,Milson:2004jx}. For a given null vector $l^\mu$ and introducing the null tetrad $\{l^\mu, k^\mu, m^\mu, \bar{m}^\mu\}$, one can associate an integer number $b$ to denote the boost weight of each of the complex scalars~\eqref{Delta0}--\eqref{Delta14} by counting each appearance of $l^\mu$ with a $+1$ and each appearance of $k^\mu$ with a $-1$. Concretely, $\Delta_0$ has $b=3$, $\Delta_1$ and $\Delta_2$ have $b=2$, $\Delta_3,\Delta_4$ and $\Delta_5$ have $b=1$ and so on until $\Delta_{14}$ with $b=-3$. The boost order of the tensor $\Z_{\lambda\rho\mu\nu}$ with respect to any null vector $l^\mu$, say $bo(l)$, is then given by the maximum $b$ of the complex scalars in the null tetrad $\{l^\mu, k^\mu, m^\mu, \bar{m}^\mu\}$. This is independent of the choice of $k^\mu$.

One immediately realises that, for a general $l^\mu$, $bo(l)=3$; but, if $l^\mu$ happens to be a PND, then $bo(l)<3$. And thus the higher orders of alignment can be simply defined by all the possibilities for $bo(l)$ in order. Hence,
$l^\mu$ is said to be a PND of multiplicity $m\in\{1,2,3,4,5,6\}$ if $bo(l) =3-m$. This leads to the classification of PNDs, as follows:
\begin{alignat}{3}
\textbf{Alignment Class I:} &\quad  bo(l)=2; &\quad  m=1.\nonumber\\
\textbf{Alignment Class II:} & \quad  bo(l)=1; & \quad m=2.\nonumber\\
\textbf{Alignment Class III:} &\quad  bo(l)=0; & \quad m=3.\nonumber\\
\textbf{Alignment Class IV:} &\quad  bo(l)=-\,1; & \quad m=4.\nonumber\\
\textbf{Alignment Class V:} & \quad bo(l)=-\,2; & \quad m=5.\nonumber\\
\textbf{Alignment Class VI:} &\quad  bo(l)=-\,3; &\quad  m=6.\nonumber
\end{alignat}
Observe that the Roman numerals denoting the alignment class agree with the value of the multiplicity $m$ of $l^\mu$. Each of the alignment classes can be expressed by invariant conditions involving only $l^\mu$ and the tensor $\Z_{\lambda\rho\mu\nu}$, which can be deduced by taking into account the corresponding complex scalars that vanish for each case. Thus, we have the following equivalences\footnote{See Eq.~\eqref{case0d} for Class I,~\eqref{case3d} for Class II,~\eqref{case7d} for Class III,~\eqref{case11b} and~\eqref{case11c} for Class IV,\eqref{case15b} for Class V and~\eqref{case17d} for Class VI.}:
\begin{alignat}{3}
   \textbf{Alignment Class I:} &\quad \Z_{\lambda\rho\mu[\nu}l_{\sigma]}l^\lambda l^\rho l^\mu = 0 &\quad \Longleftrightarrow \quad & \Delta_0 = 0\,. \\[10pt]
  \textbf{Alignment Class II:} &\quad l_{[\omega} \Z_{\lambda]\rho\mu[\nu}l_{\sigma]}l^\rho l^\mu = 0 &\quad \Longleftrightarrow \quad & \Delta_0 = \Delta_1 = \Delta_2 = 0\,. \\[10pt] 
    \textbf{Alignment Class III:} &\quad l^{[\tau}l_{[\omega} \Z^{\lambda]}{}_{\rho]\mu[\nu}l_{\sigma]}l^{\mu} = 0 &\quad \Longleftrightarrow \quad & \{\Delta_i\}^{i=0,\dots,5} = 0\,. \\[10pt]
   \textbf{Alignment Class IV:} &\quad l_{[\omega} \Z_{\lambda]\rho\mu[\nu}l_{\sigma]}l^\mu = l_{\lambda}l_{[\omega} \Z^\lambda{}_{\rho]\mu\nu} = 0 &\quad \Longleftrightarrow \quad & \{\Delta_i\}^{i=0,\dots,8} = 0\,. \\[10pt]
    \textbf{Alignment Class V:} &\quad l^{[\tau}l_{[\omega} \Z^{\lambda]}{}_{\rho]\mu\nu} = 0 &\quad \Longleftrightarrow \quad & \{\Delta_i\}^{i=0,\dots,11}  = 0\,. \\[10pt]  
    \textbf{Alignment Class VI:} &\quad l_{[\sigma} \Z_{\lambda]\rho\mu\nu} = 0 &\quad \Longleftrightarrow \quad &  \{\Delta_i\}^{i=0,\dots,13}  = 0\,.
\end{alignat}

The classification of the tensor $\Z_{\lambda\rho\mu\nu}$ is then given by two natural numbers --- that we will express in Roman numerals--- in conjunction: the first one is the maximal alignment class of any null vector, and the second one is the next to maximal alignment class\footnote{Note that these two natural numbers, describing the principal and secondary alignment classes, are denoted as (PAT, SAT) in~\cite{Ortaggio:2012jd}.}. In other words, the first natural number gives the multiplicity of the maximal aligned null vector, while the second one is the multiplicity of the next to maximal aligned null vector. Thus, the first number is always greater than or equal to the second one. But there is a further important restriction: the sum of the two numbers cannot be greater than 6. This follows because, by choosing the null tetrad with $l^\mu$ the maximally aligned and $k^\mu$ the next to maximally aligned null vectors, and using the property mentioned above of the symmetry interpretation between $\Delta_i$ and $-\,\Delta_{14-i}$ for all $i\in\{0,1,2,3,4,5,6,7\}$, if the sum of the two numbers were greater than 6, the tensor $\Z_{\lambda\rho\mu\nu}$ would necessarily vanish identically ---because all the $\Delta$-scalars would be zero.

Thereby, besides the trivial case where the tensor vanishes, the fundamental classification ends up having 11 {\em main} different types, though some of them have special situations where the second PND does not exist, in which case we will denote them with a star * added to the main type, so that in total there are 15 nontrivial types as follows\footnote{The choice of names here differs from the typical one in the classification based on ANDs~\cite{Ortaggio:2012jd}, where Types M and K will be termed as Type II, and Types B, S, and C will be called Type III. However, that terminology is adapted to the cases where the maximum boost order for null directions is 2, but it is not appropriate for higher boost orders. We have kept the standard nomenclature for Types N, D and I, though, due to its importance and intrinsic characterisation.}:
\begin{eqnarray*}
\mbox{\textrm{Type N} or null:} & (\textrm{VI},-);\\
\mbox{\textrm{Type L}:} & (\textrm{V},\textrm{I});\\
\mbox{\textrm{Type L*}:} & (\textrm{V},-);\\
\mbox{\textrm{Type F}:} & (\textrm{IV},\textrm{II});\\
\mbox{\textrm{Type H}:} & (\textrm{IV},\textrm{I});\\
\mbox{\textrm{Type H*}:} & (\textrm{IV},-);\\
\mbox{\textrm{Type D}:} & (\textrm{III},\textrm{III});\\
\mbox{\textrm{Type M}:} & (\textrm{III},\textrm{II});\\
\mbox{\textrm{Type K}:} & (\textrm{III},\textrm{I});\\
\mbox{\textrm{Type K*}:} & (\textrm{III},-);\\
\mbox{\textrm{Type B}:} & (\textrm{II},\textrm{II});\\
\mbox{\textrm{Type S} or special:} & (\textrm{II},\textrm{II},\textrm{II});\\
\mbox{\textrm{Type C}:} & (\textrm{II},\textrm{I});\\
\mbox{\textrm{Type C*}:} & (\textrm{II},-);\\
\mbox{\textrm{Type I}:} & (\textrm{I},\textrm{I}).\\
%\mbox{type I*:} & (I,-);\\
\end{eqnarray*}
The left numeral is always uniquely fixed except for Type D, where both III can be interchanged, and analogously for Types B and I. However, the second Roman numeral can be, in some of the types, chosen in many different ways. This will be analysed in the next section. However, the type S (that can be seen as a Type B-special) is of a different kind, because there are {\em three} different PNDs of Class II, and thereby there are three different choices for (II,II). This is why this case, though it could be considered as a subcase of Type B,  is included as a different one in the classification. Let us remark that all the cases with * are actually very peculiar, for being uncommon, and they require extremely specific relations between some of the scalars $\Delta_n$. This will be made plain in Sec.~\ref{subsec:Full}.

Each of the types has its own different properties, as well as several subcases and various possibilities. This is partly discussed in~\ref{superenergyappendix}, where several refinements arise naturally, and also in the next subsections where we discuss how many choices for the second numeral can be, as well as how many PNDs there can be in general.

\subsubsection{On the number and multiplicities of the PNDs}

The previous part of the classification deals with the alignment classes of given PNDs and takes care of their level of alignment with the tensor $\Z_{\lambda\rho\mu\nu}$. However, in order to get a complete view of the algebraic classification of this tensor, the possible number of PNDs should be known, as well as their alignment classes.

To that end, one needs to see how many possible null directions satisfy the relation~\eqref{PND1} or, equivalently,~\eqref{pnd}. This can be achieved by choosing any null tetrad $\{l^\mu, k^\mu, m^\mu, \bar{m}^\mu\}$, then performing an {\em arbitrary} null rotation of type~\eqref{complex_rot} so that the new $l'{}^{\mu}$, that depends on $\epsilon$, represents any possible null direction --except $k^\mu$--, and then finding which of them, that is to say, for which values of $\epsilon$ this new $l'{}^{\mu}$ is a PND. In other words, one needs to ascertain the number of solutions for the complex parameter $\epsilon$ of the equation $\Delta'_0 =0$. According to Expression~\eqref{Delta'0}, this equation reads explicitly
\begin{align}
    &\Delta _0+4 \epsilon  \Delta _1+2 \bar{\epsilon } \Delta _2+6 \epsilon ^2 \Delta _3+8 \epsilon  \bar{\epsilon }
   \Delta _4+\bar{\epsilon }^2 \Delta _5+4 \epsilon ^3 \Delta _6+12 \epsilon ^2 \bar{\epsilon } \Delta _7+4 \epsilon 
   \bar{\epsilon }^2 \Delta _8\nonumber\\
   &+\epsilon ^4 \Delta _9+8 \epsilon ^3 \bar{\epsilon } \Delta _{10}+6 \epsilon ^2
   \bar{\epsilon }^2 \Delta _{11}+2 \epsilon ^4 \bar{\epsilon } \Delta _{12}+4 \epsilon ^3 \bar{\epsilon }^2 \Delta
   _{13}+\epsilon ^4 \bar{\epsilon }^2 \Delta _{14}=0\,\label{EqtoSolve}.
\end{align}
This is a polynomial relation of total degree 6 involving the complex variable $\epsilon$ and its complex conjugate $\bar{\epsilon}$. There does not seem to be any general result in the mathematical literature for such types of polynomial equations\footnote{Notice that one can equivalently write Eq.~\eqref{EqtoSolve} as a system of two polynomial equations in the real variables $\{x,y\}$ where $\epsilon=x+i y$. However, all known results for such type of systems concerns {\em complex} roots for $\{x,y\}$, and there are no satisfactory results for the case of real solutions.}. A possible way to proceed consists of considering Eq.~\eqref{EqtoSolve}, together with its own complex conjugate equation $\bar\Delta'_0=0$, as a system of two polynomial equations in the variables $\{\epsilon,\bar\epsilon\}$. Unfortunately, all relevant results about such systems provide the number of solutions, counted with its multiplicity, for the case where the two variables are considered to be fully independent~\cite{walker1950algebraic,bernshtein1975number,sturmfels2002solving}. For cases as ours, where they are mutually complex conjugate, not even a definition of multiplicity is available.

That said, obviously the number of actual solutions will always be less or equal than the total number of solutions with the two independent complex variables. Hence, the number of the latter will provide a bound for the number of solutions of interest. Concerning the multiplicity, even though there is no mathematical accepted definition for such, we will consider in our case that the multiplicity will be the same as that arising as solutions of the system when the variables are assumed to be independent.

To fix ideas, let us consider a few simple examples. Imagine the situation is such that all the scalars are zero, except $\Delta_{14}\neq 0$. Then, Eq.~\eqref{EqtoSolve} reduces to simply
\be\label{EqforTypeN}
\epsilon^4 \bar\epsilon^2 \Delta_{14} =0\,.
\ee
In that case, renaming 
\be\label{rename}
\bar\epsilon \longrightarrow z\,,
\ee
and considering the pair $\{\epsilon,z\}$ as two independent complex variables, Eq.~\eqref{EqforTypeN} and its complex conjugate are rewritten as
\begin{equation}
\epsilon^4 z^2 \Delta_{14} =0\,, \hspace{1cm}
z^4 \epsilon^2 \bar\Delta_{14} =0\,,
\end{equation}
which have an infinite number of solutions, given by 
\begin{equation}
\left(\epsilon =0, z\right)\,,
\end{equation}
with $z$ arbitrary, and also by
\begin{equation}
\left(\epsilon, z=0\right)\,,
\end{equation}
with $\epsilon$ arbitrary. Among such a huge number of solutions, only those with $z=0$ in the first case, and only those with $\epsilon =0$ in the second case, satisfy the constraint that $z=\bar\epsilon$. Thus, the solution of the original equation is unique, given by $\epsilon =0$. Its multiplicity can be guessed by noting that Eq.~\eqref{EqforTypeN} can also be written as
\begin{equation}
    |\epsilon|^6 e^{2i\phi}\Delta_{14} =0\,,
\end{equation}
where (here and later on) $\phi$ is the phase of $\epsilon$, which leads to $|\epsilon|=0$ six times, ergo multiplicity 6.

As a second and more interesting example, let the case be such that only $\Delta_0\neq 0\neq \Delta_{14}$ are nonzero, all other $\Delta$-scalars vanish. Using the renaming~\eqref{rename}, Eq.~\eqref{EqtoSolve} collapses to simply
\be\label{EqforTypeG}
\Delta_0 +\epsilon^4 z^2 \Delta_{14} =0\,,
\ee
while the complex conjugate of~\eqref{EqtoSolve} reads
\be
\bar\Delta_0 +\epsilon^2 z^4 \bar\Delta_{14} =0\,.
\ee
The solutions to this pair of equations can be easily obtained by the method of substitution, and they are
\be\label{0-14}
\epsilon_k = \left|\frac{\Delta_0}{\Delta_{14}} \right|^{1/6} e^{i(\phi_0-\phi_{14})/2 +i (2k+1)\pi/6}\,,\hspace{5mm}
z_k =\pm \left|\frac{\Delta_0}{\Delta_{14}} \right|^{1/6} e^{i(\phi_{14}-\phi_0)/2 -i (4k+5)\pi/6}\,,
\ee
for all $k\in\{0,1,2,3,4,5\}$, where $\phi_n$ will denote the phase of any $\Delta$-scalar:
\begin{equation}
    \Delta_n = |\Delta_n| e^{i \phi_n} , \hspace{5mm}n\in\{0,1,2,3,4,5,6,7,8,9,10,11,12,13,14\}.
\end{equation}
Thus, there are 12 solutions in total, but only two proper solutions of Eq.~\eqref{EqforTypeG} (i.e. with $z_k =\bar\epsilon_k$), which are given by $k=1$ for the $-$ sign and by $k=4$ for the $+$ sign. Hence, both solutions have simple multiplicity and read
\begin{equation}
    \epsilon_{1}=i\left|\frac{\Delta_0}{\Delta_{14}} \right|^{1/6} e^{i(\phi_0-\phi_{14})/2}\,,\quad\epsilon_{4}=-\,i\left|\frac{\Delta_0}{\Delta_{14}} \right|^{1/6} e^{i(\phi_0-\phi_{14})/2}\,.\label{sol_eps_1and4}
\end{equation}
Note that the only vanishing scalars in the null tetrads provided by the complex rotations of value $\epsilon_1$ and $\epsilon_4$ are $\Delta'_{0}(\epsilon_1)$ and $\Delta'_{0}(\epsilon_4)$, respectively, but we can always choose a different null tetrad where both rotated scalars $\Delta'_{0}$ and $\Delta'_{14}$ vanish:
\begin{eqnarray}
     L_{\mu}&=&l'_{\mu}(\epsilon_{1})\,,\quad K_{\mu}=\frac{1}{4}\left|\frac{\Delta_{14}}{\Delta_{0}} \right|^{1/3}l'_{\mu}(\epsilon_{4})\,,\\
    M_{\mu}&=&m_\mu'(\epsilon_4)+\frac{l'_\mu(\epsilon_4)}{\bar{\epsilon}_1-\bar{\epsilon}_4}\,, \quad \bar{M}_{\mu}=\bar{m}_\mu'(\epsilon_4)+\frac{l'_\mu(\epsilon_4)}{\epsilon_1-\epsilon_4}\,,
\end{eqnarray}
in such a way that $L_{\mu}$ and $K_{\mu}$ would constitute the corresponding PNDs under that choice.

Another example, which will be of interest later for the Type L in the classification, arises when the only nonzero complex scalars are $\Delta_{12}, \Delta_{13}$ and $\Delta_{14}$. Eq.~\eqref{EqtoSolve} then collapses to 
\be\label{EqfortypeL}
\epsilon^3 \bar\epsilon\bigl(2\epsilon \Delta_{12} +4\bar\epsilon \Delta_{13} +\epsilon\bar\epsilon\Delta_{14}\bigr) =0\,.
\ee
Passing to the independent variables by~\eqref{rename}, we have
for Eq.~\eqref{EqfortypeL} and its complex conjugate
\begin{align}
    \epsilon^3 z\bigl(2\epsilon \Delta_{12} +4z \Delta_{13} +\epsilon z\Delta_{14}\bigr) =&\;0\,,\\
    z^3 \epsilon \bigl(2z \bar\Delta_{12}+4\epsilon \bar\Delta_{13}+ z\epsilon \bar\Delta_{14}\bigr)=&\;0\,.
\end{align}
There are obvious solutions $\left(\epsilon=0,z\right)$ with arbitrary $z$, as well as $\left(\epsilon, z=0\right)$ with arbitrary $\epsilon$. Furthermore, there are also solutions of the system
\begin{equation}
2\epsilon \Delta_{12} +4z \Delta_{13} +\epsilon z\Delta_{14} =0\,,\hspace{5mm}
2z \bar\Delta_{12}+4\epsilon \bar\Delta_{13}+ z\epsilon \bar\Delta_{14}=0\,.
\end{equation}
These can be computed easily and are given by $\left(\epsilon=0,z=0\right)$ and by
\be\label{solTypeL}
z=\bar\epsilon = 2\frac{4\Delta_{13}\bar\Delta_{13} -\Delta_{12}\bar\Delta_{12}}{\Delta_{14}\bar\Delta_{12} -2\Delta_{13}\bar\Delta_{14}}\,.
\ee
Notice that this solution is a proper solution of the original Eq.~\eqref{EqfortypeL}. In summary, there is a solution with $\epsilon =0$, providing a PND $l_\mu$ of Class V, plus a (generally) {\em unique} second PND given by $l'_\mu$ in~\eqref{complex_rot} with $\epsilon$ as in~\eqref{solTypeL}. However, there is a special situation, because the above unique extra solution is defined only if
$\Delta_{14}\bar\Delta_{12} -2\Delta_{13}\bar\Delta_{14}\neq 0$. For the very special case where
\begin{equation}
    \Delta_{14}\bar\Delta_{12} -2\Delta_{13}\bar\Delta_{14}=0\,,\label{very_special_case}
\end{equation}
this condition readily entails $|\Delta_{12}|=2|\Delta_{13}|$ and the numerator on~\eqref{solTypeL} vanishes too. In that case, there will be no further solutions in general, leading to Type L*. One can check that, nevertheless, in this special situation there are extra solutions whenever $\Delta_{12}=\bar\Delta_{12}$, given by
\begin{equation}
    \epsilon =-\,8\frac{\Delta_{13}}{\Delta_{14}} a e^{i\arccos a}\,,\label{extrasol_specialsituation}
\end{equation}
where $a$ is a nonzero real number satisfying $-1\leq a\leq 1$. Thus, the infinite values that the parameter $a$ can take within the interval $[-1,1]$ provide an infinite number of solutions. Thereby, we find a behaviour that will arise in some other extremely special situations: {\em there may be an infinite number of PNDs}. These very exceptional cases will be generally termed as ``exceptional'' within the corresponding algebraic type, and will carry a subindex `e'. For instance, the case just analysed belongs to the Type L and will be denoted by
Type L$_e$ or, alternatively, by (V,I$_{\infty}$).

In a general and generic situation, however, the scalars $\Delta_n$ will take arbitrary complex nonzero values without any relation between them, so that the original Eq.~\eqref{EqtoSolve} has to be considered. By the renaming~\eqref{rename}, we can rearrange Eq.~\eqref{EqtoSolve} as
\begin{equation}
\Delta_0'=p_0(\epsilon)+z p_1(\epsilon)+z^2p_2(\epsilon)=0\,,
\end{equation}
with 
\begin{align}
    p_0(\epsilon)&=\Delta_{0}+4 \Delta_{1} \epsilon +6  {\Delta}_{3}\epsilon ^2 +4  \Delta_6 \epsilon ^3 +{\Delta}_{9} \epsilon ^4\,,\\
     p_1(\epsilon)&=2 {\Delta}_{12} \epsilon ^4+8\Delta_{10} \epsilon ^3 +12 \Delta_{7} \epsilon ^2+8\Delta_4 \epsilon +2 {\Delta}_{2}\,,\\
     p_2(\epsilon)&= {\Delta}_{14} \epsilon ^4+4 \Delta_{13} \epsilon ^3+6\Delta_{11} \epsilon ^2 +4\Delta_8 \epsilon  +\Delta_{5}\,,
\end{align}
while its complex  conjugate equation yields
\begin{equation}
  \bar{\Delta}_0'=q_0(\epsilon)+  q_1(\epsilon) z+q_2(\epsilon) z^2+q_3(\epsilon) z^3+q_4(\epsilon) z^4\,,
\end{equation}
where 
\begin{align}
    q_0(\epsilon)&=\bar{\Delta}_{0}+2 \bar\Delta_{2} \epsilon +\bar{\Delta}_{5} \epsilon ^2\,,\\
    q_1(\epsilon)&=4 \bar{\Delta}_{1}+8\bar \Delta_4 \epsilon +4  \bar{\Delta}_{8} \epsilon ^2\,,\\
    q_2(\epsilon)&=6 \epsilon ^2 \bar{\Delta}_{11}+12 \bar{\Delta}_{7} \epsilon +6 \bar\Delta_{3}\,,\\
    q_3(\epsilon)&=4 \bar\Delta_{6}+8  \bar\Delta_{10}  \epsilon +4 \bar{\Delta}_{13} \epsilon ^2\,,\\
    q_4(\epsilon)&=\bar\Delta_{9}+2 \bar\Delta_{12} \epsilon +\bar\Delta_{14} \epsilon ^2\,.
\end{align}
Then, by taking $\Delta_0'=\bar{\Delta}_0'=0$ as a system of two equations for the variable $z$, we can define the Sylvester matrix as~\cite{sturmfels2002solving}:
\begin{equation}
 M(\epsilon)=   \left(
\begin{array}{cccccc}
 q_0 & 0 & p_0 & 0 & 0 & 0 \\
 q_1 & q_0 & p_1 & p_0 & 0 & 0 \\
 q_2 & q_1 & p_2 & p_1 & p_0 & 0 \\
 q_3 & q_2 & 0 & p_2 & p_1 & p_0 \\
 q_4 & q_3 & 0 & 0 & p_2 & p_1 \\
 0 & q_4 & 0 & 0 & 0 & p_2 \\
\end{array}
\right)\,,
\end{equation}
whose determinant, called the resultant, becomes
\begin{align}
\textrm{det}(M(\epsilon))=&\,    p_2^2 \left[p_0^2 \left(2 q_0 q_4-2 q_1 q_3+q_2^2\right)+p_0 p_1 (3 q_0 q_3-q_1 q_2)+p_1^2 q_0 q_2\right]+p_2 \big[p_0^3 \left(q_3^2-2 q_2 q_4\right)+p_0^2 p_1 (3 q_1 q_4-q_2 q_3)\nonumber\\
&+p_0 p_1^2 (q_1 q_3-4 q_0 q_4)-p_1^3 q_0 q_3\big]+q_4 \left(p_0^4 q_4-p_0^3 p_1 q_3+p_0^2 p_1^2 q_2-p_0 p_1^3 q_1+p_1^4 q_0\right)\nonumber\\
&+p_2^3 \left[p_0 \left(q_1^2-2 q_0 q_2\right)-p_1 q_0 q_1\right]+p_2^4 q_0^2\,.
\end{align}
This is a polynomial in the variable $\epsilon$ of degree 20. It is known that the solutions of the system are included in the solutions of the resultant equated to zero. Therefore, we have derived an upper bound for the number of solutions –counted with its multiplicity– of the system: 20. This number is actually exact in generic situations due to the Bernstein's theorem~\cite{walker1950algebraic,bernshtein1975number,sturmfels2002solving}, which is applied to the above system in~\ref{App:Bernstein}.

Nevertheless, this bound of 20 solutions applies to the case where the two variables $\epsilon$ and $z$ are fully independent. We need to extract the solutions with $\bar z =\epsilon$, and there is no known way to quantify this. An important remark is in order here: in the analysis of Eq.~\eqref{Delta'0} in terms of the two independent variables $\epsilon$ and $z$, if $\left(\epsilon_0,z_0\right)$ happens to be a solution of this equation, then $\left(\epsilon=\bar z_0, z=\bar\epsilon_0\right)$ is also necessarily a solution. Hence, the solutions that {\em do  not} satisfy the constraint $\bar z =\epsilon$ come in pairs, and thus the number of proper solutions that satisfy this constraint in generic situations will be 20 minus an even number. Note however, that depending on their multiplicities, the total number of {\em different} solutions may well be odd.

%In this regard, we have performed an analysis of the solutions by giving arbitrary numerical values to the scalars $\Delta_n$ and solving Eq.~\eqref{EqtoSolve} numerically with MATHEMATICA. We have never found more than 13 different solutions (without taking into account multiplicities), and so we conjecture that this is may be the maximum number in generic situations.

For the definition of nongeneric situations, as well as their general characterisations in terms of the scalars $\{\Delta_i\}_{i=0}^{14}$, consult~\ref{App:Bernstein}. Let us remark that the property of substracting an even number will also apply to these nongeneric situations. More on this in the next subsection, where the different cases arising for each type in the algebraic classification are identified and studied.

\subsubsection{The full classification including exceptional cases and the possible number of PNDs for each type}\label{subsec:Full}

The best way to complete the algebraic classification by finding the possibilities for the second Roman numeral is to analyse the different cases in order, from maximum to minimum alignment. Thus, we start with Type N.

\paragraph{Type N}

In this case, there is a PND of Class VI and, therefore, all the complex scalars vanish, except $\Delta_{14}$. This case was already studied in the previous subsection by analysing Eq.~\eqref{EqforTypeN}, where we proved that such PND is unique. Thus, only Type $$(\textrm{VI},-)\,,$$ exists.

\paragraph{Types L and L*}

These are defined by the existence of a PND of Class V. Choosing this PND as $l_\mu$ in a null tetrad implies that $\Delta_n=0$ for all $n=0,\dots,11$. Again this situation was already analysed in the previous subsection, see Eq.~\eqref{EqfortypeL}, where we proved that generally there is a second simple PND given by~\eqref{solTypeL}. This provides the case $$(\textrm{V},\textrm{I}).$$
However, we also proved that under the constraint~\eqref{very_special_case} there are two special situations in which either there is no second PND, or there is an infinite number of them; the latter if the constraint $\Delta_{12}=\bar\Delta_{12}$ holds as well. These types are then 
$$(\textrm{V},-)\,, \hspace{1cm} (\textrm{V},\textrm{I}_\infty)\,,$$
respectively.

\paragraph{Type F}

This type is given by the existence of a PND of Class IV, and a second PND of Class II. We can choose a preferred tetrad $\{l^\mu,k^\mu,m^\mu,\bar m^\mu\}$, with $l^\mu$ being the Class-IV PND and $k^\mu$ the Class-II PND. In such a preferred tetrad, one has
\begin{equation}
  \Delta_0 =\Delta_1 =\Delta_2=\Delta_3=\Delta_4=\Delta_5=\Delta_6=\Delta_7=\Delta_8=\Delta_{12}=\Delta_{13}=\Delta_{14}=0\,.  
\end{equation}
To see if there can be any other PNDs, we can consider Eq.~\eqref{EqtoSolve} with the previous restrictions, that is
\begin{equation}
    \epsilon^2\left(\epsilon^2 \Delta_9+8\epsilon\bar\epsilon \Delta_{10}+6\bar\epsilon^2 \Delta_{11}\right)=0\,,\label{EqTypeF}
\end{equation}
which, removing the factor $\epsilon^2$, can be rewritten as either
\begin{equation}
    \epsilon^{2}\Delta_9+8\epsilon\bar\epsilon \Delta_{10}+6\bar\epsilon^{2}\Delta_{11}=0\,,\label{Fextra}
\end{equation}
or
\begin{equation}
    |\epsilon|^{2}\left(e^{2i\phi}\Delta_9 +8\Delta_{10}+6e^{-2i\phi} \Delta_{11}\right)=0\,.\label{Fextra'}
\end{equation}
It is easily seen that there are no solutions for this equation with $\epsilon\neq 0$, unless very specific restrictions exist between the scalars $\Delta_9$, $\Delta_{10}$ and $\Delta_{11}$. This general case, characterised solely by the trivial solution $\epsilon = 0$, constitutes the Type
$$
(\textrm{IV},\textrm{II})\,.
$$
However, there are exceptional situations with solutions of Eq.~\eqref{Fextra'} for the phase $\phi$, whereas $|\epsilon|$ remains free, giving rise to an infinite number of extra solutions. In particular, two possibilities arise, depending on whether the extra infinite PNDs are of Class II or of Class I. The former takes place when the left-hand side of Eq.~\eqref{Fextra} has the form of a perfect square of type $(A\epsilon + B\bar \epsilon)^2$, with $|A|=|B|$. This can happen only if
\begin{equation}
    8\Delta_{10}^2=3\Delta_9 \Delta_{11}\,, \hspace{1cm} |\Delta_9|=6|\Delta_{11}|\,.
\end{equation}
In this case, the form of the solution is
\begin{equation}
    \epsilon=|\epsilon| e^{i\left[\phi_{11}-\phi_{9}+2\left(2k+1\right)\pi\right]/4}\,,\quad \textrm{where $k=0,1$}\,,
\end{equation}
so that the phase of $\epsilon$ is fixed but $|\epsilon|$ remains free, leading then to an infinite number of PNDs. To check that they are of Class II, we perform the transformation~\eqref{complex_rot} with these solutions for $\epsilon$, in order to verify that $\Delta'_0=\Delta'_1=\Delta'_2=0$. Using the formulas~\eqref{Delta'0}-\eqref{Delta'14} for the rotated tetrad, we certainly find
\begin{align}
    \Delta'_0&=\Delta'_1=\Delta'_2=\Delta'_{12}=\Delta'_{13}=\Delta'_{14}=0\,,\label{1rotated_scalars_Fe}\\
    \Delta'_{n}&\neq 0\,,\quad\textrm{for all} \; n = 3,...,11.\label{2rotated_scalars_Fe}
\end{align}
This exceptional type can then be denoted as F$_e$, or as
$$
(\textrm{IV},\textrm{II}_\infty)\,.
$$
On the other hand, the second possibility that gives rise to an infinite number of PNDs, but now of Class I, arises for example for the particular case
\begin{equation}
    \Delta_{11}=0\,, \hspace{1cm} |\Delta_9|=8|\Delta_{10}|\,,
\end{equation}
which leads to the nontrivial solutions
\begin{equation}
    \epsilon =|\epsilon| e^{i\left[\phi_{10}-\phi_9 +\left(2k+1\right)\pi\right]/2}\,, \quad \textrm{where $k=0,1$}\,,
\end{equation}
and, once again, arbitrary $|\epsilon|$. This is an infinite number of extra PNDs, providing another special Type F$_e$, which will be denoted by
$$
(\textrm{IV},\textrm{II})_{\textrm{I}_{\infty}}.
$$

In summary, there exist three different cases within the Type F: the generic case $(\textrm{IV},\textrm{II})$ with one PND of Class IV and another one of Class II, as well as two exceptional cases $(\textrm{IV},\textrm{II}_\infty)$ and $(\textrm{IV},\textrm{II})_{\textrm{I}_{\infty}}$; the first one with one PND of Class IV and infinite of Class II, and the second one with one PND of Class IV, another one of Class II and infinite of Class I. Hence, it is worthwhile to stress that in the cases $(\textrm{IV},\textrm{II})$ and $(\textrm{IV},\textrm{II})_{\textrm{I}_{\infty}}$ the PNDs of Class IV and II are uniquely defined.

\paragraph{Types H and H*}

For these types, there is a PND of Class IV, but there is no PND of Class II. We choose a preferred tetrad $\{l^\mu,k^\mu,m^\mu,\bar m^\mu\}$ with $l^\mu$ being the PND of Class IV, so that
\begin{equation}
  \Delta_0 =\Delta_1 =\Delta_2=\Delta_3=\Delta_4=\Delta_5=\Delta_6=\Delta_7=\Delta_8=0\,.  
\end{equation}
With these restrictions, Eq.~\eqref{EqtoSolve} reads
\be\label{H}
\epsilon^2\left(\epsilon^2 \Delta_9+8\epsilon\bar\epsilon \Delta_{10}+6\bar\epsilon^2 \Delta_{11}+2\epsilon^2\bar\epsilon \Delta_{12}+4\epsilon\bar\epsilon^2 \Delta_{13} +\epsilon^2\bar\epsilon^2 \Delta_{14}\right)=0\,.
\ee
First of all, we notice that there are cases devoid of nontrivial solutions for $\epsilon$; for instance, if 
\begin{equation}
    \Delta_9=\Delta_{11}=\Delta_{12}=\Delta_{13}=0\,,
\end{equation}
the previous equation reduces to
\begin{equation}
\epsilon^2 |\epsilon|^2 (8\Delta_{10} +|\epsilon|^2 \Delta_{14}) =0\,,
\end{equation}
which does not have any nontrivial solution for $\epsilon$ if $\Delta_{10}\bar\Delta_{14}$ is not real. Thus, this particular case constitutes a Type H* or
$$
(\textrm{IV},-)\,.$$
In other situations, there is at least a nontrivial solution for $\epsilon$ of Eq.~\eqref{H}. One can then adapt the tetrad such that $k^\mu$ is one PND so that, without any loss of generality, Eq.~\eqref{H} becomes
\begin{equation}
\epsilon^2 \Delta_9+8\epsilon\bar\epsilon \Delta_{10}+6\bar\epsilon^2 \Delta_{11}+2\epsilon^2\bar\epsilon \Delta_{12}+4\epsilon\bar\epsilon^2 \Delta_{13}=0\,,
\end{equation}
where we have removed the $\epsilon^2$ factor in the equation. Now the question remains on whether there can be more nontrivial solutions of this equation for $\epsilon$, and whether or not there is a finite or infinite number of them. To answer these questions, we shall show explicit examples below.

First of all, subcases with infinite and also with none $\epsilon\neq 0$ solutions arise for
\begin{equation}
\Delta_9 =\Delta_{11}=0\,, \hspace{1cm} \Delta_{12} =2 \Delta_{13}\,.
\end{equation}
Consequently, any nontrivial solution must satisfy the equation
\begin{equation}
4\Delta_{10} +\left(\epsilon +\bar \epsilon\right) \Delta_{13} =0\,,
\end{equation}
which has no solution if $\Delta_{10}\bar\Delta_{13}$ is not real, but it has an infinite number of solutions if $\Delta_{10}\bar\Delta_{13}$ is real ---because the imaginary part of $\epsilon$ remains free. The latter leads to Type H$_e$ or
$$
(\textrm{IV},\textrm{I}_\infty)\,.
$$

%$$
%|\epsilon|^2 \left(e^{2i\phi} \Delta_9+8 \Delta_{10}+6 %e^{-2i\phi} \Delta_{11}+2|\epsilon| e^{i\phi} %\Delta_{12}+4|\epsilon|e^{-i\phi} \Delta_{13}\right)=0.
%$$
On the other hand, different subcases with a finite number of nontrivial solutions for $\epsilon$ are
\begin{itemize}
    \item $\Delta_{10}=\Delta_{11}=\Delta_{13}=0$, with the unique solution $2\epsilon=-\,\bar\Delta_9/\bar\Delta_{12}$\,;
    \item $\Delta_{9}=\Delta_{11}=\Delta_{12}=0$, also with a unique solution $\epsilon =-\,2\bar\Delta_{10}/\bar\Delta_{13}$\,;
\item $\Delta_{10}=\Delta_{11}=\Delta_{12}=0$, with three distinct solutions
\begin{equation}
\epsilon_k =\frac{1}{4}\left|\frac{\Delta_9}{\Delta_{13}} \right| e^{i[\phi_{13}-\phi_9+(2k-1)\pi]/3}\,,
\quad \textrm{where $k=0,1,2$}\,.
\end{equation}
\end{itemize} 
Therefore, all these subcases lead to a general Type H, given by
$$
(\textrm{IV},\textrm{I})\,,
$$
and one should keep in mind that in some situations there are several choices for the secondary I.

\paragraph{Type D}

This type is defined by the existence of two PNDs of Class III. Choosing the preferred tetrad $\{l^\mu,k^\mu,m^\mu,\bar m^\mu\}$ with both $l^\mu$ and $k^\mu$ of Class III, we have
\begin{equation}
\Delta_0 =\Delta_1 =\Delta_2=\Delta_3=\Delta_4=\Delta_5=\Delta_9=\Delta_{10}=\Delta_{11}=\Delta_{12}=\Delta_{13}=\Delta_{14}=0\,.\end{equation}
Accordingly, Eq.~\eqref{EqtoSolve} with the previous restrictions reads
\be\label{Dextra}
4\epsilon\left(\epsilon^2 \Delta_6+3\epsilon\bar\epsilon \Delta_{7}+\bar\epsilon^2 \Delta_{8}\right)=0\,,
\ee
which, removing the $4\epsilon$ factor, can be rewritten as 
\be\label{Dextra'}
|\epsilon|^2 (e^{2i\phi}\Delta_6 +3\Delta_{7}+e^{-2i\phi} \Delta_{8})=0\,.
\ee
As is shown, Eq.~\eqref{Dextra'} acquires the same form as Eq.~\eqref{Fextra'}, leading to no solutions in general, or to an infinite number of PNDs of Class I (e.g. $\Delta_8=0$ and $|\Delta_6|=3|\Delta_7|$), or to an infinite number of Class II PNDs (if $9\Delta_7^2=4\Delta_6\Delta_8$ and $|\Delta_6|=|\Delta_8|$) with arbitrary $|\epsilon|$ in these last two exceptional situations. 

These cases with infinite extra solutions are the special Type D$_e$ and denoted by
$$
(\textrm{III},\textrm{III})_{\textrm{II}_\infty} \, ,\hspace{1cm} (\textrm{III},\textrm{III})_{\textrm{I}_{\infty}}\,.
$$
The general case with no extra PNDs is the generic Type D denoted as 
$$
(\textrm{III},\textrm{III})\,.
$$

In all cases, $(\textrm{III},\textrm{III})$, $(\textrm{III},\textrm{III})_{\textrm{II}_\infty}$ and $(\textrm{III},\textrm{III})_{\textrm{I}_{\infty}}$, the two PNDs of Class III are uniquely defined.

\paragraph{Type M}

This type is defined by the existence of a {\em unique} PND of Class III, and a second PND of Class II. We choose a preferred tetrad $\{l^\mu,k^\mu,m^\mu,\bar m^\mu\}$ with $l^\mu$ being the PND of Class III, and $k^\mu$ a Class-II one. In such a preferred tetrad one has
\begin{equation}
\Delta_0 =\Delta_1 =\Delta_2=\Delta_3=\Delta_4=\Delta_5=\Delta_{12}=\Delta_{13}=\Delta_{14}=0\,,
\end{equation}
and Eq.~\eqref{EqtoSolve} becomes
\be\label{Mextra}
\epsilon\left(4\epsilon^2 \Delta _6+12\epsilon \bar{\epsilon} \Delta _7+4 
   \bar{\epsilon }^2 \Delta _8
   +\epsilon ^3 \Delta _9+8 \epsilon ^2 \bar{\epsilon } \Delta _{10}+6 \epsilon
   \bar{\epsilon }^2 \Delta _{11}\right)=0\,.
\ee
Again, besides the trivial solution $\epsilon=0$, one can exhibit cases with an infinite number of $\epsilon\neq 0$ solutions giving PNDs of Class I ($\Delta_6=\Delta_8=\Delta_9=0$, $4\Delta_{10}=3\Delta_{11}$, $\Delta_7\bar\Delta_{11} = \bar\Delta_7\Delta_{11} $) or with no such solutions ($\Delta_6=\Delta_8=\Delta_9=0$, $4\Delta_{10}=3\Delta_{11}$, $\Delta_7\bar\Delta_{11} \neq \bar\Delta_7\Delta_{11} $) and also with a finite number of them.
They give, respectively, Types M$_e$ and M, denoted by
$$
(\textrm{III},\textrm{II})_{\textrm{I}_{\infty}}\,, \hspace{1cm} (\textrm{III},\textrm{II})\,.
$$
Note that both PNDs of Class III and Class II are uniquely determined in these cases.

The question remains if there can be an infinite number of PNDs of Class II. This will happen if the left-hand side of Eq.~\eqref{Mextra} can be factorised as $\epsilon\left(A\epsilon+B\bar\epsilon\right)^{2}\left(a+b\epsilon\right)$ with $|A|=|B|$. Specifically, this occurs if
\be\label{Condition}
8\Delta_{10}^2 = 3\Delta_9 \Delta_{11}\,, 
\hspace{4mm} 4\Delta_6\Delta_8 = 9\Delta_7^2\,, \hspace{5mm}
6\Delta_6\Delta_{11} =\Delta_8 \Delta_9\,, \hspace{4mm}2|\Delta_6|=3|\Delta_7|\,,
\ee
where, apart from the infinite number of PNDs of Class II, there exists one extra PND. Thereby, this is another exceptional Type M$_e$, denoted by
$$
(\textrm{III},\textrm{II}_\infty)\,.
$$

\paragraph{Types K and K*}

Now there is a {\em unique} PND of maximal Class III, but no PND of Class II. We choose a preferred tetrad $\{l^\mu,k^\mu,m^\mu,\bar m^\mu\}$ with $l^\mu$ being the PND of Class III, so that in this tetrad Eq.~\eqref{EqtoSolve} reads
\be\label{K}
\epsilon\left(4 \epsilon ^2 \Delta _6+12 \epsilon \bar{\epsilon } \Delta _7+4 
   \bar{\epsilon }^2 \Delta _8
   +\epsilon ^3 \Delta _9+8 \epsilon ^2 \bar{\epsilon } \Delta _{10}+6 \epsilon
   \bar{\epsilon }^2 \Delta _{11}+2 \epsilon ^3 \bar{\epsilon } \Delta _{12}+4 \epsilon ^2 \bar{\epsilon }^2 \Delta
   _{13}+\epsilon ^3 \bar{\epsilon }^2 \Delta _{14}\right)=0\,.
\ee
The first thing to know is whether there are cases without nonzero solutions for $\epsilon$ while keeping $\Delta_{14}$ and at least one of $\Delta_6,\Delta_7,\Delta_8$ different from zero. By setting $\Delta_8=\Delta_9=\Delta_{10}=\Delta_{11}=\Delta_{12}=\Delta_{13}=0$ and $\Delta_6 =3\Delta_7$, this equation can be written as
%$$
%|\epsilon|^3\left(4\Delta_6 e^{3i\phi} +12\Delta_7 e^{i\phi} +4\Delta_8 e^{-i\phi} +|\epsilon|^3 \Delta_{14} e^{2i\phi}\right)=0$$
\begin{equation}
\epsilon^2[4\Delta_6 (\epsilon+\bar\epsilon) +\epsilon^2\bar\epsilon^2 \Delta_{14}]=0\,,
\end{equation}
which does not provide any nontrivial solution if $\Delta_6\bar\Delta_{14}\neq \bar\Delta_6\Delta_{14}$. This case then represents a Type K*, or
$$
(\textrm{III},-)\,.
$$
Otherwise, if there are solutions of Eq.~\eqref{K} leading to PNDs different from $l_\mu$, we can choose one of these extra PND as the $k_\mu$ in the null tetrad, so that without any loss of generality $\Delta_{14}$ can be set to zero in Eq.~\eqref{K}:
\be\label{K1}
\epsilon\left(4 \epsilon ^2 \Delta _6+12 \epsilon  \bar{\epsilon } \Delta _7+4 
   \bar{\epsilon }^2 \Delta _8
   +\epsilon ^3 \Delta _9+8 \epsilon ^2 \bar{\epsilon } \Delta _{10}+6 \epsilon
   \bar{\epsilon }^2 \Delta _{11}+2 \epsilon ^3 \bar{\epsilon } \Delta _{12}+4 \epsilon ^2 \bar{\epsilon }^2 \Delta
   _{13}\right)=0\,.
\ee
Nevertheless, now we must keep at least one of the complex scalars $\Delta_{12}$ and $\Delta_{13}$ different from zero; otherwise, this will belong to the previous Type M. Now, following the same ideas as in previous cases, it becomes rather easy to find cases with an infinite number of extra solutions (e.g. all $\Delta_n =0$ except $\Delta_6$ and $\Delta_{12}$, with $\Delta_6\bar\Delta_{12}= \bar\Delta_6\Delta_{12} $) and with no extra solutions (e.g. all $\Delta_n =0$ except $\Delta_6$ and $\Delta_{12}$, with $\Delta_6\bar\Delta_{12}\neq \bar\Delta_6\Delta_{12} $), or with a finite number of them. These lead respectively to the Types K$_e$ and K, or equivalently
$$
(\textrm{III},\textrm{I}_\infty)\,, \hspace{1cm} (\textrm{III},\textrm{I})\,,
$$
where in the latter case there may be several different choices for the PND of Class I.

\paragraph{Types B and S}

Now there are two alignment PNDs of Class II, so that we can choose a tetrad with both $l^\mu$ and $k^\mu$ of Class II. This implies $\Delta_0=\Delta_1=\Delta_2 =\Delta_{12}=\Delta_{13}=\Delta_{14} =0$ and the main equation~\eqref{EqtoSolve} written in this preferred tetrad reads
\be\label{B}
6 \epsilon ^2 \Delta _3+8 \epsilon  \bar{\epsilon }
   \Delta _4+\bar{\epsilon }^2 \Delta _5+4 \epsilon ^3 \Delta _6+12 \epsilon ^2 \bar{\epsilon } \Delta _7+4 \epsilon 
   \bar{\epsilon }^2 \Delta _8+\epsilon ^4 \Delta _9+8 \epsilon ^3 \bar{\epsilon } \Delta _{10}+6 \epsilon ^2
   \bar{\epsilon }^2 \Delta _{11}=0\,,
\ee
where one must keep at least one of $\Delta_3,\Delta_4,\Delta_5$, and at least one of $\Delta_9,\Delta_{10},\Delta_{11}$ different from zero. 

The first question to elucidate is the possible existence of a third PND of Class II. Intuition developed so far tells us that this may be the case if there are $\epsilon\neq 0$ solutions of the previous equation with multiplicity 2. Keeping the double solution for $\epsilon =0$, this may happen if the left-hand side in Eq.~\eqref{B} can be factorised in any of the following forms: $A\epsilon^2 (\epsilon+B)^2$, $A\epsilon^2 (\bar\epsilon+B)^2$, $A\bar\epsilon^2 (\epsilon+B)^2$, $A\epsilon\bar\epsilon (\epsilon+B)^2$, $A\epsilon^2 (\epsilon +B)(\bar\epsilon+\bar B)$, $A\epsilon\bar\epsilon  (\epsilon +B)(\bar\epsilon+\bar B)$, or $(a\epsilon +b\bar\epsilon)^2(A +B \epsilon + C\epsilon^2)$, for some $A,B,C, a, b\in \mathbb{C}$ and with $|a|=|b|$. These lead to the following seven possibilities respectively:
\begin{enumerate}[label=\roman*)]
    \item \label{casei} All $\Delta_n=0$ except for $\Delta_3,\Delta_6$ and $\Delta_{9}$ with $2\Delta_6^2=3\Delta_3\Delta_{9}$. Then, Eq.~\eqref{B} has a double solution given by
\begin{equation}
    \epsilon =-\,2\frac{\Delta_6}{\Delta_9}=-\,3\frac{\Delta_3}{\Delta_6}\,.
\end{equation}
    \item
All $\Delta_n=0$ except for $\Delta_3,\Delta_7$ and $\Delta_{11}$ with $\Delta_7^2=\Delta_3\Delta_{11}$. Then, Eq.~\eqref{B} has a double solution given by
\begin{equation}
\bar\epsilon =-\,\frac{\Delta_7}{\Delta_{11}}=-\,\frac{\Delta_3}{\Delta_7}\,.
\end{equation}
    \item
All $\Delta_n=0$ except for $\Delta_5,\Delta_8$ and $\Delta_{11}$ with $2\Delta_8^2=3\Delta_5\Delta_{11}$. Then, Eq.~\eqref{B} has a double solution given by
\begin{equation}
\epsilon =-\,\frac{\Delta_8}{3\Delta_{11}}=-\,\frac{\Delta_5}{2\Delta_8}\,.
\end{equation}
    \item 
All $\Delta_n=0$ except for $\Delta_4,\Delta_7$ and $\Delta_{10}$ with $9\Delta_7^2=16\Delta_4\Delta_{10}$. Then, Eq.~\eqref{B} has a double solution given by
\begin{equation}
\epsilon =-\,\frac{3\Delta_7}{4\Delta_{10}}=-\,\frac{4\Delta_{4}}{3\Delta_7}\,.
\end{equation}
    \item
All $\Delta_n=0$ except for $\Delta_3,\Delta_6,\Delta_7$ and $\Delta_{10}$ with $\Delta_7\Delta_6=\Delta_3\Delta_{10}$ and $3\Delta_7 \bar \Delta_{10}= \bar\Delta_6 \Delta_{10}$. Then, Eq.~\eqref{B} has a double solution given by
\begin{equation}
\epsilon =-\,\frac{3\Delta_7}{2\Delta_{10}}=-\,\frac{\bar\Delta_{6}}{2\bar\Delta_{10}}\,.
\end{equation}
\item\label{Casevi}
All $\Delta_n=0$ except for $\Delta_4,\Delta_7,\Delta_8$ and $\Delta_{11}$ with $\Delta_7\Delta_8=\Delta_4\Delta_{11}$ and $\Delta_8\bar\Delta_{11} = 3\bar\Delta_7\Delta_{11}$. Then, Eq.~\eqref{B} has a double solution given by
\begin{equation}
\epsilon =-\,\frac{2\Delta_8}{3\Delta_{11}}=-\,\frac{2\bar\Delta_{7}}{\bar\Delta_{11}}\,.
\end{equation}
\item\label{Casevii}
Conditions~\eqref{Condition}, together with
\begin{equation}
8\Delta_4^2 =3\Delta_3\Delta_5\,, \hspace{1cm} 6\Delta_8\Delta_{3} =\Delta_6 \Delta_5\,,
\end{equation}
all hold. 
\end{enumerate}
To check whether or not these solutions define a PND of Class II, we need to verify that in the new tetrad~\eqref{complex_rot} the corresponding scalars $\Delta'_0,\Delta'_1$ and $\Delta'_2$ vanish. Starting from the last possibility {\rm vii)}, Eq.~\eqref{B} factorises as ($e^{i\alpha}=3\Delta_7/(2\Delta_6)$)
\be\label{IIinfinity}
\bigl(\epsilon+e^{i\alpha}\bar\epsilon\bigr)^{2}\bigl(\Delta_9\epsilon^2+6\Delta_7 \epsilon+\Delta_5\bigr)=0\,,
\ee
leading to an infinite number of solutions
\begin{equation}
    \epsilon =\pm i|\epsilon|e^{i\alpha/2}\,,
\end{equation}
with arbitrary modulus. Using now formulas~\eqref{Delta'0}-\eqref{Delta'14} with the above restrictions on the $\Delta_n$, one easily gets
\begin{equation}
    \Delta'_0=\Delta'_1=\Delta'_2=0\,, \hspace{3mm} \Delta'_3 \neq 0\,,
\end{equation}
for arbitrary values of $|\epsilon|$ and thus this case has an infinite number of PNDs of Class II. This is one of the Types B$_e$, denoted by
$$
(\textrm{II},\textrm{II}_\infty) \equiv (\textrm{II}_\infty)\,.
$$
From Expression~\eqref{IIinfinity}, one sees that there are two further PNDs of Class I (or another one of Class II).

Concerning the other cases {\rm i) -- vi)}, the formulas~\eqref{Delta'0}-\eqref{Delta'14} we have, for the first case {\rm i)}
\begin{equation}
\Delta'_0=\Delta_1'=\Delta'_2=0\,, \hspace{1cm} \Delta'_3 =\Delta_3 \neq 0\,.
\end{equation}
hence this gives indeed another PND of Class II. For completeness, the rest of scalars in the new tetrad are
\begin{equation}
\Delta'_4=\Delta'_5=\Delta'_7=\Delta'_8=\Delta'_{10}=\Delta'_{11}=\Delta'_{12}=\Delta'_{13}=\Delta'_{14}=0\,,\hspace{4mm}
\Delta'_6=-\,\Delta_6\,, \hspace{4mm} \Delta'_{9} =\Delta_{9}\,,
\end{equation}
which keeps the property $2\Delta'_6{}^2 = 3\Delta'_3 \Delta'_9$\,. It is easy to check then that, starting with the new null tetrad $\{l'_\mu, k_\mu, m'_\mu, \bar m'_\mu\}$ and getting the third PND of Class II by solving the primed version of Eq.~\eqref{B} one gets back to the original PND given by $l_\mu$.

Concerning possibility {\rm ii)}, formulas~\eqref{Delta'0}-\eqref{Delta'14}  lead to
\begin{equation}
\Delta'_0 =\Delta'_1 =\Delta'_2=0\,, \quad \Delta'_3 =\Delta'_4=0\,, \quad\Delta'_5 =6\Delta_{11} \frac{\bar\Delta_3^2}{\bar\Delta_7^2}\neq 0\,,
\end{equation}
so that this gives indeed another PND of Class II. For completeness, the rest of scalars in the new tetrad are
\begin{align}
\Delta'_6&=\Delta'_7=\Delta'_9=\Delta'_{10}=\Delta'_{12}=\Delta'_{13}=\Delta'_{14}=0\,,\\
\Delta'_8&=-\,3\Delta_{11}\frac{\bar\Delta_3}{\bar\Delta_7}\,, \hspace{4mm} \Delta'_{11} =\Delta_{11}\,.
\end{align}
Notice that $2\Delta'_8{}^2=3\Delta'_5\Delta'_{11}$ leading to possibility {\rm iii)}. One expects, therefore, that possibility {\rm iii)} will also define a PND of Class II, which should actually provide new $\Delta'_n$ defining the possibility {\rm ii)}. Indeed, using formulas~\eqref{Delta'0}-\eqref{Delta'14} under possibility {\rm iii)} we obtain
\begin{align}
\Delta'_0&=\Delta'_1 =\Delta'_2=\Delta'_4=\Delta'_5=\Delta'_6=\Delta'_8=\Delta'_9=\Delta'_{10}=\Delta'_{12}=\Delta'_{13}=\Delta'_{14}=0\,, \\
\Delta'_3 &=\frac{\bar\Delta_5^2}{4\bar\Delta_8^2}\Delta_{11}\neq 0\,, \hspace{4mm} \Delta'_7= -\,\frac{\bar\Delta_5}{2\bar\Delta_8}\Delta_{11}\,,\hspace{4mm}
\Delta'_{11} = \Delta_{11}\,,
\end{align}
with $\Delta'_7{}^2 = \Delta'_3\Delta'_{11}$ as required.

Consider now possibility {\rm iv)}. A similar calculation provides
\begin{align}
\Delta'_0&=\Delta'_1 =\Delta'_2=\Delta'_4=\Delta'_5=\Delta'_8=\Delta'_9=\Delta'_{11}=\Delta'_{12}=\Delta'_{13}=\Delta'_{14}=0\,,\\
\Delta'_3&=\frac{3}{4} \frac{\Delta_7\bar\Delta_7}{\bar\Delta_{10}}\,,
\hspace{3mm} \Delta'_6=-\frac{3\bar\Delta_7}{2\bar\Delta_{10}}\Delta_{10}\,, \hspace{3mm} \Delta'_7 =-\,\frac{1}{2} \Delta_7\,, \hspace{3mm} \Delta'_{10} = \Delta_{10}\,,
\end{align}
which satisfies $\Delta'_7\Delta'_6=\Delta'_3\Delta'_{10}$ and $3\Delta'_7 \bar \Delta'_{10}= \bar\Delta'_6 \Delta'_{10}$ leading to possibility {\rm v)}. Again, we expect then that possibility {\rm v)} will lead to a new PND of possibility {\rm iv)}. This can be checked as before, because in possibility {\rm v)} the new $\Delta'_n$ are
\begin{align}
\Delta'_0&=\Delta'_1 =\Delta'_2=\Delta'_3=\Delta'_5=\Delta'_6=\Delta'_8=\Delta'_9=\Delta'_{11}=\Delta'_{12}=\Delta'_{13}=\Delta'_{14}=0\,,\\
\Delta'_4&=\frac{9}{4} \frac{\Delta_7^2}{\bar\Delta_{10}}\,, \hspace{3mm} \Delta'_7 =-\,2 \Delta_7\,,\hspace{3mm} \Delta'_{10} = \Delta_{10}\,,
\end{align}
having $16\Delta'_4 \Delta'_{10} = 9 \Delta'_7{}^2$, that is, the properties defining case {\rm iv)}.

Finally, in possibility {\rm vi)} we compute
\begin{align}
\Delta'_0&=\Delta'_1 =\Delta'_2=\Delta'_3=\Delta'_5=\Delta'_6=\Delta'_9=\Delta'_{10}=\Delta'_{12}=\Delta'_{13}=\Delta'_{14}=0\,,\\
\Delta'_4&=\Delta_4\,,
\hspace{3mm} \Delta'_7=- \,\Delta_7\,, \hspace{3mm} \Delta'_8=\-\Delta_8\,, \hspace{3mm} \Delta'_{11} = \Delta_{11}\,,
\end{align}
defining an extra PND and keeping the type possibility {\rm vi)}.

Summarising, for all possibilities {\rm i)--vi)} there is always a third PND of Class II --and no more. This is a special situation, because the notation $(\textrm{II},\textrm{II})$ would be ambiguous, as we do not know which two PNDs, among the three ones of Class II, are there. And there are three different possible choices for two Class-II PNDs among three. This is the reason that we introduce a special notation for this particular type, called Type S, given by
$$
(\textrm{II},\textrm{II},\textrm{II})\,.
$$
Furthermore, the six possibilities studied may be different, in the sense that the three PNDs in each possibility may be of different kinds in the refined classification based on the superenergy tensor developed in~\ref{superenergyappendix}. One can easily see, however, that possibilities {\rm i), ii) and iii)} are equivalent, and possibilities {\rm, iv), v) and vi)} are also equivalent between them. The former has two PNDs of Class IIa and one of Class IId\footnote{See Table~\ref{TableZ1} for a description of Class IIa and Class IId.}, while the latter has two PNDs of Class II (with $\Delta_3=0$) and one PND of Class IIa. Thus, there exist two different types $(\textrm{II},\textrm{II},\textrm{II})$, given by
$$
(\textrm{IIa},\textrm{IIa},\textrm{IId})\hspace{1cm}  \mbox{and} \hspace{1cm} 
(\textrm{II}_{\Delta_{3}=0},\textrm{II}_{\Delta_{3}=0},\textrm{IIa})\,.
$$

Going back to Eq.~\eqref{B}, when the above possibilities {\rm i) -- vii)} do not hold, there are only two PNDs of Class II. Thus the only remaining question is to discern if there can be an infinite number of extra PNDs (all necessarily of Class I). For this task, it is enough to show an example where this can happen. In particular, assume that all $\Delta_n=0$, except for $\Delta_3, \Delta_{10}$. Then, Eq.~\eqref{B} collapses to simply
\begin{equation}
2\epsilon^2\left(3\Delta_3 +4\epsilon\bar\epsilon \Delta_{10}\right)=0\,,
\end{equation}
which has an infinite number of solutions for $\epsilon$ if $\Delta_3/\Delta_{10}$ is real and negative, as the phase of $\epsilon$ remains free. In consequence, there is another Type B$_e$ and the general Type B, or also
$$
(\textrm{II},\textrm{II})_{\textrm{I}_{\infty}}\,, \hspace{1cm} (\textrm{II},\textrm{II})\,,
$$
where in the latter case there may be a finite number of extra PNDs of Class I.

\paragraph{Types C and C*}

Types C are defined by having a unique PND of Class II, and this is the maximal alignment for all PNDs. Thus, all other PNDs, if they exist, can only be of Class I. Choosing the null tetrad with $l_\mu$ along the PND of Class II, Eq.~\eqref{EqtoSolve} reads
\be\label{C}
6 \epsilon ^2 \Delta _3+8 \epsilon  \bar{\epsilon }
   \Delta _4+\bar{\epsilon }^2 \Delta _5+4 \epsilon ^3 \Delta _6+12 \epsilon ^2 \bar{\epsilon } \Delta _7+4 \epsilon 
   \bar{\epsilon }^2 \Delta _8
   +\epsilon ^4 \Delta _9+8 \epsilon ^3 \bar{\epsilon } \Delta _{10}+6 \epsilon ^2
   \bar{\epsilon }^2 \Delta _{11}+2 \epsilon ^4 \bar{\epsilon } \Delta _{12}+4 \epsilon ^3 \bar{\epsilon }^2 \Delta
   _{13}+\epsilon ^4 \bar{\epsilon }^2 \Delta _{14}=0\,.
\ee
Cases devoid of nontrivial solutions for $\epsilon$ are easily found. For instance, set all $\Delta_n=0$, except for $\Delta_3$ and $\Delta_{14}$. Then the above equation is reduced to
\be\label{C1}
    \epsilon^2\left(6\Delta_3 + \epsilon^2\bar\epsilon^2 \Delta_{14}\right)=0\,,
\ee
without $\epsilon\neq 0$ solution if $\Delta_3/\Delta_{14}$ is not real and negative. Thus, such cases lead to Type C* or
$$
(\textrm{II},-)\,.
$$
On the other hand, in the same situation, if $\Delta_3/\Delta_{14}$ is real and negative then there exists an infinite number of solutions for $\epsilon$ as only the norm $|\epsilon|$ is fixed. This leads to Type C$_e$ or
$$
(\textrm{II},\textrm{I}_\infty)\,.
$$
The remaining case with a finite number of nontrivial solutions of Eq.~\eqref{C} are simply called Type C, or denoted by
$$
(\textrm{II},\textrm{I})\,.
$$

\paragraph{Type I}

Now, all possible PNDs are of the basic Class I, and there is at least one of these. Choosing such a PND as $l_\mu$ in the null tetrad, we have $\Delta_0=0$ with $|\Delta_1|^2+ |\Delta_2|^2 \neq 0$, and this is the unique restriction on Eq.~\eqref{EqtoSolve}. In this regard, we have been unable to find any possibility with just a unique PND, or with none, and we believe that they do not exist. Hence, Types I* and $\emptyset$, corresponding to $(\textrm{I},-)$ and to $(-,-)$, respectively, are missing.

On the other hand, by following the same ideas as in previous types, it is quite straightforward to find cases with an infinite number of extra PNDs, or with only a finite number (an example given by Eq.~\eqref{0-14}, for the couple of solutions~\eqref{sol_eps_1and4}). These are the cases I$_e$ and I; namely,
$$
(\textrm{I},\textrm{I}_\infty) \,, \hspace{1cm} (\textrm{I},\textrm{I})\,,
$$
respectively.

\begin{figure}[H]
    \centering
 \resizebox{\textwidth}{!}{
\begin{tikzpicture}[node distance=3cm and 2cm, auto]

    % Nodes (Types)
    \node (I) {\fbox{Type I}};
    \node (C) [below of=I] {\fbox{Type C}};

    \node (B) [below of=C, yshift=-2.5cm,xshift=-1cm] {\fbox{Type B}};
    \node (K) [right of=B] {\fbox{Type K}};
    \node (M) [below of=B,yshift=-2.5cm] {\fbox{Type M}};
    \node (D) [below of=M] {\fbox{Type D}};
    \node (H) [right of=M] {\fbox{Type H}};
    % \node (H) [right of=L] {\fbox{Type H}};
    \node (L) [right of=D] {\fbox{Type L}};
     \node (F) [right of=L] {\fbox{Type F}};
      \node (O) [below of=L,yshift=-0.5cm] {\fbox{Type O}};

 \node (CAst) [right of=C,xshift=5cm] {\fbox{Type C*}};
  \node (CAstHelp) [above of=CAst,yshift=-2.5cm] {\(\Delta_0 =\Delta_{1}=\Delta_{2}= 0\)};
  \node (CAstHelp2) [below of=CAst,yshift=2.3cm] {$\Delta_{14}\neq0$};

      \node (KAst) [below of=CAstHelp2,yshift=-2cm] {\fbox{Type K*}};
        \node (HAst) [right of=F,xshift=-1cm] {\fbox{Type H*}};
         \node (LAst) [right of=F,xshift=1cm] {\fbox{Type L*}};
          \node (N) [right of=O,xshift=3cm] {\fbox{Type N}};
 
\node (S) [left of=M,xshift=-2cm] {\fbox{Type S}};

%%arrows
% Arrow and Label for Delta_0 = 0
    \node (Delta0) [above of=I,yshift=-2.5cm] {\(\Delta_0 =\Delta_{14}= 0\)};

     \draw[->] (I) -- (C) node[midway, above,sloped] {\(\Delta_1= 0\)};
    \draw[->] (I) -- (C) node[midway, below,sloped] {\(\Delta_2 = 0\)};
    \draw[->] (C) -- (B) node[midway, above,sloped] {\(\Delta_{12}=\Delta_{13} = 0\)};
     \draw[->] (C) -- (K) node[midway,above,sloped] {\(\Delta_{3}=\Delta_{4} =\Delta_{5}= 0\)};
      \draw[->] (B) -- (M) node[midway,below,sloped] {\(\Delta_{3}=\Delta_{4} =\Delta_{5}= 0\)};
       \draw[->] (K) -- (M) node[midway,above,sloped] {\(\Delta_{12}=\Delta_{13} = 0\)};
        \draw[->] (K) -- (H) node[midway,above,sloped] {\(\Delta_6=\Delta_7=\Delta_8=0\)};

     %    \draw[->] (B) -- (S) node[midway,above,sloped] {\(l'{}^\lambda l'_{[\alpha}\Z_{\sigma]\lambda\tau[\rho} l'_{\beta]} l'{}^\tau=0\)};
          \draw[->] (B) -- (S) node[midway,above,sloped] {See cases~\ref{casei}-\ref{Casevi}};
          \draw[->] (S) -- (M) node[midway,above,sloped] {\(\Delta_3=\Delta_4=\Delta_5=0\)};

       %  \draw[->] (K) -- (L) node[midway,below,sloped] {\(\Delta_9=\Delta_{10}=\Delta_{11}=0\)};
        
       %  \draw[->] (K) -- (H) node[midway,above,sloped] {\(\Delta_6=\Delta_7=\Delta_8=0\)};
         \draw[->] (M) -- (D) node[midway,above,sloped] {\(\Delta_9=\Delta_{10}=0\)};
          \draw[->] (M) -- (D) node[midway,below,sloped] {\(\Delta_{11}=0\)};

  \draw[->] (H) -- (L) node[midway,above,sloped] {\(\Delta_9=\Delta_{10}=0\)};
          \draw[->] (H) -- (L) node[midway,below,sloped] {\(\Delta_{11}=0\)};
           \draw[->] (H) -- (F) node[midway,above,sloped] {\(\Delta_{12}=\Delta_{13}=0\)};
           \draw[->] (L) -- (O) node[midway,below,sloped] {\(\Delta_{12}=\Delta_{13}=0\)};
           %\draw[->] (H) -- (F) node[midway,below,sloped] {\(\Delta_{12}=\Delta_{13}=0\)};

         %  \draw[->] (N) -- (O) node[midway,above,sloped] {\(\Delta_{14}=0\)};
            \draw[->] (D) -- (O) node[midway,below,sloped] {\(\Delta_{6}=\Delta_{7}=\Delta_{8}=0\)};
              \draw[->] (F) -- (O) node[midway,above,sloped] {\(\Delta_{9}=\Delta_{10}=\Delta_{11}=0\)};

\draw[->] (CAstHelp2) -- (KAst) node[midway, above,sloped] {\(\Delta_3=\Delta_4=\Delta_5= 0\)};

\draw[->] (KAst) -- (HAst) node[midway, above,sloped] {\(\Delta_6=\Delta_7=\Delta_8= 0 \)};
\draw[->] (KAst) -- (HAst) node[midway, below,sloped] {\(\Delta_9=\Delta_{11}=\Delta_{12}=\Delta_{13}= 0\)};

\draw[->] (KAst) -- (LAst) node[midway, above,sloped] {\(\Delta_6=\Delta_7=\Delta_8=\Delta_9=\Delta_{10}= \Delta_{11}=0\)};
\draw[->] (HAst) -- (O) node[midway, below,sloped] {\(\Delta_{10}=\Delta_{14}- 0\)};
\draw[->] (LAst) -- (N) node[midway, above,sloped] {\(\Delta_{12}=\Delta_{13}= 0\)};
\draw[->] (N) -- (O) node[midway, above,sloped] {\(\Delta_{14}= 0\)};
              
\end{tikzpicture}

}
    \caption{Flow diagram of the algebraic classification of the tensor $\Z_{\lambda\rho\mu\nu}$. The null tetrad is chosen as in Table~\ref{TableZ1PND}, while for simplicity in the presentation the extra constraints related to the cases marked with $*$ are not shown in the diagram, but they can be found for each case in Sec.~\ref{subsec:Full}.}
    \label{fig:diagram0}
\end{figure}
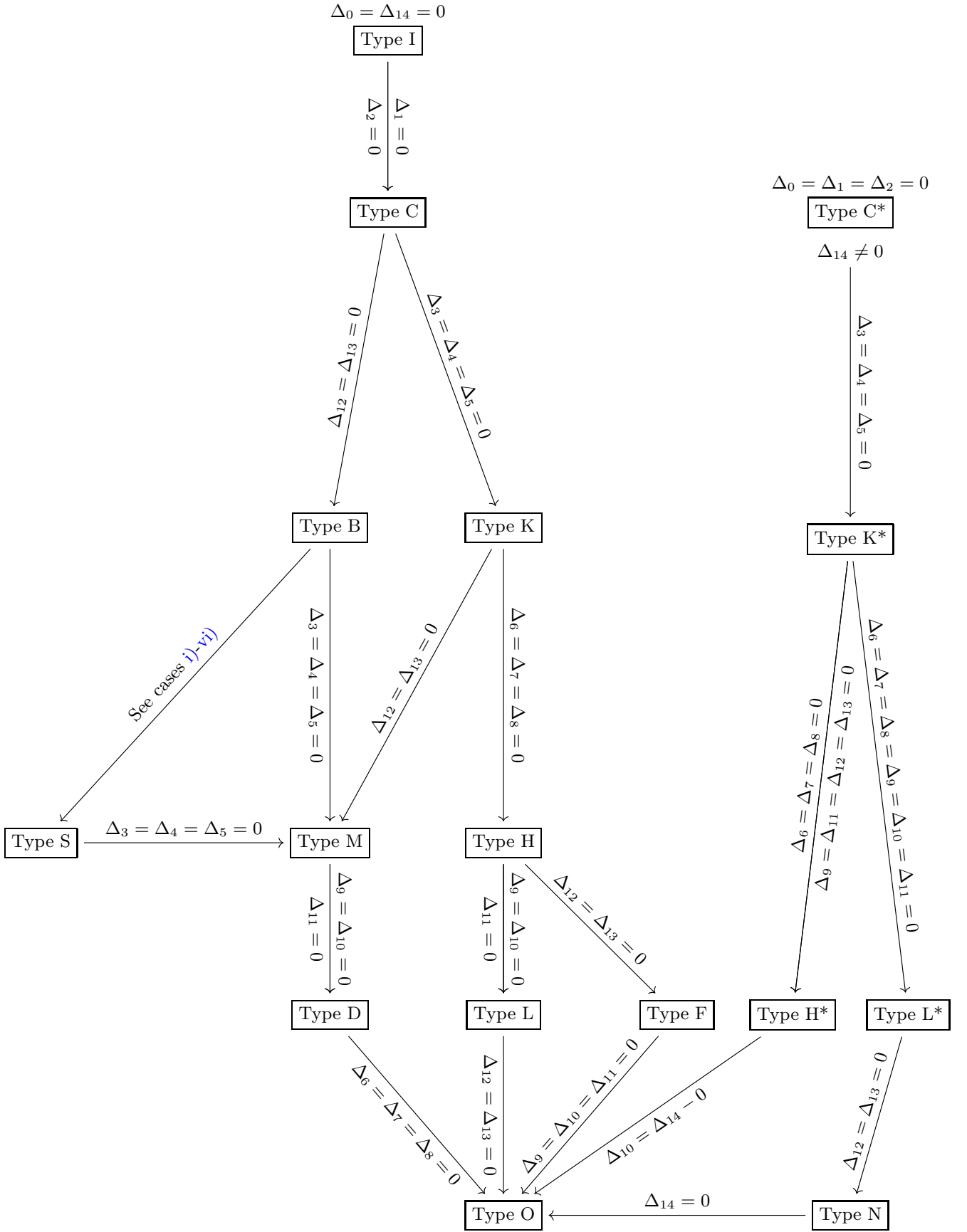

With the algebraic classification of the tensor $\Z_{\lambda\rho\mu\nu}$ settled, we display in Table~\ref{TableZ1PND} a summary of all of the algebraic types obtained in this section, while we show their possible degenerations in Figure~\ref{fig:diagram0}.

\begin{table}[H]
\centering
\renewcommand{\arraystretch}{2}
\resizebox{16cm}{!}{\begin{tabular}{|c|c| c| c|c| }
\hline
\textbf{Type}&\textbf{Main case}& \textbf{Exceptional cases} & \textbf{Complex scalars} & \textbf{Intrinsic characterisation} \\
\hline
\multirow{2}{*}{I}& \multirow{2}{*}{$(\textrm{I},\textrm{I})$} & \multirow{2}{*}{$(\textrm{I},\textrm{I}_{\infty})$} & \multirow{2}{*}{$\Delta_0=\Delta_{14}=0$}  & $\Z_{\lambda\rho\mu[\nu}l_{\sigma]}l^\lambda l^\rho l^\mu=0$
\\
&& & &$\Z_{\lambda\rho\mu[\nu}k_{\sigma]}k^\lambda k^\rho k^\mu=0$
\\
\hline
\multirow{2}{*}{C}& \multirow{2}{*}{$(\textrm{II},\textrm{I})$} & \multirow{2}{*}{$(\textrm{II},\textrm{I}_\infty)$}  & \multirow{2}{*}{$ \{\Delta_i\}^{i=0,\cdots,2}=\Delta_{14}=0$ }& $l_{[\omega} \Z_{\lambda]\rho\mu[\nu}l_{\sigma]}l^\rho l^\mu=0$
\\
&& & & $\Z_{\lambda\rho\mu[\nu}k_{\sigma]}k^\lambda k^\rho k^\mu=0$
\\
\hline
\multirow{2}{*}{C*} & \multirow{2}{*}{$(\textrm{II},-)$} & \multirow{2}{*}{-}  & \multirow{1}{*}{e.g. only $ \Delta_3,\Delta_{14}\neq 0$} & $l_{[\omega} \Z_{\lambda]\rho\mu[\nu}l_{\sigma]}l^\rho l^\mu=0$\\
&& & \textrm{and} $\Delta_3/\Delta_{14}\in \mathbb{R}^{-}$ &  $\Z_{\lambda\rho\mu[\nu}k_{\sigma]}k^\lambda k^\rho k^\mu\neq 0\,\,\forall \, k^{\mu}\neq l^{\mu}$
\\
\hline
\multirow{2}{*}{B}& \multirow{2}{*}{$(\textrm{II},\textrm{II})$} & $(\textrm{II},\textrm{II}_\infty)$ & \multirow{2}{*}{$ \{\Delta_i\}^{i=0,\cdots,2}=\{\Delta_i\}^{i=12,\cdots,14}=0$} &  $l_{[\omega} \Z_{\lambda]\rho\mu[\nu}l_{\sigma]}l^\rho l^\mu=0$ \\
&& $(\textrm{II},\textrm{II})_{\textrm{I}_{\infty}}$& & $k_{[\omega} \Z_{\lambda]\rho\mu[\nu}k_{\sigma]}k^\rho k^\mu=0$
\\
\hline
\multirow{3}{*}{S}& \multirow{3}{*}{$(\textrm{II},\textrm{II},\textrm{II})$} & \multirow{3}{*}{-}& \multirow{2}{*}{$\{\Delta_i\}^{i=0,\cdots,2}=\{\Delta_i\}^{i=12,\cdots,14}=0$} &  $l_{[\omega} \Z_{\lambda]\rho\mu[\nu}l_{\sigma]}l^\rho l^\mu=0$ 
\\
&& \multirow{2}{*}{} & \multirow{2}{*}{\textrm{and cases}~\ref{casei}-\ref{Casevi}} &$k_{[\omega} \Z_{\lambda]\rho\mu[\nu}k_{\sigma]}k^\rho k^\mu=0$\\
&& & & $l'_{[\omega} \Z_{\lambda]\rho\mu[\nu}l'_{\sigma]}l'^\rho l'^\mu=0$
\\
\hline
\multirow{2}{*}{K}& \multirow{2}{*}{$(\textrm{III},\textrm{I})$} &  \multirow{2}{*}{$(\textrm{III},\textrm{I}_\infty)$}  &\multirow{2}{*}{ $ \{\Delta_i\}^{i=0,\cdots,5}=\Delta_{14}=0$} & $l^{[\tau}l_{[\omega} \Z^{\lambda]}{}_{\rho]\mu[\nu}l_{\sigma]}l^{\mu}=0$ 
\\
&& & &$\Z_{\lambda\rho\mu[\nu}k_{\sigma]}k^\lambda k^\rho k^\mu=0$
\\
\hline
\multirow{2}{*}{K*}& \multirow{2}{*}{$(\textrm{III},-)$} & \multirow{2}{*}{-} &  $  \{\Delta_i\}^{i=0,\cdots,5}=0$ & $l^{[\tau}l_{[\omega} \Z^{\lambda]}{}_{\rho]\mu[\nu}l_{\sigma]}l^{\mu}=0$ \\
&&& and e.g. $\{\Delta_i\}^{i=8,\cdots,13}=0\,,$\,$\Delta_6 =3\Delta_7, \,\Delta_6\bar\Delta_{14}\notin\mathbb{R}$ & $\Z_{\lambda\rho\mu[\nu}k_{\sigma]}k^\lambda k^\rho k^\mu\neq 0\,\,\forall \, k^{\mu}\neq l^{\mu}$\\
\hline
\multirow{2}{*}{M}& $\multirow{2}{*}{$(\textrm{III},\textrm{II})$} $& \multirow{1}{*}{$(\textrm{III},\textrm{II}_\infty)$}   &  \multirow{2}{*}{$ \{\Delta_i\}^{i=0,\cdots,5}=\{\Delta_i\}^{i=12,\cdots,14}=0$} & $l^{[\tau}l_{[\omega} \Z^{\lambda]}{}_{\rho]\mu[\nu}l_{\sigma]}l^{\mu}=0$ 
\\
&& $(\textrm{III},\textrm{II})_{\textrm{I}_{\infty}}$ & & $k_{[\omega} \Z_{\lambda]\rho\mu[\nu}k_{\sigma]}k^\rho k^\mu = 0$
\\
\hline
\multirow{2}{*}{D}& \multirow{2}{*}{$(\textrm{III},\textrm{III})$} & \multirow{1}{*}{$(\textrm{III},\textrm{III})_{\textrm{II}_\infty} $} & \multirow{2}{*}{$ \{\Delta_i\}^{i=0,\cdots,5}=\{\Delta_i\}^{i=9,\cdots,14}=0$} &$l^{[\tau}l_{[\omega} \Z^{\lambda]}{}_{\rho]\mu[\nu}l_{\sigma]}l^{\mu}=0$
\\
&&$(\textrm{III},\textrm{III})_{\textrm{I}_{\infty}}$ & & $k^{[\tau} k_{[\omega} \Z^{\lambda]}{}_{\rho]\mu[\nu}k_{\sigma]}k^{\mu}=0$
\\
\hline
\multirow{3}{*}{H}& $\multirow{3}{*}{$(\textrm{IV},\textrm{I})$} $&  \multirow{3}{*}{$(\textrm{IV},\textrm{I}_\infty)$} & \multirow{3}{*}{$ \{\Delta_i\}^{i=0,\cdots,8}=\Delta_{14}=0$} & $l_{[\omega} \Z_{\lambda]\rho\mu[\nu}l_{\sigma]}l^\mu=0$
\\
&& & & $l_{\lambda}l_{[\omega} \Z^\lambda{}_{\rho]\mu\nu}=0$
\\
&& & & $\Z_{\lambda\rho\mu[\nu}k_{\sigma]}k^\lambda k^\rho k^\mu=0$
\\
\hline
\multirow{3}{*}{H*} & \multirow{3}{*}{$(\textrm{IV},-)$} & \multirow{3}{*}{-} & $  \{\Delta_i\}^{i=0,\cdots,8}=0$ & $l_{[\omega} \Z_{\lambda]\rho \mu[\nu}l_{\sigma]}l^\mu=0$ \\
 && & \multirow{2}{*}{\textrm{and e.g.} $\Delta_9=\Delta_{11}=\Delta_{12}=\Delta_{13}=0\,,\Delta_{10}\bar{\Delta}_{14}\notin \mathbb{R}$} & $l_{\lambda}l_{[\omega} \Z^\lambda{}_{\rho]\mu\nu}=0$ \\
 && & & $\Z_{\lambda\rho\mu[\nu}k_{\sigma]}k^\lambda k^\rho k^\mu\neq 0\,\,\forall \, k^{\mu}\neq l^{\mu}$
 \\
\hline
\multirow{3}{*}{F}& \multirow{3}{*}{$(\textrm{IV},\textrm{II})$}& \multirow{2}{*}{$(\textrm{IV},\textrm{II}_\infty)$}  & \multirow{3}{*}{$ \{\Delta_i\}^{i=0,\cdots,8}=\{\Delta_i\}^{i=12,\cdots,14}=0$} & $l_{[\omega} \Z_{\lambda]\rho\mu[\nu}l_{\sigma]}l^\mu=0$
\\
&&\multirow{3}{*}{$(\textrm{IV},\textrm{II})_{\textrm{I}_{\infty}}$}& & $l_{\lambda}l_{[\omega} \Z^\lambda{}_{\rho]\mu\nu}=0$
\\
&& & &  $k_{[\omega} \Z_{\lambda]\rho\mu[\nu}k_{\sigma]}k^\rho k^\mu=0$
\\
\hline
\multirow{2}{*}{L} & \multirow{2}{*}{$(\textrm{V},\textrm{I})$}& \multirow{2}{*}{$(\textrm{V},\textrm{I}_\infty)$} & \multirow{2}{*}{$ \{\Delta_i\}^{i=0,\cdots,11}=\Delta_{14}=0$}&  $l^{[\tau}l_{[\omega} \Z^{\lambda]}{}_{\rho]\mu\nu}=0$ \\
&& & & $\Z_{\lambda\rho\mu[\nu}k_{\sigma]}k^\lambda k^\rho k^\mu=0$\\
\hline
\multirow{2}{*}{L*} & \multirow{2}{*}{$(\textrm{V},-)$} & \multirow{2}{*}{-} &  $ \{\Delta_i\}^{i=0,\cdots,11}=0$  & $l^{[\tau}l_{[\omega} \Z^{\lambda]}{}_{\rho]\mu\nu}=0$ \\
&& & $\Delta_{14}\bar\Delta_{12} -2\Delta_{13} \bar\Delta_{14} =0,$ \, \textrm{and}\, $\Delta_{12}\neq \bar\Delta_{12}$& $\Z_{\lambda\rho\mu[\nu}k_{\sigma]}k^\lambda k^\rho k^\mu \neq 0\,\,\forall \, k^{\mu}\neq l^{\mu}$  \\
\hline
 N&$(\textrm{VI},-)$ & - &  $ \{\Delta_i\}^{i=0,\cdots,13}=0$& $l_{[\sigma} \Z_{\lambda]\rho\mu\nu}=0$  \\
\hline
O&-& - &  $ \{\Delta_i\}^{i=0,\cdots,14}=0$& $\Z_{\lambda\rho\mu\nu}=0$  \\
\hline
\end{tabular}}
\caption{Algebraic types for the tensor $\Z_{\lambda\rho\mu\nu}$. The complex scalars are shown in the preferred null tetrad chosen such that in general the left and right numerals refer to null vectors $l^\mu$ and $k^\mu$, respectively, while for simplicity in the presentation the extra constraints related to the exceptional cases with infinite PNDs are not shown in the table, but they can be found for each case in Sec.~\ref{subsec:Full}.}
\label{TableZ1PND}
\end{table}

%%%%%%%%%%%%%%%%%%%%%%%
\subsection{Refined classification based on the superenergy tensor}\label{sec:supere}

As previously stressed, the tensor $\Z_{\lambda\rho\mu\nu}$ is symmetric in the first pair of indices and antisymmetric in the last one, while it additionally fulfils the algebraic symmetries~\eqref{AlgSym1Z} and~\eqref{AlgSym2Z}. Therefore, its superenergy tensor is given by (formula (19) in~\cite{Senovilla:1999xz}, conveniently adapted):
\begin{eqnarray}
    T_{\alpha\beta\lambda\mu\tau\nu}(\Z) &=& \Z_{\alpha\lambda\tau\rho}\Z_{\beta\mu\nu}{}^\rho + \Z_{\beta\lambda\tau\rho}\Z_{\alpha\mu\nu}{}^\rho +\Z_{\alpha\mu\tau\rho}\Z_{\beta\lambda\nu}{}^\rho +\Z_{\alpha\lambda\nu\rho}\Z_{\beta\mu\tau}{}^\rho \nonumber\\
&&-\,g_{\alpha\beta} \left(\Z_{\sigma\lambda\tau\rho}\Z^\sigma{}_{\mu\nu}{}^\rho  +\Z_{\sigma\lambda\nu\rho}\Z^\sigma{}_{\mu\tau}{}^\rho\right)- g_{\lambda\mu} \left(\Z_{\alpha\sigma\tau\rho}\Z_{\beta}{}^\sigma{}_\nu{}^\rho  +\Z_{\alpha\sigma\nu\rho}\Z_{\beta}{}^\sigma{}_\tau{}^\rho \right) \nonumber\\
&&-\,\frac{1}{2} g_{\tau\nu} \left( \Z_{\alpha\lambda\sigma\rho}\Z_{\beta\mu}{}^{\sigma\rho} +  \Z_{\alpha\mu\sigma\rho}\Z_{\beta\lambda}{}^{\sigma\rho} \right)+g_{\alpha\beta} g_{\lambda\mu} \Z_{\sigma\gamma\tau\rho} \Z^{\sigma\gamma}{}_\nu{}^\rho\nonumber\\
&&+\,\frac{1}{2} g_{\alpha\beta} g_{\tau\nu} \Z_{\sigma\lambda\gamma\rho} \Z^\sigma{}_\mu{}^{\gamma\rho}+\frac{1}{2}g_{\lambda\mu}  g_{\tau\nu} \Z_{\alpha\sigma\gamma\rho} \Z_\beta{}^{\sigma\gamma\rho}-\frac{1}{4} g_{\alpha\beta}g_{\lambda\mu}  g_{\tau\nu} \Z_{\delta\sigma\gamma\rho} \Z^{\delta\sigma\gamma\rho}\,.\label{seZ}
\end{eqnarray}

This tensor has the following direct properties
\begin{equation}
    T_{\alpha\beta\lambda\mu\tau\nu}(\Z) =T_{(\alpha\beta)(\lambda\mu)(\tau\nu)}(\Z) = T_{\lambda\mu\alpha\beta\tau\nu}(\Z) , \hspace{1cm} T_{\alpha\beta\lambda\mu\rho}{}^\rho(\Z)  =0\,.
\end{equation}

%\paragraph{a) Type I}

A more refined algebraic classification of the tensor $\Z_{\lambda\rho\mu\nu}$ can then be achieved by using this superenergy tensor, as outlined in~\cite{Senovilla:2010ej,Senovilla:2006gf}. Any superenergy tensor has the {\em dominant property}, meaning in our case that \begin{equation}
    T_{\alpha\beta\lambda\mu\tau\nu}(\Z)u_1^\alpha u^\beta_2 u^\lambda_3 u^\mu_4 u^\nu_5 u^\tau_6 \geq 0\,,
\end{equation}
for arbitrary future pointing $u^\mu_a$ ($a\in \{1,2,3,4,5,6\}$), in such a way that the equality can only occur if at least one of the $u^\alpha_a$ is null. Thereby, one defines the PND of $\Z_{\lambda\rho\mu\nu}$ as the null $l^\mu$ such that
\begin{equation}
  T_{\alpha\beta\lambda\mu\tau\nu}(\Z)l^\alpha l^\beta l^\lambda l^\mu l^\tau l^\nu =0 \,.\label{pnd}  
\end{equation}
These PNDs are sometimes called aligned null directions (AND), and the classification using the superenergy tensor~\eqref{seZ} is greatly related to the one based on null alignment (see e.g.~\cite{Milson:2004jx}), as the aligned null directions are the PNDs. It must be stressed that relation~\eqref{pnd} is fully equivalent to either~\eqref{PND1} or~\eqref{PND2}.

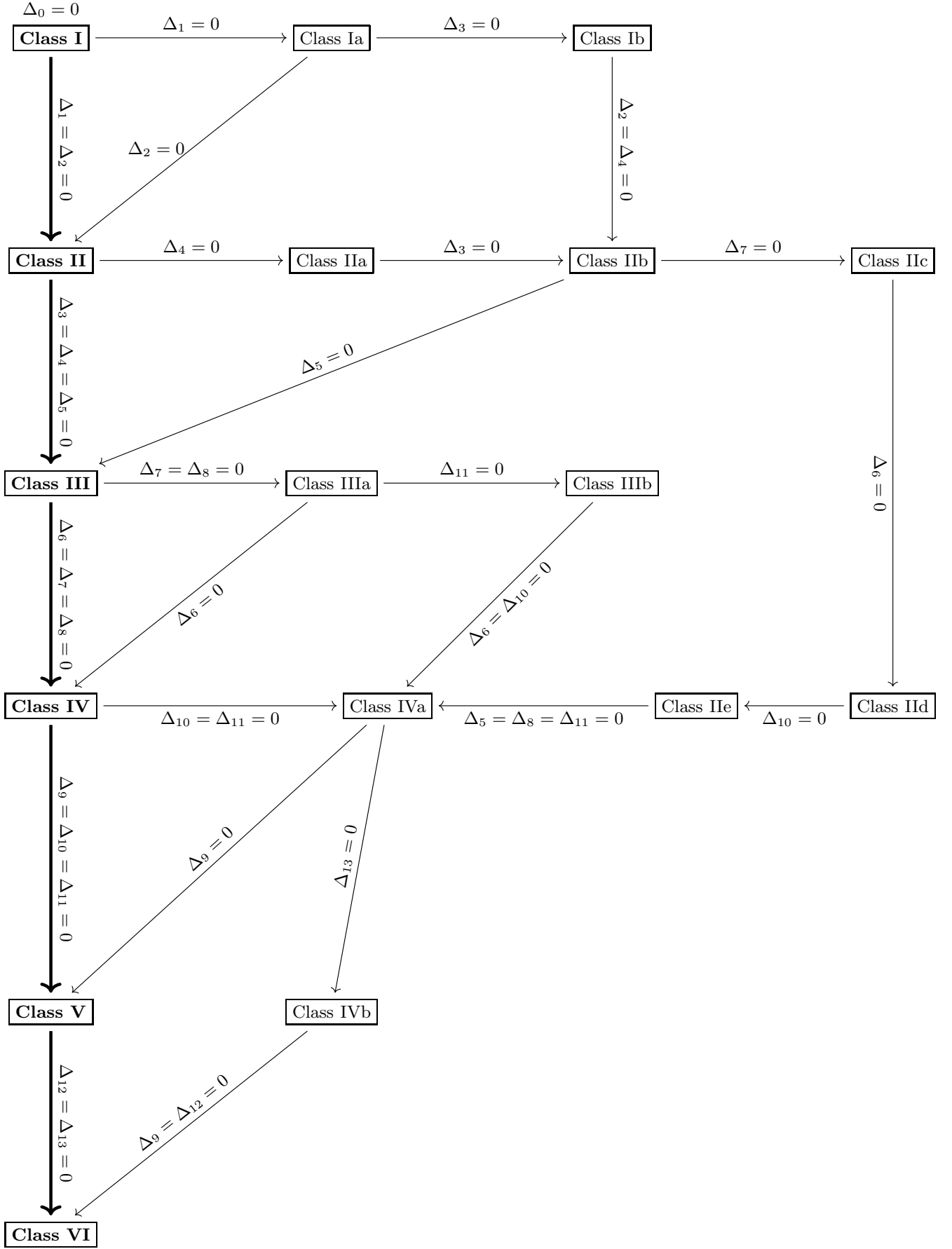
\begin{figure}[H]
    \centering
 \resizebox{1\textwidth}{!}{

\begin{tikzpicture}[node distance=2.5cm and 2cm, auto]

    % Nodes (Main Types aligned to the left)
    \node (I) {\fbox{\textbf{Class I}}};
    \node (II) [below of=I, yshift=-1.5cm] {\fbox{\textbf{Class II}}};
    \node (III) [below of=II, yshift=-1.5cm] {\fbox{\textbf{Class III}}};
    \node (IV) [below of=III, yshift=-1.5cm] {\fbox{\textbf{Class IV}}};
    \node (V) [below of=IV, yshift=-3cm] {\fbox{\textbf{Class V}}};
    \node (VI) [below of=V, yshift=-1.5cm] {\fbox{\textbf{Class VI}}};

 % Arrow and Label for Delta_0 = 0
    \node (Delta0) [above of=I,yshift=-2cm] {\(\Delta_0 = 0\)};

    % Subclasses with adjusted positions
    \node (Ia) [right of=I, xshift=2.5cm] {\fbox{Class Ia}};
    \node (Ib) [right of=Ia, xshift=2.5cm] {\fbox{Class Ib}};
    \node (IIa) [right of=II, xshift=2.5cm] {\fbox{Class IIa}};
    \node (IIb) [right of=IIa, xshift=2.5cm] {\fbox{Class IIb}};
    \node (IIIa) [right of=III, xshift=2.5cm] {\fbox{Class IIIa}};
    \node (IIIb) [right of=IIIa, xshift=2.5cm] {\fbox{Class IIIb}};
    \node (IIc) [right of=IIb, xshift=2.5cm] {\fbox{Class IIc}};
    \node (IId) [below of=IIc, yshift=-5.5cm] {\fbox{Class IId}};
    \node (IIe) [left of=IId,xshift=-1cm] {\fbox{Class IIe}};
    \node (IVa) [right of=IV, xshift=3.5cm] {\fbox{Class IVa}};
    \node (IVb) [right of=V, xshift=2.5cm] {\fbox{Class IVb}};

    % Arrows (Connections)
    \draw[->] (I) -- (Ia) node[midway, above] {\(\Delta_1 = 0\)};
    \draw[->] (Ia) -- (Ib) node[midway, above] {\(\Delta_3 = 0\)};
    \draw[->] (Ia) -- (II) node[midway, left] {\(\Delta_2 = 0\)};
    \draw[->] (II) -- (IIa) node[midway, above] {\(\Delta_4 = 0\)};
    \draw[->] (IIa) -- (IIb) node[midway, above] {\(\Delta_3 = 0\)};
    \draw[->] (Ib) -- (IIb) node[midway, sloped, above] {\(\Delta_2 = \Delta_4 = 0\)};
    \draw[->] (IIb) -- (III) node[midway, sloped,above] {\(\Delta_5 = 0\)};
    \draw[->] (III) -- (IIIa) node[midway, above] {\(\Delta_7 = \Delta_8 = 0\)};
    \draw[->] (IIIa) -- (IIIb) node[midway, above] {\(\Delta_{11} = 0\)};
    \draw[->] (IIb) -- (IIc) node[midway, above, sloped] {\(\Delta_7 = 0\)};
    \draw[->] (IIc) -- (IId) node[midway,sloped,below] {\(\Delta_6 = 0\)};
    \draw[->] (IId) -- (IIe) node[midway, below, sloped] {\(\Delta_{10} = 0\)};
     \draw[->] (IIIa) -- (IV) node[midway, below, sloped] {\(\Delta_{6} = 0\)};
 %   \draw[->] (IIb) -- (IIe) node[midway, above, sloped] {\(\Delta_{10} = 0\)};
    \draw[->] (IIe) -- (IVa) node[midway, below, sloped] {\(\Delta_5 = \Delta_8 =\Delta_{11}= 0\)};
    \draw[->] (IIIb) -- (IVa) node[midway, sloped, below] {\(\Delta_{6} = \Delta_{10} = 0\)};
     \draw[->] (IV) -- (IVa) node[midway, sloped, below] {\(\Delta_{10}=\Delta_{11} = 0\)};
    \draw[->,ultra thick] (IV) -- (V) node[midway, sloped, above] {\(\Delta_9=\Delta_{10} = \Delta_{11} = 0\)};
    \draw[->] (IVa) -- (V) node[midway, sloped, above] {\(\Delta_9 = 0\)};
    \draw[->] (IVa) -- (IVb) node[midway, sloped, above] {\(\Delta_{13} = 0\)};
    \draw[->] (IVb) -- (VI) node[midway, above, sloped] {\(\Delta_9=\Delta_{12} = 0\)};
    \draw[->,ultra thick] (V) -- (VI) node[midway, above, sloped] {\(\Delta_{12}=\Delta_{13} = 0\)};
    \draw[->, ultra thick] (I) -- (II) node[midway, above, sloped] {\(\Delta_{1}=\Delta_{2} = 0\)};
     \draw[->,ultra thick] (II) -- (III) node[midway, above, sloped] {\(\Delta_3=\Delta_{4}=\Delta_{5} = 0\)};
      \draw[->,ultra thick] (III) -- (IV) node[midway, above, sloped] {\(\Delta_{6}=\Delta_{7}=\Delta_8 = 0\)};

\end{tikzpicture}

}
    \caption{Flow diagram of the alignment classes of the tensor $\Z_{\lambda\rho\mu\nu}$ derived from its superenergy tensor.}
    \label{fig:diagram}
\end{figure}

The refined classification simply analyses the level of alignment of any particular PND by finding the actual number of contractions with $l^\mu$ needed to get the zero on the right-hand side of~\eqref{pnd}. This is efficiently achieved by removing, in an orderly manner, instances of the given PND from the original equation~\eqref{pnd} step by step. For simplicity in the presentation, we derive in detail the aforementioned classification in~\ref{superenergyappendix}, choosing $l^\mu$ as the given PND, which allows us to find seventeen different alignment classes. The main results can be summarised in Table~\ref{TableZ1}, while in Figure~\ref{fig:diagram} we show a flow diagram, specifying how all of these classes are related. 

Once the refined classes have been identified, a more elaborate classification can be achieved. The basic idea is to consider each of the 15 main types and particularise the two (or exceptionally three) Roman numerals to the different possibilities arising in Table~\ref{TableZ1}. The full classification considers all combinations of possibilities derived from that table, is too long but straightforward to get, and thus we will just explain how to derive it by exhibiting illustrative examples. 

The most obvious refinement arises for Type S, and has already been identified leading to the more specific cases
$$
(\textrm{IIa},\textrm{IIa},\textrm{IId})\hspace{1cm}  \mbox{and} \hspace{1cm} 
(\textrm{II}_{\Delta_{3}=0},\textrm{II}_{\Delta_{3}=0},\textrm{IIa})\,.
$$
Types N and L* cannot be refined, but Type L can as
$$
(\textrm{V},\textrm{I}), \hspace{3mm}(\textrm{V},\textrm{Ia}), \hspace{3mm} (\textrm{V},\textrm{Ib}).
$$
Type F, for instance, will lead to 18 subtypes by combining the three classes IV, IVa and IVb with the classes II, IIa, IIb, IIc, IId, and IIe. And the type with more subcases is Type B, with a total of 36 subpossibilities. And so on and so forth. The notation for each case is also obvious.

\begin{table}[H]
\centering
\renewcommand{\arraystretch}{1.8}
\begin{tabular}{|c|c| c| c| c|}
\hline
\textbf{Alignment Class}& \textbf{$bo(l)$}& \textbf{Superenergy} & \textbf{Complex scalars} & \textbf{Intrinsic characterisation} \\
\hline
 I & 2& $T_{\alpha\beta\lambda\mu\tau\nu}l^\alpha l^\beta l^\lambda l^\mu l^\nu l^\tau=0$& $\Delta_0=0$ & $   \Z_{\lambda\rho\mu[\nu}l_{\sigma]}l^\lambda l^\rho l^\mu=0$ \\
 \hline
Ia &2 &$T_{\alpha\beta\lambda\mu\tau\nu}l^\alpha l^\beta l^\lambda l^\mu l^\tau=0$ & $\Delta_0=\Delta_1=0 $ & $ \Z_{\lambda\rho\mu\nu}l^\lambda l^\rho l^\mu = \Z_{\lambda\rho[\mu\nu}l_{\sigma]}l^\lambda l^\rho=0$\\
 \hline
 II&1 & $ T_{\alpha\beta\lambda\mu\tau\nu} l^\beta l^\lambda l^\mu l^\tau l^\nu=0$ & $ \Delta_0=\Delta_1=\Delta_2=0$ &$l_{[\omega} \Z_{\lambda]\rho\mu[\nu}l_{\sigma]}l^\rho l^\mu=0$\\
 \hline
 Ib& 2&$ T_{\alpha\beta\lambda\mu\tau\nu} l^\alpha l^\beta l^\lambda l^\mu=0$ & $\Delta_0= \Delta_1=\Delta_3=0 $ &$\Z_{\lambda\rho\mu\nu}l^\lambda l^\rho=0$ \\
 \hline
 IIa & 1 & $ T_{\alpha\beta\lambda\mu\tau\nu} l^\alpha l^\beta  l^\tau l^\nu =0$ & $\Delta_0= \Delta_1=\Delta_2=\Delta_4=0$& $\Z_{\lambda\rho\mu[\nu}l_{\sigma]}l^\rho l^\mu=0$\\
 \hline
 IIb& 1& $ T_{\alpha\beta\lambda\mu\tau\nu}l^\alpha l^\beta l^\lambda l^\tau =0$ & $ \{\Delta_i\}^{i=0,\cdots,4}=0$ &  $\Z_{\lambda\rho\mu\nu}l^{\lambda}l^{\rho}=l_{[\omega} \Z_{\lambda]\rho[\mu\nu}l_{\sigma]}l^\rho=0$ \\
 \hline
 III &0 &$ T_{\alpha\beta\lambda\mu\tau\nu}  l^\alpha l^\lambda l^\tau l^\nu =0$ & $ \{\Delta_i\}^{i=0,\cdots,5}=0$& $l^{[\tau}l_{[\omega} \Z^{\lambda]}{}_{\rho]\mu[\nu}l_{\sigma]}l^{\mu}=0$ \\
 \hline
 IIc&1 & $T_{\alpha\beta\lambda\mu\tau\nu}l^\alpha l^\beta l^\tau=0$& $ \{\Delta_i\}^{i=0,\cdots,4}=\Delta_7=0$& $\Z_{\lambda\rho\mu\nu}l^{\rho}l^{\mu}=\Z_{\lambda\rho[\mu\nu}l_{\sigma]}l^{\lambda}=0$ \\
 \hline
 IId &  1  & $T_{\alpha\beta\lambda\mu\tau\nu}l^\alpha l^\beta l^\lambda=0$& $ \{\Delta_i\}^{i=0,\cdots,4}=\Delta_6=\Delta_7=0$&$ l_{\lambda}l_{[\omega} \Z^\lambda{}_{\rho]\mu\nu}=0$ \\
 \hline
IIIa&0    &$T_{\alpha\beta\lambda\mu\tau\nu}l^\lambda l^\tau l^\nu=0$ & $ \{\Delta_i\}^{i=0,\cdots,5}= \Delta_7=\Delta_8=0$&  $l_{[\omega}\Z_{\lambda]\rho\mu[\nu}l_{\sigma]}l^\mu=0$\\
 \hline
 IV & -1 & $T_{\alpha\beta\lambda\mu\tau\nu}l^\alpha l^\lambda  l^\tau=0$& $\{\Delta_i\}^{i=0,\cdots,8}=0$& $l_{[\omega}\Z_{\lambda]\rho\mu[\nu}l_{\sigma]}l^\mu=l_{\lambda}l_{[\omega} \Z^\lambda{}_{\rho]\mu\nu}=0$ \\
 \hline
 IIe& 1  & $T_{\alpha\beta\lambda\mu\tau\nu}l^\alpha l^\beta=0$&$ \{\Delta_i\}^{i=0,\cdots,4}=\Delta_6=\Delta_7=\Delta_{10}=0$ & $\Z_{\lambda\rho\mu\nu}l^\lambda=0$\\
 \hline
IIIb & 0  & $T_{\alpha\beta\lambda\mu\tau\nu}l^\tau l^\nu=0$&$ \{\Delta_i\}^{i=0,\cdots,5}= \Delta_7=\Delta_8=\Delta_{11}=0$  &$\Z_{\lambda\rho\mu[\nu}l_{\sigma]}l^\mu=0$ \\
 \hline
IVa &-1& $T_{\alpha\beta\lambda\mu\tau\nu}l^\lambda l^\tau =0$& $ \{\Delta_i\}^{i=0,\cdots,8}=\Delta_{10}=\Delta_{11}=0$ & $l_{[\omega}\Z_{\lambda]\rho\mu\nu}l^{\nu}=l_{[\omega}\Z_{\lambda]\rho[\mu\nu}l_{\sigma]}=0$\\
 \hline
V & -2 & $T_{\alpha\beta\lambda\mu\tau\nu}l^\alpha l^\lambda  =0$& $\{\Delta_i\}^{i=0,\cdots,11}=0$ & $l^{[\tau}l_{[\omega}\Z^{\lambda]}{}_{\rho]\mu\nu}=0$\\
 \hline
IVb & -1 & $ T_{\alpha\beta\lambda\mu\tau\nu} l^\nu  =0$ &$ \{\Delta_i\}^{i=0,\cdots,8}=\Delta_{10}=\Delta_{11}=\Delta_{13}=0$ &$\Z_{\lambda\rho\mu\nu}l^{\nu}=\Z_{\lambda\rho[\mu\nu}l_{\sigma]}=0$ \\
 \hline
VI & -3&$ T_{\alpha\beta\lambda\mu\tau\nu} l^\alpha  =0$ & $ \{\Delta_i\}^{i=0,\cdots,13}=0$ & $l_{[\sigma}\Z_{\lambda]\rho\mu\nu}=0$\\
\hline
\end{tabular}
\caption{Alignment classes for the tensor $\Z_{\lambda\rho\mu\nu}$ derived from its superenergy tensor.}
\label{TableZ1}
\end{table}
%\iffalse Types renamed as follows:
%\begin{tabular}{|c|c|}
%Type I& Class I (bo=2)\\
%Type I'& Class Ia (bo=2)\\
%Type Ii& Class II   (bo=1)\\
%Type I''&Class Ib (bo=2) \\
%Type Ii'& Class IIa (bo=1)\\
%Type Ii''& Class IIb  (bo=1)\\
%Type II& Class III (bo=0)\\
%Type Ii'''& Class IIc  (bo=1)\\
%Type Ii'''' & Class IId  (bo=1)\\
%Type II' & Class IIIa (bo=0)\\
%Type III & Class IV (bo=-1)\\
%Type Ii''''' & Class IIe (bo=1)\\
%Type II'' & Class  IIIb (bo=0)\\
%Type III' & Class IVa (bo= -1)\\
%Type IIIi & Class V (bo=-2)\\
%Type IIIi' & Class IVb  (bo=-1)\\
%Type N & Class VI (bo=-3)\\
%\end{tabular}}
%\fi

\section{Algebraic types of Reissner-Nordström-like solutions with dynamical torsion and nonmetricity}\label{sec:application}

Once the algebraic classification in general metric-affine geometries is clear, it is possible to characterise any solution of the field equations of MAG according to its algebraic types. Hence, we consider Reissner-Nordström-like solutions endowed with dynamical torsion and nonmetricity, which in fact represent the broadest family of static and spherically symmetric black hole solutions with spin, dilation and shear charges in MAG.

The MAG model associated with the solutions is described by the gravitational action~\cite{Bahamonde:2022kwg}:
\begin{align}\label{full_action}
S = &\,\frac{1}{64\pi}\int
\Bigl[
-\,4R-6d_{1}\tilde{R}_{\lambda\left[\rho\mu\nu\right]}\tilde{R}^{\lambda\left[\rho\mu\nu\right]}-9d_{1}\tilde{R}_{\lambda\left[\rho\mu\nu\right]}\tilde{R}^{\mu\left[\lambda\nu\rho\right]}+2d_{1}\big(\tilde{R}_{[\mu\nu]}+\hat{R}_{[\mu\nu]}\bigr)\bigl(\tilde{R}^{[\mu\nu]}+\hat{R}^{[\mu\nu]}\bigr)
\Bigr.
\nonumber\\
\Bigl.
&+18d_{1}\tilde{R}_{\lambda\left[\rho\mu\nu\right]}\tilde{R}^{(\lambda\rho)\mu\nu}-3d_{1}\tilde{R}_{(\lambda\rho)\mu\nu}\tilde{R}^{(\lambda\rho)\mu\nu}+6d_{1}\tilde{R}_{(\lambda\rho)\mu\nu}\tilde{R}^{(\lambda\mu)\rho\nu}+2\left(2e_{1}-f_{1}\right)\tilde{R}^{\lambda}\,_{\lambda\mu\nu}\tilde{R}^{\rho}\,_{\rho}\,^{\mu\nu}
\Bigr.
\nonumber\\
\Bigl.
&+8f_{1}\tilde{R}_{(\lambda\rho)\mu\nu}\tilde{R}^{(\lambda\rho)\mu\nu}-2f_{1}\bigl(\tilde{R}_{(\mu\nu)}-\hat{R}_{(\mu\nu)}\bigr)\bigl(\tilde{R}^{(\mu\nu)}-\hat{R}^{(\mu\nu)}\bigr)+3\left(1-2a_{2}\right)T_{[\lambda\mu\nu]}T^{[\lambda\mu\nu]}
\Bigr]d^4x\sqrt{-g}\,.
\end{align}
As can be seen, the model constitutes an extension of GR in the presence of dynamical torsion and nonmetricity, whose field strength tensors are given by deviations from the first and third Bianchi identities of GR. In terms of building blocks of the curvature tensor, the action reads
\begin{align}
S = &\,\frac{1}{64\pi}\int
\Big[
-4R-9d_{1}{\nearrow\!\!\!\!\!\!\!\tilde{R}}^{(T)}_{\lambda[\rho\mu\nu]}{\nearrow\!\!\!\!\!\!\!\tilde{R}}^{(T)}{}^{\lambda[\rho\mu\nu]}+2d_{1}\tilde{R}^{(T)}_{[\mu\nu]}\tilde{R}^{(T)}{}^{[\mu\nu]}-\frac{d_{1}}{8}\ast\tilde{R}^2+\frac{1}{8}\left(d_{1}+32e_{1}\right)\tilde{R}^{\lambda}{}_{\lambda\mu\nu}\tilde{R}^{\rho}{}_{\rho}{}^{\mu\nu}  \nonumber\\
&+8f_{1}^{}  {}^{(1)}\tilde{Z}_{\lambda\rho\mu\nu}{}^{(1)}\tilde{Z}^{\lambda\rho\mu\nu}+\frac{1}{3}\left(4f_{1}-3d_{1}\right){\nearrow\!\!\!\!\!\!\!\tilde{R}}^{(Q)}_{\lambda [\rho\mu\nu]}{\nearrow\!\!\!\!\!\!\!\tilde{R}}^{(Q)}{}^{\lambda  [\rho\mu\nu]}+\frac{1}{6}\left(3d_{1}+16f_{1}\right)\hat{R}^{(Q)}_{[\mu\nu]}\hat{R}^{(Q)}{}^{[\mu\nu]}+d_{1}\tilde{R}^{(T)}_{[\mu\nu]}\tilde{R}^{\lambda}{}_{\lambda}{}^{\mu\nu}\nonumber\\
&+6d_{1}{\nearrow\!\!\!\!\!\!\!\tilde{R}}^{(T)}_{\lambda[\rho\mu\nu]}{\nearrow\!\!\!\!\!\!\!\tilde{R}}^{(Q)}{}^{\lambda[\rho\mu\nu]}+2d_{1}\tilde{R}^{(T)}_{[\mu\nu]}\hat{R}^{(Q)}{}^{[\mu\nu]}+\frac{1}{2} d_{1}\hat{R}^{(Q)}_{[\mu\nu]}\tilde{R}^{\lambda}{}_{\lambda}{}^{\mu\nu}+3\left(1-2a_{2}\right)T_{[\lambda\mu\nu]}T^{[\lambda\mu\nu]}
\Big]d^4x\sqrt{-g}\,.
\end{align}
Thereby, it introduces $\{{\nearrow\!\!\!\!\!\!\!\tilde{R}}^{(T)}_{\lambda[\rho\mu\nu]},\tilde{R}^{(T)}_{[\mu\nu]},\ast\tilde{R},\tilde{R}^{\lambda}{}_{\lambda\mu\nu},\Z_{\lambda\rho\mu\nu},{\nearrow\!\!\!\!\!\!\!\tilde{R}}^{(Q)}_{\lambda [\rho\mu\nu]},\hat{R}^{(Q)}_{[\mu\nu]}\}$ as field strength tensors for torsion and nonmetricity, the latter including nontrivial trace and traceless parts.

By setting the form of the metric, torsion and nonmetricity tensors relative to a static and spherically symmetric space-time~\cite{Hohmann:2019fvf}:
\begin{equation}\label{sph_metric}
    ds^{2}=\Psi_{1}(r)dt^{2} -\frac{dr^{2}}{\Psi_{2}(r)}-r^{2}d\vartheta^{2}-r^{2}\sin^{2}\vartheta \, d\varphi^{2}\,,
\end{equation}
we can consider null vectors
\begin{align}
    l^\mu&=\bigg\{\frac{1}{\sqrt{2}}\left(\frac{\Psi_{2}(r)}{\Psi^{3}_{1}(r)}\right)^{1/4},-\,\frac{1}{\sqrt{2}}\left(\frac{\Psi^{3}_{2}(r)}{\Psi_{1}(r)}\right)^{1/4},0,0\bigg\}\,, \quad m^\mu=\bigg\{0,0,\frac{i}{\sqrt{2}r},\frac{\csc\vartheta}{\sqrt{2}r}\bigg\}\,,\\
    k^\mu&=\bigg\{\frac{1}{\sqrt{2}\left(\Psi_{1}(r)\Psi_{2}(r)\right)^{1/4}},\frac{\left(\Psi_{1}(r)\Psi_{2}(r)\right)^{1/4}}{\sqrt{2}},0,0\bigg\}\,, \quad \bar{m}^\mu=\bigg\{0,0,-\,\frac{i}{\sqrt{2}r},\frac{\csc\vartheta}{\sqrt{2}r}\bigg\}\,,
\end{align}
where $l^{\mu}$ and $k^{\mu}$ correspond to radially ingoing and outgoing null geodesics of the static and spherically symmetric space-time, respectively. Then, given the fact that the method of PNDs applied to the tensor $\Z_{\lambda\rho\mu\nu}$ provides a completely new algebraic classification, with a much richer collection of algebraic types in comparison with the well-known algebraic types of the rest of the field strength tensors of the model, it is worthwhile to study its algebraic structure in a static and spherically symmetric space-time.

First of all, it turns out that the only nontrivial complex scalars of the tensor $\Z_{\lambda\rho\mu\nu}$ in a general static and spherically symmetric space-time are $\Delta_{1}$, $\Delta_{7}$ and $\Delta_{13}$. This immediately tells us that $l_\mu$ and $k_\mu$ constitute PNDs of Class I for this tensor, unless some of the mentioned complex scalars vanishes. To see if there are any other PNDs, we simply analyse Eq.~\eqref{EqtoSolve} for the rotated principal scalar, which is reduced to
\begin{equation}
    \Delta'_0=4\epsilon\left(\Delta_{1}+3\epsilon\bar{\epsilon}\Delta_{7}+\epsilon^{2}\bar{\epsilon}^{2}\Delta_{13}\right)=0\, .
\end{equation}
Specifically, there will be no $\epsilon\neq 0$ solutions, and therefore no further PNDs for the tensor $\Z_{\lambda\rho\mu\nu}$, unless one of the following conditions hold:
\be\label{1st}
\frac{\pm\left(9\Delta_{7}^{2}-4\Delta_{1}\Delta_{13}\right)^{1/2}-3\Delta_{7}}{2\Delta_{13}} \in \mathbb{R}^{+}\,, \hspace{1cm} \Delta_{13}\neq 0\,,
\ee
 or 
\be\label{2nd}
-\,\frac{\Delta_1}{\Delta_7}\in \mathbb{R}^{+}\,, \hspace{1cm}\Delta_{13}=0\,, \, \, \, \Delta_7\neq 0\,.
\ee
Thereby, if neither~\eqref{1st} nor~\eqref{2nd} holds, then the tensor $\Z_{\lambda\rho\mu\nu}$ in a general static and spherically symmetric space-time is of Type I, with only two PNDs of Class I (i.e. case $(\textrm{I},\textrm{I})$).

By contrast, if~\eqref{1st} holds, there exist infinite nontrivial solutions of Eq.~\eqref{EqtoSolve}, where the modulus $|\epsilon|$ is fixed but the phase remains arbitrary. In this case, on top of the PNDs $l_\mu$ and $k_\mu$ associated with the trivial solution $\epsilon = 0$, there is then an infinite number of different PNDs. In general, all of them will be of Class I, thus leading to a Type I$_e$ of the kind
$$
(\textrm{I},\textrm{I}_\infty)\,,
$$
unless the further constraint
\begin{equation}
    9\Delta_{7}^{2}-4\Delta_{1}\Delta_{13}=0\,, \hspace{1cm} -\,\frac{\Delta_1}{\Delta_7}\in \mathbb{R}^+\,,
\end{equation}
is satisfied, in which case there is {\em double} solution for the norm
\begin{equation}
|\epsilon|^2  = -\,\frac{2\Delta_1}{3\Delta_7} = -\,\frac{3\Delta_7}{2\Delta_{13}}\,.
\end{equation}
In this particular case, it is straightforward to check by formulas~\eqref{Delta'0}-\eqref{Delta'14} that the above value of $|\epsilon|$ implies
\begin{equation}
\Delta'_0 =\Delta'_1 =\Delta'_2=0\,, \hspace{3mm} \Delta'_3 =-\,\bar\epsilon \Delta_7 \neq 0\,,
\end{equation}
which means that the infinite PNDs are of Class II. Thus, in this case the tensor $\Z_{\lambda\rho\mu\nu}$ is of Type B$_e$, version
$$
(\textrm{II},\textrm{II}_\infty) \equiv (\textrm{II}_\infty)\,,
$$
with two extra PNDs of Class I.%, associated with the trivial solution $\epsilon = 0$.

On the other hand, if~\eqref{2nd} holds, then $\Delta'_{8}=\Delta'_{9}=\Delta'_{10}=\Delta'_{11}=\Delta'_{12}=\Delta'_{13}=\Delta'_{14}=0\,,\Delta'_{7} \neq 0$, which means that $k_\mu$ is a PND of Class III, and for any value of $\epsilon$ such that
\begin{equation}
|\epsilon |=+\sqrt{\frac{-\Delta_1}{3\Delta_7}}\,,
\end{equation}
then $l'_\mu$ defines an infinite number of extra PNDs of Class I. This is a Type K$_e$, or
$$
(\textrm{III},\textrm{I}_\infty)\,.
$$

If $\Delta_{13}= 0$, but~\eqref{2nd} does not hold, then the PND $k_\mu$ is of Class III, and the only different PND is $l_\mu$. In this case, the tensor $\Z_{\lambda\rho\mu\nu}$ is of Type K, or
$$
(\textrm{III},\textrm{I})\,.
$$

Finally, if $\Delta_1=\Delta_{13}=0$, there are no PNDs different from $l_\mu$ and $k_\mu$, but both of them are of Class III, leading to Type D, or
$$
(\textrm{III},\textrm{III})\,,
$$
whereas, if $\Delta_{7}=\Delta_{13}=0$, then $k_\mu$ is actually of Class V and the tensor $\Z_{\lambda\rho\mu\nu}$ becomes Type L, that is
$$
(\textrm{V},\textrm{I})\,.
$$

Once the algebraic structure of the tensor $\Z_{\lambda\rho\mu\nu}$ in a static and spherically symmetric space-time is clear, it is then straightforward to determine its algebraic type for the Reissner-Nordstr\"{o}m-like solutions of the model. In this case, the metric functions read
\begin{equation}
    \Psi(r)\equiv\Psi_{1}(r)=\Psi_{2}(r)=1-\frac{2 m}{r}+\frac{d_{1}\kappa_{\rm s}^2-4e_{1}\kappa_{\rm d}^2-2f_{1}\kappa_{\rm sh}^2}{r^2}\,,
\end{equation}
where, on top of the mass $m$, the constants $\kappa_{\rm s}$, $\kappa_{\rm d}$ and $\kappa_{\rm sh}$ represent the spin, dilation and shear charges of the solution. On the other hand, the complex scalar $\Delta_{13}$ vanishes, whereas $\Delta_{1}$ and $\Delta_{7}$ acquire the following values:
\begin{align}\label{Delta1RN}
\Delta_{1} &=\left\{
    \begin{array}{cc}
    -\displaystyle\,\frac{i\kappa_{\rm s}\left[2\kappa_{\rm sh}d_{1}+c_{2}\left(d_{1}-8f_{1}\right)r+2c_{3}\left(d_{1}-8f_{1}\right)r^{-\,\frac{\left(d_{1}-8f_{1}\right)}{\left(d_{1}+8f_{1}\right)}}\right]}{2\left(d_{1}-8f_{1}\right)r^{2}\Psi(r)}\,, & \hspace{0.23cm} \text{if} \; d_{1} \neq \pm 8f_{1}\,; \\
    \\[10pt]
    -\,\displaystyle\frac{i\kappa_{\rm s}\left[\kappa_{\rm sh}\left(1+\log{(r)}\right)+c_{2}r+2c_{3}\right]}{2r^{2}\Psi(r)}\,, & \text{if} \; d_{1} = 8f_{1}\,;\\
    \\[10pt]
    -\,\displaystyle\frac{i\kappa_{\rm s}\left(\kappa_{\rm sh}+c_{2}r\right)}{2r^{2}\Psi(r)}\,, & \hspace{0.31cm}\text{if} \; d_{1} = -\,8f_{1}\,; \\
    \end{array}
    \right.
\end{align}
and 
\begin{equation}
  \Delta_{7}=\frac{\kappa_{\rm sh}}{6r^{2}}\,,\quad  \forall \, d_{1}, f_{1}\in\mathbb{R}\,.
\end{equation}
Therefore, the algebraic type of the tensor $\Z_{\lambda\rho\mu\nu}$ for the Reissner-Nordstr\"{o}m-like solutions is Type K$_e=(\textrm{III},\textrm{I}_\infty)$, except at the points where the complex scalar $\Delta_1$ in Expression~\eqref{Delta1RN} vanishes; at those points, the algebraic type becomes Type D. Similarly, if the spin charge $\kappa_{\rm s}$ vanishes, then $\Delta_1 = 0$ and the algebraic type is always Type D, provided that the shear charge is nonzero. For a vanishing shear charge, but nonzero spin charge, the complex scalar $\Delta_{7}$ vanishes and the algebraic type is Type L $=(\textrm{V},\textrm{I})$, except at the points where $\Delta_{7}$ also vanishes, which corresponds to the trivial Type O.

In addition, for the Reissner-Nordström-like solutions, the Riemannian Weyl and traceless Ricci tensors fulfil the constraints
%note the Riemannian traceless Ricci tensor has $V^{(3)}_{*} > 0$ for a general static and spherically symmetric space-time if $g_{tt} \neq -\,1/g_{rr}$
\begin{align}
    &^{(1)}W_{\lambda \rho \mu [\nu}k_{\omega]}k^{\rho}k^{\mu}={}^{(1)}W_{\lambda \rho \mu [\nu}l_{\omega]}l^{\rho}l^{\mu}=0\,,\\
    &U^{(3)}_{*}=V^{(3)}_{*}=0\,, \quad  W^{(3)}_{*} = \frac{64\left(d_{1}\kappa_{\rm s}^2-4e_{1}\kappa_{\rm d}^2-2f_{1}\kappa_{\rm sh}^2\right)^{4}}{r^{16}}\,,
\end{align}
describing, respectively, algebraic types $[(1\,1)\,1]$ and $[(1,1)\,(1\,1)]$, since the traceless Ricci tensor can be described by a diagonal matrix with two eigenvalues $\lambda_{\pm}=\pm\bigl(d_{1}\kappa_{\rm s}^{2}-4e_{1}\kappa_{\rm d}^{2}-2f_{1}\kappa_{\rm sh}^{2}\bigr)/r^{4}$ and four eigenvectors $\{(1,0,0,0),(0,1,0,0),(0,0,1,0),(0,0,0,1)\}$.

Furthermore, for the field strength tensors ${\nearrow\!\!\!\!\!\!\!\tilde{R}}^{(T)}_{\lambda[\rho\mu\nu]}$ and ${\nearrow\!\!\!\!\!\!\!\tilde{R}}^{(Q)}_{\lambda[\rho\mu\nu]}$, we have
\begin{align}
    \tilde{U}^{(1)}_{*}&=\tilde{V}^{(1)}_{*}=0\,, \quad  \tilde{W}^{(1)}_{*}=\frac{1024\kappa_{\rm sh}^{4}}{81r^{8}}\,,\\
    \tilde{U}^{(2)}_{*}&=\tilde{V}^{(2)}_{*}=\tilde{W}^{(2)}_{*}=0\,,
\end{align}
leading to algebraic types $[2\,(1\,1)]$ and $[(2\,1\,1)]$, respectively, since the former is characterised by two eigenvalues $\lambda_{\pm}=\pm \,2\kappa_{\rm s}/(3r^{2})$ and the latter only by $\lambda = 0$, but both of them give rise to three eigenvectors $\{(1,1,0,0),(0,0,1,0),(0,0,0,1)\}$.

Finally, the field strength tensors $\tilde{R}^{(T)}_{[\mu\nu]}$, $\tilde{R}^{\lambda}{}_{\lambda\mu\nu}$ and $\hat{R}^{(Q)}_{[\mu\nu]}$ satisfy
\begin{align}
    &\bigl(\tilde{R}^{(T)}_{[\mu\nu]}l_{\lambda}-\tilde{R}^{(T)}_{[\mu\lambda]}l_{\nu}\bigr)l^{\mu}=\bigl(\tilde{R}^{(T)}_{[\mu\nu]}k_{\lambda}-\tilde{R}^{(T)}_{[\mu\lambda]}k_{\nu}\bigr)k^{\mu}=0\,,\\
    &\bigl(\hat{R}^{(Q)}_{[\mu\nu]}l_{\lambda}-\hat{R}^{(Q)}_{[\mu\lambda]}l_{\nu}\bigr)l^{\mu}=\bigl(\hat{R}^{(Q)}_{[\mu\nu]}k_{\lambda}-\hat{R}^{(Q)}_{[\mu\lambda]}k_{\nu}\bigr)k^{\mu}=0\,,\\
    &\bigl(\tilde{R}^\rho\,_{\rho\mu\nu}l_{\lambda}-\tilde{R}^\rho\,_{\rho\mu\lambda}l_{\nu}\bigr)l^{\mu}=\bigl(\tilde{R}^\rho\,_{\rho\mu\nu}k_{\lambda}-\tilde{R}^\rho\,_{\rho\mu\lambda}k_{\nu}\bigr)k^{\mu}=0\,,
\end{align}
so that they are doubly aligned with the PNDs $l^{\mu}$ and $k^{\mu}$.

\section{Conclusions}\label{sec:conclusions}

In this work, we have derived the algebraic classification of the gravitational field in general metric-affine geometries, which are characterised by the presence of curvature, torsion and nonmetricity. For this task, we have considered the irreducible decomposition of the curvature tensor under the pseudo-orthogonal group, which in general displays eleven fundamental parts: three of them constituting the generalisations of the Ricci scalar and of the Weyl and Ricci tensors in metric-affine geometry, as well as eight additional quantities that represent field strength tensors for torsion and nonmetricity. Thereby, a study on the algebraic structure of all of these quantities has a relevant interest in the search and analysis of solutions of the field equations of MAG, which in turn can describe a wide variety of systems, such as black holes and stars with intrinsic hypermomentum, gravitational waves and cosmological scenarios.

Taking into account the algebraic symmetries of the eleven fundamental parts of the curvature tensor, they can be sorted into four different categories, each one characterised by its own type of algebraic classification. Specifically, three of these categories match the well-known algebraic classifications of the Weyl, Ricci and Faraday tensors (see Tables~\ref{tab:Algebraictypes1},~\ref{tab:Algebraictypes2} and~\ref{tab:Algebraictypes3}), whereas the last one is related to one of the field strengths of the traceless nonmetricity tensor and provides a completely new algebraic classification. Then, we formally classify this quantity by means of its PNDs and their levels of alignment, finding a total of sixteen algebraic types, whose main properties and possible degenerations are shown in Table~\ref{TableZ1PND} and Figure~\ref{fig:diagram0}. In fact, as pointed out in~\cite{Senovilla:1999xz}, several refinements can also arise when establishing the alignment classes of the PNDs from the superenergy tensor of this quantity, which are displayed in detail in Table~\ref{TableZ1} and Figure~\ref{fig:diagram}.

As an immediate application, we determine the algebraic types for the Reissner-Nordstr\"{o}m-like solutions of MAG, showing that indeed the aforementioned field strength of the traceless nonmetricity tensor presents a rich algebraic structure, in contrast with the Riemannian Weyl and Ricci tensors, as well as with the rest of field strenghts of the torsion and nonmetricity tensors of the solution. In any case, despite of the complexity of the solution, the gravitational field turns out to be algebraically special, which could be relevant to address the corresponding extension to stationary and axisymmetric space-times, by providing a significant simplification of the field equations of the model in such space-times. Further research in this direction will be addressed in future works.

\bigskip
\bigskip
\noindent
\section*{Acknowledgements}
This research was initiated during a visit by J.M.M.S. to the Tokyo Institute of Technology.
The authors would like to thank Mart\'in Sombra for helpful discussions. S.B. is supported by “Agencia Nacional de Investigación y Desarrollo” (ANID), Grant “Becas Chile postdoctorado
al extranjero” No. 74220006. The work of J.G.V. is supported by the Institute for Basic Science (IBS-R003-D1). J.G.V. also acknowledges the JSPS Postdoctoral Fellowships for Research in Japan and KAKENHI Grant-in-Aid for Scientific Research No. JP22F22044.
J.M.M.S. is supported by Basque Government Grant No. IT1628-22, and by Grant No. PID2021-123226NB-I00 funded by the Spanish MCIN/AEI/10.13039/501100011033 together with ``ERDF A way of making Europe''.

\newpage

\appendix

\renewcommand{\thesection}{Appendix \Alph{section}}
% Redefine equation numbering to keep it in the form "A1, A2, ..."
\renewcommand{\theequation}{\Alph{section}\arabic{equation}}
\setcounter{equation}{0} % Reset equation counter

\section{Explicit computations of the alignment classes of $\Z_{\lambda\rho\mu\nu}$ based on its superenergy tensor}\label{superenergyappendix}

In this appendix, we carry out all the computations for the alignment classification of the tensor $\Z_{\lambda\rho\mu\nu}$ using its superenergy tensor~\eqref{seZ}. In general, the main alignment classes based on this method arise by considering all the possible contractions of the null vector $l^\mu$ and the tensor $\Z_{\lambda\rho\mu\nu}$. For this reason, we shall divide the presentation into six different subsections.

%%%%%%%%%%%%%%%%%
\subsection{Contraction of $T_{\alpha\beta\lambda\mu\tau\nu}(\Z)$ with six copies of $l^\mu$: Class I}

The first possible contraction is the superenergy tensor contracted with 6 copies of $l^\mu$, which simply becomes
\begin{equation}
  T_{\alpha\beta\lambda\mu\tau\nu}l^\alpha l^\beta l^\lambda l^\mu l^\nu l^\tau = 4\bigl(\Z_{\alpha\lambda\tau\rho}l^\alpha l^\lambda l^\tau\bigr)\bigl(\Z_{\beta\mu\nu}{}^\rho l^\beta l^\mu l^\nu\bigr)=-\,8 \Delta _{0}{} \bar{\Delta }_{0}{} =0 \,, \label{cond1}
\end{equation}
leading to
\begin{equation}\label{A2}
    \Delta_0=0\,.
\end{equation}
Then, if such a PND exists, the tensor $\Z_{\lambda\rho\mu\nu}$ is said to be of {\bf Class I}. For this case, the maximum $bo(l)$ is 2. 

Let us analyse Eq.~\eqref{cond1} further. This condition implies that the vector $\Z_{\beta\mu\tau}{}^\rho l^\beta l^\mu l^\tau$ is null, and as it is also orthogonal to $l_\rho$, it must be proportional to it yielding~\eqref{PND1}. Conversely, in general one has
\begin{eqnarray}
    \Z_{\lambda\rho\mu\nu} l^\lambda l^\rho l^\mu=\bigl(\Delta _{1}{} + \bar{\Delta }_{1}{}\bigr) l_{\nu}  - \bar{\Delta }_{0}{} m_{\nu}  - \Delta _{0}{} \bar{m}_{\nu}\,.%=(\Delta _{1}{} + \bar{\Delta }_{1}{}) l_{\nu  }\,,
\end{eqnarray}
%that means that this quantity is proportional to $l_\rho$. 
so that the combination
\begin{eqnarray}
    \Z_{\lambda\rho\mu[\nu}  l_{\sigma]}l^\lambda l^\rho l^\mu=\Delta _{0}{} l_{[\nu} \bar{m}_{\sigma] }+\bar{\Delta }_{0}{} l_{[\nu} m_{\sigma] } =0\label{case0d}
\end{eqnarray}
is equivalent to Eq.~\eqref{cond1} or to~\eqref{A2} and represents the intrinsic characterisation of PND for the tensor $\Z_{\lambda\rho\mu\nu}$.

By the ``symmetry'' mentioned in Sec.~\eqref{ap:classi} between the null vectors $l^\mu$ and $k^\mu$, one immediately knows that
\begin{equation}
    \Delta_{14}=0\,,
\end{equation}
is the corresponding characterisation for $k^\mu$ to be a PND, that is to say
\begin{equation}
     \Z_{\lambda\rho\mu[\nu}  k_{\sigma]}k^\lambda k^\rho k^\mu=0\,.
\end{equation}

\subsection{Contraction of $T_{\alpha\beta\lambda\mu\tau\nu}(\Z)$ with five copies of $l^\mu$: Class Ia and Class II}

The next step consists of removing one null vector $l^\mu$ from Expression~\eqref{pnd}. By doing that, there are two independent possibilities, which we shall explain and categorise separately.

\subsubsection{Class Ia}

The first possible contraction is
\begin{equation}
T_{\alpha\beta\lambda\mu\tau\nu}l^\alpha l^\beta l^\lambda l^\mu l^\tau =-\, l^{\alpha  } l^{\beta  } l^{\lambda  } l^{\mu  } l_{\nu  } \,^{(1)}{}\tilde{Z}_{\alpha  \beta  }{}^{\tau  \sigma  } \,^{(1)}{}\tilde{Z}_{\lambda  \mu  \tau  \sigma  } + 4 l^{\alpha  } l^{\tau  } l^{\beta  } l^{\lambda  } l^{\mu  } \,^{(1)}{}\tilde{Z}_{\alpha  \beta  \nu  \sigma  }\, ^{(1)}{}\tilde{Z}_{\lambda  \mu  \tau  }{}^{\sigma  }=0 \,, \label{cond2a} 
\end{equation}
from where one can notice that $\Delta_0=0$ (by contracting it with $l^\nu$). Then, by assuming this, the above expression becomes
\begin{equation}
T_{\alpha\beta\lambda\mu\tau\nu}l^\alpha l^\beta l^\lambda l^\mu l^\tau=-\,8 \Delta _{1}{} \bar{\Delta }_{1}{} l_{\nu  }=0\,. \label{cond2b} 
\end{equation}
 Therefore, this  case (even though it does not imply that $l^\mu$ is a multiple PND) will be labeled as {\bf Class Ia} and is equivalent to having
\begin{equation}
\Delta_0= \Delta_1 =0\,,\label{cond2c}
\end{equation}
meaning that the corresponding $bo$ is 2.
Since Eq.~\eqref{cond2a} includes the specific contraction $\Z_{\alpha\lambda\sigma\rho}l^\alpha l^\lambda$, it is advantageous to compute the explicit dependence on $\Delta_3$ for this term using the representation of the tensor $\Z_{\lambda\rho\mu\nu}$ in terms of complex scalars and null vectors given by~\eqref{tensorZ1}:
\begin{equation}
    \Z_{\alpha\lambda\sigma\rho}l^\alpha l^\lambda =(l_\sigma u_\rho - l_\rho u_\sigma )+f^{(1)}_{\sigma\rho}\,,\quad u_\rho=\Delta _{3}{} m_{\rho  } + \bar{\Delta }_{3}{} \bar{m}_{\rho  }\,, \quad l^\sigma u_\sigma =0\,,\label{appp1}
\end{equation}
where $f^{(1)}_{\sigma\rho}$ is a tensor depending on $\Delta_0$, $\Delta_1$ and their conjugates.
Then, it turns out that the further contractions $\Z_{\beta\mu\nu\rho} l^\beta l^\mu l^\nu$ and $\Z_{\alpha\lambda[\sigma\rho}l_{\beta]}l^\alpha l^\lambda$ depend solely on these two scalars as
\begin{align}
     \Z_{\beta\mu\nu\rho} l^\beta l^\mu l^\nu=&\,\bigl(\Delta _{1}{} + \bar{\Delta }_{1}{}\bigr) l_{\rho  }  - \bar{\Delta }_{0}{} m_{\rho  }  - \Delta _{0}{} \bar{m}_{\rho  }\,,\\
     \Z_{\alpha\lambda[\sigma\rho}l_{\beta]}l^\alpha l^\lambda =&-2 \Delta _{0}{} l_{[\sigma  }k_{\rho  }  \bar{m}_{\beta]  }-2 \bar{\Delta} _{0}{} l_{[\sigma  }k_{\rho  }  m_{\beta]  } + 2\bigl( \Delta _{1}{} - \bar{\Delta} _{1}{} \bigr)l_{[\sigma  } m_{\rho  } \bar{m}_{\beta  ]}\,,
\end{align}
in such a way that Expression~\eqref{cond2c} is equivalent to vanishing these two independent contractions.

Therefore, the intrinsic characterisation of this class is simply given by the constraints
\begin{align}
     \Z_{\beta\mu\nu\rho} l^\beta l^\mu l^\nu&=0\,,\\
     \Z_{\alpha\lambda[\sigma\rho}l_{\beta]}l^\alpha l^\lambda &=0\,.\label{typeIi}
\end{align}

\subsubsection{Class II}

The second possible contraction with five copies of $l_\mu$ is
\begin{equation}
    T_{\alpha\beta\lambda\mu\tau\nu} l^\beta l^\lambda l^\mu l^\tau l^\nu= 4 l^{\tau  } l^{\sigma  } l^{\beta  } l^{\lambda  } l^{\mu  } \,^{(1)}{}\tilde{Z}_{\alpha  \beta  \lambda  \omega  } \,^{(1)}{}\tilde{Z}_{\mu  \tau  \sigma  }{}^{\omega  }  -2 l_{\alpha  } l^{\tau  } l^{\beta  } l^{\lambda  } l^{\mu  } \,^{(1)}{}\,\tilde{Z}_{\beta  }{}^{\sigma  }{}_{\lambda  }{}^{\omega  } \,^{(1)}{}\tilde{Z}_{\mu  \sigma  \tau  \omega  }=0\,,
\end{equation}
where by contracting it with $l^\alpha$ one gets $\Delta_0=0$ and then the above expression becomes
\begin{equation}
       T_{\alpha\beta\lambda\mu\tau\nu} l^\beta l^\lambda l^\mu l^\tau l^\nu=-\,2l_\alpha\,  \Z_{\sigma\lambda\tau\rho}l^\lambda l^\tau \Z^\sigma{}_{\mu\nu}{}^\rho l^\mu l^\nu =-\,4 \bigl(\Delta _{1}{} \bar{\Delta }_{1}{} + \Delta _{2}{} \bar{\Delta }_{2}{}\bigr) l_{\alpha  }=0\label{help3}\,,
\end{equation}
which clearly means $\Delta_1=\Delta_2=0$. This is the second intermediate case of Type I, and now it does state the multiplicity of the PND $l^\mu$ and the maximum $bo(l)$ is 1. We will call this {\bf Class II} and its $\Delta$-scalar characterisation would then read as
\begin{equation}
    \Delta_0= \Delta_1 =\Delta_2 =0\,.\label{cond3Delta}
\end{equation}

Now, by using Expression~\eqref{tensorZ1} and assuming~\eqref{cond3Delta}, one finds
\begin{equation}
    \Z_{\sigma\lambda\tau\rho}l^\lambda l^\tau =l_\sigma P_\rho +Q_\sigma l_\rho+f^{(2)}_{\sigma\rho}\,,\quad  l^\rho P_\rho =0\,, \hspace{3mm} l^\sigma Q_\sigma =0\,,\label{appA2}
\end{equation}
where
\begin{align}
    P_\mu&=\bigl(\Delta _{7}{} + \bar{\Delta }_{7}{}\bigr) l_{\mu  }  - \bar{\Delta }_{4}{} m_{\mu  } - \Delta _{4}{} \bar{m}_{\mu  }\,,\label{Pis}\\
    Q_\mu&=\bigl(\Delta _{7}{} + \bar{\Delta }_{7}{}\bigr) l_{\mu  } -\bigl( \Delta _{3}{} +  \bar{\Delta }_{4}{}\bigr) m_{\mu  } - \bigl( \bar{\Delta }_{3}{} +  \Delta _{4}{}\bigr) \bar{m}_{\mu  }\,,
\end{align}
and $f^{(2)}_{\sigma\rho}$ is a tensor depending on $\Delta_0$, $\Delta_1$, $\Delta_2$ and their conjugates. Then, from~\eqref{appA2} one can intrinsically write the equivalent form of~\eqref{cond3Delta} as
\begin{equation}
    l^\lambda l_{[\alpha}\Z_{\sigma]\lambda\tau[\rho} l_{\beta]} l^\tau =\Delta _{0}{} l_{[\alpha  }k_{\sigma ] }  l_{[\rho  } \bar{m}_{\beta]  } %+\bar{\Delta }_{0}{}  l_{[\alpha  }k_{\sigma]  } l_{[\rho  } m_{\beta  ]} 
    + \Delta _{1}{}  m_{[\alpha  }  l_{\sigma]  }l_{[\rho  }\bar{m}_{\beta]  } %+ \bar{\Delta }_{1}{}  \bar{m}_{[\alpha  }l_{\sigma]  } l_{[\rho  }m_{\beta]  } 
    + \Delta _{2}{} \bar{m}_{[\alpha  }l_{\sigma]  }   l_{[\rho  }\bar{m}_{\beta]  }%+ \bar{\Delta }_{2}{}  m_{[\alpha  }l_{\sigma]  }l_{{\rho  }  m_{\beta } }\\
    +\textrm{c.c.}=0\,,\label{case3d}
\end{equation}
where c.c. stands for complex conjugate.

\subsection{Contraction of $T_{\alpha\beta\lambda\mu\tau\nu}(\Z)$ with four copies of $l^\mu$: Class Ib, Class IIa, Class IIb and Class III}

The next step consists of removing two null vectors $l^\mu$ from Expression~\eqref{pnd}. In that case, there are four independent possibilities which again we shall explain and categorise separately. Notice that, for all of these cases, the condition $\Delta_0=\Delta_1=0$ always holds. This can be straightforwardly seen by taking all the possible independent contractions with four copies and contracting them with $l_\mu$.

\subsubsection{Class Ib}

The first contraction with four copies of $l_\mu$ is
\begin{equation}
  T_{\alpha\beta\lambda\mu\tau\nu} l^\alpha l^\beta l^\lambda l^\mu=  -\,\bigl(l^{\alpha  } l^{\beta  } \, ^{(1)}{}\tilde{Z}_{\alpha  \beta  }{}^{\rho  \sigma  }\bigr)\bigl(l^{\lambda  } l^{\mu  }  \,^{(1)}{}\tilde{Z}_{\lambda  \mu  \rho  \sigma  }\bigr)g_{\nu  \tau  } + 4\bigl( l^{\alpha  } l^{\beta  } \,^{(1)}{}\tilde{Z}_{\alpha  \beta  \nu  \rho  }\bigr)\bigl(l^{\lambda  } l^{\mu  }  \,^{(1)}{}\tilde{Z}_{\lambda  \mu  \tau  }{}^{\rho  }\bigr)=0\,,\label{cond3a}
\end{equation}
from where one easily notices that $\Delta_0=\Delta_1=0$ by contracting it with $l^\tau l^\nu$ and $l^\tau$, respectively. By using those conditions, we arrive at
\begin{equation}
    T_{\alpha\beta\lambda\mu\tau\nu} l^\alpha l^\beta l^\lambda l^\mu =-\,8 \Delta _{3}{} \bar{\Delta }_{3}{} l_{\nu  } l_{\tau  }=0\label{cond3b}\,.
\end{equation}
Thus, this case requires 
\begin{equation}
    \Delta_0= \Delta_1 =\Delta_3 =0\,,\label{Deltacon4}
\end{equation}
and it will be labelled as {\bf Class Ib}. The maximum $bo(l)$ (boost order of $l$) is 2 once again.

From Eq.~\eqref{cond3a}, one notices that the intrinsic characterisation of this case just simplifies as
\be\label{typeIii}
\Z_{\alpha\lambda\sigma\rho}l^\alpha l^\lambda =-\,2 \Delta _{0}{} k_{[\sigma  } \bar{m}_{\rho  ]}+ 2 \Delta _{1}{} (k_{[\sigma  } l_{\rho ] } + m_{[\sigma  } \bar{m}_{\rho ] })+2 \Delta _{3}{} l_{[\sigma  } m_{\rho]  } +\textrm{c.c.}=0\,,
\ee
which is equivalent to the condition~\eqref{Deltacon4}.

\subsubsection{ Class IIa}

The next possible contraction with four copies of $l_\mu$ reads
\begin{equation}
   T_{\alpha\beta\lambda\mu\tau\nu} l^\alpha l^\beta  l^\tau l^\nu = -\,2\bigl(l^{\alpha  }l^{\beta  }  \,^{(1)}{}\tilde{Z}^{\omega  }{}_{\alpha\beta  }{}^{\nu  }\bigr)\bigl( l^{\rho  } l^{\sigma  } \,^{(1)}{}\tilde{Z}_{ \omega\rho   \sigma  \nu  }\bigr)g_{\lambda  \mu  }+ 4\bigl(l^{\alpha  }l^{\beta  }   \,^{(1)}{}\tilde{Z}_{\lambda  \alpha  \beta  \omega  }\bigr)\bigl(l^{\rho  } l^{\sigma  }\,^{(1)}{}\tilde{Z}_{\mu  \rho  \sigma  }{}^{\omega  }\bigr)=0\,,\label{cond4a}
\end{equation}
where again we notice that $\Delta_0=0$ (by contracting it with $l^\lambda l^\mu$) and also $\Delta _{1}{} =\Delta_2=0$ (by contracting it with $l^\mu$). With these conditions, the above equation reduces to
\begin{equation}
  T_{\alpha\beta\lambda\mu\tau\nu} l^\alpha l^\beta  l^\tau l^\nu =-\,8 \Delta _{4}{} \bar{\Delta }_{4}{} l_{\lambda  } l_{\mu  }=0\,.\label{cond4b}
  \end{equation}
 Then, putting all of the conditions together, we find that
\begin{equation}
    \Delta_0= \Delta_1 =\Delta_2 =\Delta_4=0\,.\label{condDelta5a}
\end{equation}
We will denote this case as {\bf Class IIa}. For this case the maximum $bo(l)$ is 1.

From Eq.~\eqref{cond4a}, we notice that the quantity needed to characterise the previous conditions intrinsically is
\be\label{app5}
\Z_{\sigma\lambda\tau\rho}l^\lambda l^\tau = J_\sigma l_\rho+f^{(3)}_{\sigma\rho}, \hspace{1cm} l^\rho J_\rho =0\,,
\ee
where
\begin{equation}
    J_\mu=\bigl(\Delta _{7}{} + \bar{\Delta }_{7}{}\bigr) l_{\mu  }  - \Delta _{3}{} m_{\mu  } - \bar{\Delta }_{3}{} \bar{m}_{\mu  }\,,
\end{equation}
and $f^{(3)}_{\sigma\rho}$ is a tensor depending on $\Delta_0$, $\Delta_1$, $\Delta_2$, $\Delta_4$ and their conjugates. Then, from Expression~\eqref{app5}, we find the intrinsic characterisation
\be\label{typeIii''}
l^\lambda l^\tau \Z_{\sigma\lambda\tau[\rho} l_{\beta]}=- \,\Delta _{0}{} k_{\sigma  }  \bar{m}_{[\rho  }l_{\beta ] }    + \Delta _{1}{} m_{\sigma  } \bar{m}_{[\rho  } l_{\beta  ]} - \Delta _{2}{} \bar{m}_{\sigma  }l_{[\rho  } \bar{m}_{\beta ] }- \Delta _{4}{} l_{\sigma  } \bar{m}_{[\rho  }l_{\beta  ]} +\textrm{c.c.}=0\,,
\ee
which is equivalent to the condition~\eqref{condDelta5a}.

\subsubsection{Class IIb}

The next possibility is given by
\begin{eqnarray}
    T_{\alpha\beta\lambda\mu\tau\nu} l^\alpha l^\beta l^\lambda l^\tau  &=&l^{\alpha  } l^{\beta  }\Big( \frac{1}{2} l_{\mu  } l_{\nu  } \,^{(1)}{}\tilde{Z}_{\alpha  }{}^{\lambda  \omega  \rho  } \,^{(1)}{}\tilde{Z}_{\beta  \lambda  \omega  \rho  }-2  l^{\lambda  } l_{\mu  }\, ^{(1)}{}\tilde{Z}_{\alpha  \omega  \nu  \rho  } \,^{(1)}{}\tilde{Z}_{\beta  }{}^{\omega  }{}_{\lambda  }{}^{\rho  } -  l^{\lambda  } l_{\nu  } \,^{(1)}{}\tilde{Z}_{\beta  \lambda  }{}^{\omega  \rho  }\, ^{(1)}{}\tilde{Z}_{\mu  \alpha  \omega  \rho  } \nonumber\\
&& +\, 2 l^{\omega  }  l^{\lambda  } \,^{(1)}{}\tilde{Z}_{\lambda  \omega  \nu  }{}^{\rho  } \,^{(1)}{}\tilde{Z}_{\mu  \alpha  \beta  \rho  } + 2  l^{\omega  }  l^{\lambda  }\, ^{(1)}{}\tilde{Z}_{\beta  \lambda  \omega  }{}^{\rho  } \,^{(1)}{}\tilde{Z}_{\mu  \alpha  \nu  \rho  }\Big)=0\,. \label{Class IIb} 
\end{eqnarray}
By contracting this expression with $l^\mu l^\nu$, one notices that $\Delta_0=0$, whereas by contracting it with $l^\mu$ and $l^\nu$, one finds $\Delta_1=0$ and $\Delta_2=0$, respectively. Then, by assuming these three conditions, the expression becomes
\begin{equation}
 T_{\alpha\beta\lambda\mu\tau\nu} l^\alpha l^\beta l^\lambda l^\tau  =-\,4\bigl(\Delta _{3}{} \bar{\Delta }_{3}{} + \Delta _{4}{} \bar{\Delta }_{4}{}\bigr) l_{\mu  } l_{\nu  }=0\,,
\end{equation}
where one notices that $\Delta_3=\Delta_4=0$. Hence, the corresponding $\Delta$-scalar version of Eq.~\eqref{Class IIb} reads
\begin{equation}
    \Delta_0= \Delta_1 =\Delta_2=\Delta_3=\Delta_4=0\,,\label{cond6Delta}
\end{equation}
and we will label this case as {\bf Class IIb}. The maximum $bo(l)$ is 1 now.

Let us now find the intrinsic characterisation of this case. From Eq.~\eqref{Class IIb}, we notice that we need to write down the quantity $l^\alpha \Z_{\alpha\beta\lambda \mu}$. Thus, by considering as in the previous cases the explicit form~\eqref{tensorZ1} for the tensor $\Z_{\lambda\rho\mu\nu}$, we find
\be
l^\alpha \Z_{\alpha\beta\lambda \mu} = A\,  l_\beta (l_\lambda k_\mu -l_\mu k_\lambda) +l_\beta (l_\lambda R_\mu -l_\mu R_\lambda) +2 l_\beta h_{[\lambda\mu]} +h_{\beta\mu} l_{\lambda} -h_{\beta\lambda} l_\mu+f^{(1)}_{\beta\lambda\mu}\,,  \label{Zlll}
\ee
where
\begin{equation}
    h_{\mu\nu}= - \,\Delta _{6}{} m_{\mu  } m_{\nu  } % - \bar{\Delta }_{6}{} \bar{m}_{\mu  } \bar{m}_{\nu  } 
    - \Delta _{7}{} m_{\nu  } \bar{m}_{\mu  } +\textrm{c.c.} \,,\quad %- \bar{\Delta }_{7}{} m_{\mu  } \bar{m}_{\nu  }  \,,\\
      R_{\mu}=\Delta _{10}{} m_{\mu  }+\textrm{c.c.}\,, \quad A= -\,\bigl( \Delta _{7}{} +\bar{\Delta }_{7}{}\bigr)\,.\label{help1}% + \bar{\Delta }_{10}{} \bar{m}_{\mu  }\,.%\quad A= -( \Delta _{7}{} +\bar{\Delta }_{7}{})\,.
\end{equation}
are two quantities fully orthogonal to both $l_\mu$ and $k_\mu$, while $f^{(1)}_{\beta\lambda\mu}$ is a tensor depending on the complex scalars $\Delta_0$, $\Delta_1$, $\Delta_2$, $\Delta_3$, $\Delta_4$ and their conjugates. Then, the following contractions solely depend on the aforementioned scalars and provide the intrinsic characterisation of this case:
\begin{eqnarray}
 l^{\alpha  } l^{\beta  }\, ^{(1)}{}\tilde{Z}_{\alpha  \beta  \mu  \nu  }&=&  -\,2 \Delta _{0}{} k_{[\mu  } \bar{m}_{\nu]  } + 2 \Delta _{1}{} \bigl(k_{[\mu  } l_{\nu ] }  - \bar{m}_{[\mu  } m_{\nu  ]}\bigr) -2 \Delta _{3}{}  m_{[\mu  }l_{\nu  ]}  +\textrm{c.c.}=0\,,\\
  l_{[\gamma}l^\alpha \Z_{\beta]\alpha[\lambda \mu} l_{\nu]}&=&2 \Delta _{0}{} k_{[\gamma  }  l_{\beta ] }  \bar{m}_{[\lambda  }l_{\mu  }k_{\nu ] } + 2 \Delta _{1}{} \bigl( l_{[\gamma  } m_{\beta ] }  \bar{m}_{[\lambda  } l_{\mu  }k_{\nu  ]} - l_{[\gamma  }k_{\beta  ]}  \bar{m}_{[\lambda  }m_{\mu  }l_{\nu  ]}\bigr)-2 \Delta _{2}{}  l_{[\gamma  }\bar{m}_{\beta  ]}\bar{m}_{[\lambda  }k_{\mu  } l_{\nu ] } \nonumber\\
    &&+\bigl(\Delta _{3}{}  - \bar{\Delta }_{4}{}\bigr) l_{[\gamma  }m_{\beta ] }\bar{m}_{[\lambda  }   m_{\mu  }l_{\nu  ]}  + \bigl(\bar{\Delta }_{3}{}  - \Delta _{4}{}\bigr) l_{[\gamma  }\bar{m}_{\beta  ]}   m_{[\lambda  } \bar{m}_{\mu  }l_{\nu  ]}+\textrm{c.c.}=0\,,
\end{eqnarray}
%\begin{eqnarray}
 % l^\alpha \Z_{\alpha\beta\lambda [\mu}l^\lambda l_{\nu]}&=&- \Delta _{0}{} k_{\beta  } \bar{m}_{[\mu  } l_{\nu  ]} + \Delta _{1}{} m_{\beta  } \bar{m}_{[\mu  } l_{\nu  ]} + \Delta _{2}{} \bar{m}_{\beta  } \bar{m}_{[\mu  } l_{\nu  ]} + \Delta _{4}{} l_{\beta  } l_{[\mu  } \bar{m}_{\nu ] }+\textrm{c.c.}=0, \, \, \, \mbox{and}  \\
%l_{[\gamma}l^\alpha \Z_{\beta]\alpha\lambda \mu} l^\mu &=& - \Delta _{0}{}\bar{m}_{\lambda  } k_{[\gamma  } l_{\beta]  } +(\Delta _{1}{} + \bar{\Delta }_{1}{}) l_{\lambda  }k_{[\gamma  } l_{\beta ] }  - \Delta _{1}{} \bar{m}_{\lambda  }l_{[\gamma  } m_{\beta ] }   - \Delta _{2}{} \bar{m}_{\lambda  }l_{[\gamma  } \bar{m}_{\beta  ]} \nonumber\\
%&&+ (\Delta _{3}{} + \bar{\Delta }_{4}{})  m_{\beta  }l_{[\gamma  } l_{\lambda ] }  +\textrm{c.c.}=0\,, 
%\end{eqnarray}
which together are therefore equivalent to Expression~\eqref{cond6Delta}.

\subsubsection{Class III}

The final possibility with four copies of $l_\mu$ is defined by
\begin{eqnarray}
    T_{\alpha\beta\lambda\mu\tau\nu}  l^\alpha l^\lambda l^\tau l^\nu &=&l^{\alpha  } \bigl(2 l^{\varphi  } l^{\lambda  } l_{\mu  } \,^{(1)}{}\tilde{Z}_{\beta  \rho   \omega\alpha   } ^{(1)}{}\tilde{Z}_{\lambda  }{}^{\rho  }{}_{\varphi  }{}^{\omega  }-l_{\beta  } l^{\lambda  } l_{\mu  }\,^{(1)}{}\tilde{Z}^{\varphi  \rho \omega }{}_{\alpha  }\, ^{(1)}{}\tilde{Z}_{\varphi  \rho  \lambda  \omega  }  -2 l^{\varphi  } l^{\rho  } l^{\lambda  }\, ^{(1)}{}\tilde{Z}_{\beta  \mu   \omega\alpha   } \,^{(1)}{}\tilde{Z}_{\lambda  \varphi  \rho  }{}^{\omega  }    \nonumber\\
    && +2 l^{\varphi  } l_{\beta  } l^{\lambda  }\,^{(1)}{}\tilde{Z}_{\mu  \rho  \omega \alpha   }\, ^{(1)}{}\tilde{Z}_{\lambda  }{}^{\rho  }{}_{\varphi  }{}^{\omega  }  - 2 l^{\varphi  } l^{\rho  } l^{\omega  } \,^{(1)}{}\tilde{Z}_{\beta  \varphi  \lambda \alpha   } \,^{(1)}{}\tilde{Z}_{\mu  \rho  \omega  }{}^{\lambda  }\nonumber\\
    && +2 l^{\varphi  } l^{\rho  } l^{\omega  } \,^{(1)}{}\tilde{Z}_{\beta  \lambda  \varphi  \alpha  } \,^{(1)}{}\tilde{Z}_{\mu  \rho  \omega  }{}^{\lambda  } \bigr)=0\,.\label{pnd2''''}
\end{eqnarray}
It is easy to see that, by contracting the above expression with $l^\mu l^\beta$, one finds $\Delta_0=0$. Then, by contracting it with $l^\mu$, one gets $\Delta_1=\Delta_2=0$. Thereby, by replacing these three conditions in Eq.~\eqref{pnd2''''}, we arrive at
\begin{equation}
    T_{\alpha\beta\lambda\mu\tau\nu}  l^\alpha l^\lambda l^\tau l^\nu  = -\,2 \bigl(\Delta _{3}{} \bar{\Delta }_{3}{} + 2 \Delta _{4}{} \bar{\Delta }_{4}{} + \Delta _{5}{} \bar{\Delta }_{5}{}\bigr) l_{\beta  } l_{\mu  }=0\,,
\end{equation}
meaning that $\Delta_3=\Delta_4=\Delta_5=0$. Putting all together, this case is represented by
\begin{equation}
    \Delta_0= \Delta_1 =\Delta_2=\Delta_3=\Delta_4=\Delta_5=0\,.\label{deltacond7}
\end{equation}
The maximum $bo(l)$ is now zero, so this will mean {\bf Class III} in the terminology of ANDs.

By looking into Eq.~\eqref{pnd2''''}, one notices that we need to compute the form of $l^\tau \Z_{\sigma\gamma\rho\tau}$, in order to find the intrinsic characterisation of this case:
\be\label{Zl1}
l^\mu \Z_{\alpha\beta \lambda\mu} = l_\lambda \bigl[A \bigl(l_\alpha k_\beta +l_\beta k_\alpha\bigr)+c_{\alpha\beta}+l_\alpha v_\beta +l_\beta v_\alpha\bigr] +l_\alpha y_{\beta\lambda} +l_\beta y_{\alpha\lambda}+f^{(2)}_{\alpha\beta\lambda}\,,
\ee
where
\begin{align}
c_{\mu\nu}&=-\,\bigl( \Delta _{6}{} + \bar{\Delta }_{8}{}\bigr) m_{\mu  } m_{\nu  } - \bigl( \Delta _{7}{} + \bar{\Delta }_{7}{}\bigr) m_{\nu  } \bar{m}_{\mu  }+\textrm{c.c.}\,,\label{cctensor}\\
y_{\mu\nu}&= - \,\Delta _{7}{} m_{\mu  } \bar{m}_{\nu  }  - \Delta _{8}{} \bar{m}_{\mu  } \bar{m}_{\nu  }+\frac{1}{2} \Delta _{11}{} l_{\mu  } \bar{m}_{\nu  } +\textrm{c.c.}\,,\\
    v_\mu&= \bigl(\Delta _{10}{} + \bar{\Delta }_{11}{}\bigr) m_{\mu  }- \frac{1}{2} \Delta _{13}{} l_{\mu  }+\textrm{c.c.}\,,\label{vvector}
\end{align}
are three quantities fully orthogonal to $l^{\mu}$ and $k^{\mu}$, $A$ is given by Expression~\eqref{help1} and $f^{(2)}_{\alpha\beta\lambda}$ is a tensor depending on the complex scalars $\Delta_0, \Delta_1, \Delta_2, \Delta_3, \Delta_4$, $\Delta_5$ and their conjugates. Then, it turns out that the intrinsic characterisation of this case is given by
\begin{align}
   %  l^\alpha \Z_{\alpha\beta\lambda [\mu}l^\lambda l_{\nu]}&=&- \Delta _{0}{} k_{\beta  }  \bar{m}_{[\mu  } l_{\nu  ]}+ \Delta _{1}{} m_{\beta  } \bar{m}_{[\mu  } l_{\nu  ]} + \Delta _{2}{}  \bar{m}_{\beta  } \bar{m}_{[\mu  }l_{\nu  ]} + \Delta _{4}{} l_{\beta  } l_{[\mu  } \bar{m}_{\nu]  }+\textrm{c.c.}=0\,, \, \, \,\textrm{and}\\
l^{[\gamma} l_{[\delta} \Z^{\alpha]}{}_{\beta] \lambda[\mu} l^{\lambda} l_{\nu]} =&- \Delta _{0}{} k^{[\gamma  } l^{\alpha ] } k_{[\delta  } l_{\beta]  } \bar{m}_{[\mu  }l_{\nu]  }  + \Delta _{1}{}  m^{[\gamma  }l^{\alpha ] } k_{[\delta  } l_{\beta]  }  \bar{m}_{[\mu  }l_{\nu]  } -\Delta _{1}{}  k^{[\gamma }l^{\alpha ] }  l_{[\delta  } m_{\beta ] }   \bar{m}_{[\mu  }l_{\nu]  }  \nonumber\\
&+ \Delta _{2}{} \bar{m}^{[\gamma  } l^{\alpha  ]} k_{[\delta  } l_{\beta ] } \bar{m}_{[\mu  }l_{\nu  ]} - \Delta _{2}{}k^{[\gamma  } l^{\alpha  ]}    l_{[\delta  } \bar{m}_{\beta]  } \bar{m}_{[\mu  }l_{\nu  ]} - \Delta _{3}{} l_{[\delta  } m_{\beta ] } l^{[\gamma  } m^{\alpha]  }  \bar{m}_{[\mu  }l_{\nu  ]}\nonumber\\
&- \Delta _{4}{} l^{[\gamma  } \bar{m}^{\alpha]  }l_{[\delta  }  m_{\beta]  }  \bar{m}_{[\mu  }l_{\nu  ]}- \Delta _{4}{} l^{[\gamma  }  m^{\alpha]  }l_{[\delta  }\bar{m}_{\beta]  }    \bar{m}_{[\mu  }l_{\nu  ]}- \Delta _{5}{} l^{[\gamma  }\bar{m}^{\alpha  ]} l_{[\delta  }  \bar{m}_{\beta]  } \bar{m}_{[\mu  } l_{\nu  ]} +\textrm{c.c.} =0\,,\label{case7d}
\end{align}
which is equivalent to the condition~\eqref{deltacond7}.

\subsection{Contraction of $T_{\alpha\beta\lambda\mu\tau\nu}(\Z)$ with three copies of $l^\mu$: Class IIc, Class IId, Class IIIa and Class IV}

We remove now three null vectors $l^\mu$ from Expression~\eqref{pnd} and, by doing that, there are four independent possibilities that we shall explain and categorise separately. Notice that all of these cases always satisfy $\Delta_0=\Delta_1=\Delta_2=\Delta_3=\Delta_4=0$ plus other extra conditions.

\subsubsection{Class IIc}

The first possibility with three copies of $l_\mu$ is
\begin{eqnarray}
T_{\alpha\beta\lambda\mu\tau\nu}l^\alpha l^\beta l^\tau &=&l^{\alpha  } \Big(\frac{1}{2} l^{\beta  } l_{\nu  } g_{\lambda  \mu  }\, ^{(1)}{}\tilde{Z}_{\alpha  }{}^{\sigma  \rho  \omega  } \,^{(1)}{}\tilde{Z}_{\beta  \sigma  \rho  \omega  } -2 l^{\sigma  } l^{\beta  } g_{\lambda  \mu  } \,^{(1)}{}\tilde{Z}_{\alpha  \rho  \nu  \omega  }\, ^{(1)}{}\tilde{Z}_{\beta  }{}^{\rho  }{}_{\sigma  }{}^{\omega  } + 2 l^{\sigma  } l^{\beta  }\,^{(1)}{}\tilde{Z}_{\alpha\mu    \nu  \rho  } \,^{(1)}{}\tilde{Z}_{\lambda  \beta  \sigma  }{}^{\rho  } \nonumber\\
&&+ \,2 l^{\sigma  } l^{\beta  } \,^{(1)}{}\tilde{Z}_{ \alpha\lambda   \nu  \rho  } \,^{(1)}{}\tilde{Z}_{\mu  \beta  \sigma  }{}^{\rho  }  - l^{\beta  } l_{\nu  } \,^{(1)}{}\tilde{Z}_{ \alpha\lambda   \sigma  \rho  } \,^{(1)}{}\tilde{Z}_{\mu  \beta  }{}^{\sigma  \rho  }\Big)=0\,.\label{part8}
\end{eqnarray}
By contracting this equation with $l^\lambda l^\mu, l^\mu l^\nu$ and $l^\lambda$, gives $\Delta_0=\Delta_1=\Delta_2=\Delta_3=\Delta_4=0$, respectively. Putting all together, the above condition is just
\begin{equation}
    -\,8 \Delta _{7}{} \bar{\Delta }_{7}{} l_{\lambda  } l_{\mu  } l_{\nu  }=0\,,
\end{equation}
which implies the following $\Delta$ characterisation:
\begin{equation}
    \Delta_0= \Delta_1 =\Delta_2=\Delta_3=\Delta_4=\Delta_7=0\,.\label{delta8case}
\end{equation}
Thus, the maximum $bo(l)$ is again 1. We will labelled this as {\bf Class IIc}.

By looking into Eq.~\eqref{part8}, one immediately notices that the contraction $l^\alpha \Z_{\alpha\beta\lambda \mu}$ obtained in~\eqref{Zlll} is also important to find the intrinsic characterisation of this case, which arises by the further contraction
\begin{eqnarray}
      l^{\alpha  } \, ^{(1)}{}\tilde{Z}_{\alpha  \lambda  \beta  \mu  }l^{\beta  }&=&- \,\Delta _{0}{} k_{\lambda  } \bar{m}_{\mu  }+ \Delta _{1}{} \bigl(k_{\lambda  } l_{\mu  } + m_{\lambda  } \bar{m}_{\mu  }\bigr)+ \Delta _{2}{} \bar{m}_{\lambda  } \bar{m}_{\mu  }- \Delta _{3}{} l_{\mu  } m_{\lambda  } \nonumber\\
    &&-\,2 \Delta _{4}{} \bar{m}_{(\lambda  }l_{\mu ) }+\frac{1}{2}\bigl(\Delta _{7}{}+\bar{\Delta} _{7}{}\bigr) l_{\lambda  } l_{\mu  } +\textrm{c.c.} =0\,,\,\,
    \end{eqnarray}
    and
    \begin{eqnarray}
   l^{\alpha  } \,^{(1)}{}\tilde{Z}_{\alpha  \lambda  [\beta  \mu  }l_{\sigma]} &=&-\,2 \Delta _{0}{} k_{\lambda  }  \bar{m}_{[\beta  }l_{\mu  } k_{\sigma]  }  -2 \Delta _{1}{} \bigl( m_{\lambda  } \bar{m}_{[\beta  } k_{\mu  } l_{\sigma]  }+ k_{\lambda  }   \bar{m}_{[\beta  }m_{\mu  }l_{\sigma  ]}\bigr)+ 2 \Delta _{2}{}  \bar{m}_{\lambda  }\bar{m}_{[\beta  } l_{\mu  } k_{\sigma]  } + 2 \Delta _{3}{}  m_{\lambda  } \bar{m}_{[\beta  }m_{\mu  }l_{\sigma  ]}\nonumber\\
   && -\,2 \Delta _{4}{} \bigl( l_{\lambda  } \bar{m}_{[\beta  } l_{\mu  } k_{\sigma  ]} - \bar{m}_{\lambda  }  \bar{m}_{[\beta  } m_{\mu  }l_{\sigma  ]}\bigr)+\bigl(\bar{\Delta }_{7}{}- \Delta _{7}{} \bigr) l_{\lambda  }  \bar{m}_{[\beta  }   m_{\mu  } l_{\sigma  ]}+\textrm{c.c.}=0\,,\label{case8d}
\end{eqnarray}
both representing an equivalent result to the condition~\eqref{delta8case}. 

\subsubsection{Class IId}

The second case with three null vectors $l_\mu$ is
\begin{eqnarray}
T_{\alpha\beta\lambda\mu\tau\nu}l^\alpha l^\beta l^\lambda &=&l^{\alpha  } \Big(\frac{1}{2} l^{\beta  } l_{\mu  } g_{\nu  \tau  }\, ^{(1)}{}\tilde{Z}_{\alpha  }{}^{\lambda  \varphi  \iota  }\, ^{(1)}{}\tilde{Z}_{\beta  \lambda  \varphi  \iota  } -2 l^{\beta  } l_{\mu  } \,^{(1)}{}\tilde{Z}_{\alpha  \lambda  \nu  \varphi  } \,^{(1)}{}\tilde{Z}_{\beta  }{}^{\lambda  }{}_{\tau  }{}^{\varphi  }  - l^{\beta  } l^{\lambda  } g_{\nu  \tau  } \,^{(1)}{}\tilde{Z}_{  \alpha  \mu\varphi  \iota  }\,^{(1)}{}\tilde{Z}_{\beta  \lambda  }{}^{\varphi  \iota  }  \nonumber\\
&&+ 2 l^{\beta  } l^{\lambda  } \, ^{(1)}{}\tilde{Z}_{\alpha\mu    \nu  \varphi  }\,^{(1)}{}\tilde{Z}_{\beta  \lambda  \tau  }{}^{\varphi  } + 2 l^{\beta  } l^{\lambda  } \,^{(1)}{}\tilde{Z}_{ \alpha\mu   \tau  \varphi  } \,^{(1)}{}\tilde{Z}_{\beta  \lambda  \nu  }{}^{\varphi  }\Big)=0\,.\label{case9a}
\end{eqnarray}
It is easy to see from this equation that $\Delta_0=\Delta_1=\Delta_2=\Delta_3=\Delta_4=0$, since its contractions with $l^\mu l^\tau l^\nu, l^\mu l^\tau,l^\nu l^\tau, l^\mu$ and $l^\tau$, respectively, give rise to such conditions. Then, Eq.~\eqref{case9a} simply provides
\begin{eqnarray}
    T_{\alpha\beta\lambda\mu\tau\nu}l^\alpha l^\beta l^\lambda &=&-\,4 \bigl(\Delta _{6}{} \bar{\Delta }_{6}{} + \Delta _{7}{} \bar{\Delta }_{7}{}\bigr) l_{\mu  } l_{\nu  } l_{\tau  }=0\,,
\end{eqnarray}
and therefore the $\Delta$-characterisation reads:
\begin{equation}
    \Delta_0= \Delta_1 =\Delta_2=\Delta_3=\Delta_4=\Delta_6=\Delta_7=0\,.\label{case9b}
\end{equation}
The maximum $bo(l)$ is again 1 and we will label this as {\bf Class IId}.

Once again, it is clear from Eq.~\eqref{case9a} that the contraction $l^\alpha \Z_{\alpha\beta\lambda \mu}$ is essential to find an intrinsic characterisation for this case, which turns out to read
\begin{eqnarray}
   l_{\alpha  }l_{[\rho} \,^{(1)}{}\tilde{Z}^\alpha{}_{ \lambda]  \beta  \mu  }&=& 2 \Delta _{0}{}   l_{[\lambda  } k_{\rho  ]}k_{[\beta  }\bar{m}_{\mu]  }+ 2 \Delta _{1}{} \bigl( l_{[\lambda  }k_{\rho ] } l_{[\beta  }k_{\mu]  }   -   m_{[\lambda  }l_{\rho ] } \bar{m}_{[\beta  } k_{\mu ] }+ l_{
[\lambda  }k_{\rho]  } \bar{m}_{[\beta  } m_{\mu ] } \bigr) -2 \Delta _{2}{} \bar{m}_{[\lambda  }l_{\rho]  } \bar{m}_{[\beta  }k_{\mu  ]}  \nonumber\\
  &&+\, 2 \Delta _{3}{} \bigl(l_{[\lambda  } k_{\rho ] } m_{[\beta  }l_{\mu]  }  -  m_{[\lambda  }l_{\rho ] }  k_{[\beta  } l_{\mu]  } - l_{[\lambda  } m_{\rho  ]}  \bar{m}_{[\beta  }m_{\mu]  }\bigr)+ 2 \Delta _{4}{} \bigl(\bar{m}_{[\lambda  }l_{\rho ] } l_{[\beta  } k_{\mu  ]}  + \bar{m}_{[\lambda  }l_{\rho]  } \bar{m}_{[\beta  }m_{\mu]  }  \bigr) \nonumber\\
  &&+\,2 \Delta _{6}{}  m_{[\lambda  }l_{\rho ] } m_{[\beta  } l_{\mu ] }    -2 \Delta _{7}{} \bar{m}_{[\lambda  }l_{\rho]  }l_{[\beta  }  m_{\mu ] }+\textrm{c.c.} =0\,,\label{case9dd}
 \end{eqnarray}
 and, as expected, is equivalent to the condition~\eqref{case9b}. 

\subsubsection{Class IIIa}

The third possible contraction with three copies of $l_\mu$ is
\begin{eqnarray}
T_{\alpha\beta\lambda\mu\tau\nu}l^\lambda l^\tau l^\nu &=&l^{\lambda  } \bigl(2 l^{\sigma  } l_{\mu  } \,^{(1)}{}\tilde{Z}_{\alpha  \nu   \omega \lambda  } \,^{(1)}{}\tilde{Z}_{\beta  }{}^{\nu  }{}_{\sigma  }{}^{\omega  }-l^{\sigma  } l_{\mu  } g_{\alpha  \beta  }\,^{(1)}{}\tilde{Z}^{\nu  \omega\rho   }{}_{\lambda  }  \,^{(1)}{}\tilde{Z}_{\nu  \omega  \sigma  \rho  } -2 l^{\sigma  } l^{\nu  }\, ^{(1)}{}\tilde{Z}_{\alpha  \mu   \omega   \lambda}\, ^{(1)}{}\tilde{Z}_{\beta  \sigma  \nu  }{}^{\omega  } \nonumber\\
&&  - \,2 l^{\sigma  } l^{\nu  }\,^{(1)}{}\tilde{Z}_{\alpha  \sigma  \nu  }{}^{\omega  } \,^{(1)}{}\tilde{Z}_{\beta  \mu   \omega \lambda  }  \bigr)+2 l^{\sigma  } l^{\nu  } l^{\lambda  }g_{\alpha  \beta  }\,^{(1)}{}\tilde{Z}_{\mu  \omega  \rho\lambda    }\, ^{(1)}{}\tilde{Z}_{\sigma  }{}^{\omega  }{}_{\nu  }{}^{\rho  } =0\,.\label{cond10a}
\end{eqnarray}
Then, by contracting the above equation with $l^{\alpha  } l^{\beta  } l^{\mu  }, l^{\alpha  } l^{\beta  } $ and $l^\alpha$, one finds $\Delta_0=\Delta_1=\Delta_2=\Delta_3=\Delta_4=\Delta_5=0$. By using those conditions, the above equation is reduced to
\begin{equation}
T_{\alpha\beta\lambda\mu\tau\nu}l^\lambda l^\tau l^\nu =-\,4 \bigl(\Delta _{7}{} \bar{\Delta }_{7}{} + \Delta _{8}{} \bar{\Delta }_{8}{}\bigr) l_{\alpha  } l_{\beta  } l_{\mu  }=0\,,
\end{equation}
from where we conclude 
\begin{equation}
    \Delta_0= \Delta_1 =\Delta_2=\Delta_3=\Delta_4=\Delta_5=\Delta_7=\Delta_8=0\,.\label{con10b}
\end{equation}
The maximum $bo(l)$ is then zero and we will name this case as {\bf Class IIIa}.

From Eq.~\eqref{cond10a}, we notice that this time the contraction $l^\mu \Z_{\alpha\beta \lambda\mu}$ obtained in~\eqref{Zl1} is important to provide an intrinsic characterisation for this case
\begin{eqnarray}
  l_{[\gamma}\Z_{\alpha]\beta \lambda[\mu}l^\lambda l_{\nu]}&=&\Delta _{0}{}k_{[\gamma  } l_{\alpha]  } k_{\beta  }  \bar{m}_{[\mu  }l_{\nu ] }   - \Delta _{1}{} \bigl( k_{[\gamma  } l_{\alpha]  } m_{\beta  } \bar{m}_{[\mu  }l_{\nu  ]} -l_{[\gamma  } m_{\alpha ] } k_{\beta  }  \bar{m}_{[\mu  }l_{\nu ] } \bigr) + \Delta _{2}{} \bigl( l_{[\gamma  }k_{\alpha ] }  \bar{m}_{\beta  } \bar{m}_{[\mu  }  l_{\nu ] }-  \bar{m}_{[\gamma  }l_{\alpha  ]}  k_{\beta  } \bar{m}_{[\mu  }l_{\nu]  }\bigr)\nonumber\\
  &&+\,\Delta _{3}{} m_{[\gamma  }l_{\alpha]  }  m_{\beta  }\bar{m}_{[\mu  }l_{\nu ] }    - \Delta _{4}{} \bigl(l_{[\gamma  }k_{\alpha]  } l_{\beta  } \bar{m}_{[\mu  } l_{\nu ] }  + l_{[\gamma  } \bar{m}_{\alpha]  }m_{\beta  } \bar{m}_{[\mu  }  l_{\nu]  } + l_{[\gamma  }  m_{\alpha]  } \bar{m}_{\beta  } \bar{m}_{[\mu  }l_{\nu ] }\bigr)\nonumber\\
  &&+\,\Delta _{5}{} l_{[\gamma  } \bar{m}_{\alpha]  } \bar{m}_{\beta  } l_{[\mu  } \bar{m}_{\nu ] }- \Delta _{7}{} m_{[\gamma  }l_{\alpha]  } l_{\beta  }  \bar{m}_{[\mu  }l_{\nu ] }  + \Delta _{8}{} l_{[\gamma  }\bar{m}_{\alpha]  }l_{\beta  }\bar{m}_{[\mu  }  l_{\nu ] }  +\textrm{c.c.} =0\,,
\end{eqnarray} 
which is equivalent to the condition~\eqref{con10b}. 

\subsubsection{Class IV}

The last possibility with three copies of $l_\mu$ is
\begin{eqnarray}
T_{\alpha\beta\lambda\mu\tau\nu}l^\alpha l^\lambda  l^\tau &=&-\, \frac{1}{4} l_{\beta  } l_{\mu  } l_{\nu  }\, ^{(1)}{}\tilde{Z}_{\alpha  \lambda  \sigma  \tau  } \,^{(1)}{}\tilde{Z}^{\alpha  \lambda  \sigma  \tau  }+l^{\alpha  } \bigl(-l^{\sigma  } l^{\lambda  }\, ^{(1)}{}\tilde{Z}_{\beta  \mu    \tau\alpha  } \,^{(1)}{}\tilde{Z}_{\lambda  \sigma  \nu  }{}^{\tau  }  + l^{\lambda  } l_{\mu  }\, ^{(1)}{}\tilde{Z}_{\beta  \sigma    \tau \alpha } \,^{(1)}{}\tilde{Z}_{\lambda  }{}^{\sigma  }{}_{\nu  }{}^{\tau  } \nonumber\\
&&- \,l_{\beta  } l_{\mu  }\,\Z^{\lambda  \sigma \tau }{}_{\alpha  } \Z_{\lambda  \sigma  \nu  \tau  }   + l_{\beta  } l^{\lambda  }\, ^{(1)}{}\tilde{Z}_{\mu  \sigma   \tau \alpha  }\, ^{(1)}{}\tilde{Z}_{\lambda  }{}^{\sigma  }{}_{\nu  }{}^{\tau  }\bigr)+l^{\alpha  } \bigl(- l^{\lambda  } l_{\mu  } \,^{(1)}{}\tilde{Z}_{\alpha  }{}^{\sigma  }{}_{\lambda  }{}^{\tau  } \,^{(1)}{}\tilde{Z}_{\beta  \sigma  \nu  \tau  } \nonumber\\
&&+ \,\frac{1}{2} l_{\mu  } l_{\nu  } \,^{(1)}{}\tilde{Z}_{\alpha  }{}^{\lambda  \sigma  \tau  } \,^{(1)}{}\tilde{Z}_{\beta  \lambda  \sigma  \tau  } - \frac{1}{2} l^{\lambda  } l_{\nu  } \,^{(1)}{}\tilde{Z}_{\alpha  \lambda  }{}^{\sigma  \tau  } \,^{(1)}{}\tilde{Z}_{\beta  \mu  \sigma  \tau  } + l^{\sigma  } l^{\lambda  } \,^{(1)}{}\tilde{Z}_{\alpha  \lambda  \sigma  }{}^{\tau  } \,^{(1)}{}\tilde{Z}_{\beta  \mu  \nu  \tau  }  \nonumber\\
&&+\, l^{\sigma  } l^{\lambda  }\,^{(1)}{}\tilde{Z}_{ \alpha\mu   \nu  \tau  }\, ^{(1)}{}\tilde{Z}_{\beta  \lambda  \sigma  }{}^{\tau  } - l_{\beta  } l^{\lambda  } \,^{(1)}{}\tilde{Z}_{\alpha  }{}^{\sigma  }{}_{\lambda  }{}^{\tau  } \,^{(1)}{}\tilde{Z}_{\mu  \sigma  \nu  \tau  } + \frac{1}{2} l_{\beta  } l_{\nu  } \,^{(1)}{}\tilde{Z}_{\alpha  }{}^{\lambda  \sigma  \tau  } \,^{(1)}{}\tilde{Z}_{\mu  \lambda  \sigma  \tau  } \nonumber\\
&&+\, l^{\sigma  } l^{\lambda  } \,^{(1)}{}\tilde{Z}_{\alpha \beta   \nu  \tau  } \,^{(1)}{}\tilde{Z}_{\mu  \lambda  \sigma  }{}^{\tau  } - \frac{1}{2} l^{\lambda  } l_{\nu  } \,^{(1)}{}\tilde{Z}_{ \alpha\beta   \sigma  \tau  } \,^{(1)}{}\tilde{Z}_{\mu  \lambda  }{}^{\sigma  \tau  }\bigr)=0\,.
\end{eqnarray}
It is easy to see, by contracting this equation with $l^\beta l^\mu l^\nu,l^\beta l^\mu, l^\beta l^\nu ,l^\beta,l^\nu$, that $\Delta_0=\Delta_1=\Delta_2=\Delta_3=\Delta_4=\Delta_5=0$, respectively. Then, the equation acquires the simple form
\begin{equation}
T_{\alpha\beta\lambda\mu\tau\nu}l^\alpha l^\lambda  l^\tau =-\,2\bigl(\Delta _{6}{} \bar{\Delta }_{6}{} + 2 \Delta _{7}{} \bar{\Delta }_{7}{} + \Delta _{8}{} \bar{\Delta }_{8}{}\bigr) l_{\beta  } l_{\mu  } l_{\nu  }=0\,,\label{case11a}
\end{equation}
which means that the $\Delta$-characterisation becomes
\begin{equation}
    \Delta_0= \Delta_1 =\Delta_2=\Delta_3=\Delta_4=\Delta_5=\Delta_6=\Delta_7=\Delta_8=0\,.
\end{equation}
Accordingly, the maximum $bo(l)$ in this case is $-1$ and thus we will name this case as {\bf Class IV}. The intrinsic characterisation for this case can be written as
\begin{eqnarray}
    l_{[\gamma}\Z_{\alpha]\beta \lambda[\mu}l^\lambda l_{\nu]}&=&\Delta _{0}{} k_{[\gamma  }l_{\alpha  ]}  k_{\beta  } \bar{m}_{[\mu  } l_{\nu ] } + \Delta _{1}{} \bigl(l_{[\gamma  }m_{\alpha  ]}k_{\beta  } \bar{m}_{[\mu  }  l_{\nu]  }   - k_{[\gamma  } l_{\alpha ] } m_{\beta  } \bar{m}_{[\mu  } l_{\nu  ]}\bigr) \nonumber\\
    &&+\, \Delta _{2}{} \bigl(\bar{m}_{[\gamma  } l_{\alpha]  }k_{\beta  } l_{[\mu  }  \bar{m}_{\nu]  }-  l_{[\gamma  } k_{\alpha  ]} \bar{m}_{\beta  } l_{[\mu  }\bar{m}_{\nu]  } \bigr)- \Delta _{3}{} l_{[\gamma  }m_{\alpha ] }   m_{\beta  } \bar{m}_{[\mu  }l_{\nu  ]}\nonumber\\
    && + \,\Delta _{4}{} \bigl(k_{[\gamma  } l_{\alpha]  } l_{\beta  }\bar{m}_{[\mu  } l_{\nu ] }   - l_{[\gamma  }m_{\alpha  ]}   \bar{m}_{\beta  } \bar{m}_{[\mu  } l_{\nu ] }+ \bar{m}_{[\gamma  } l_{\alpha]  }  m_{\beta  } \bar{m}_{[\mu  }l_{\nu ] }\bigr) + \Delta _{5}{} l_{[\gamma  }  \bar{m}_{\alpha ] } \bar{m}_{\beta  } l_{[\mu  }\bar{m}_{\nu]  } \nonumber\\
    &&+ \,\Delta _{7}{}  l_{[\gamma  } m_{\alpha  ]} l_{\beta  } \bar{m}_{[\mu  }l_{\nu  ]} + \Delta _{8}{}  l_{[\gamma  }\bar{m}_{\alpha  ]}l_{\beta  }  \bar{m}_{[\mu  }l_{\nu]  } +\textrm{c.c.} =0\,,\label{case11b}\\
     l_{\alpha  }l_{[\rho} \,^{(1)}{}\tilde{Z}^\alpha{}_{ \lambda]  \beta  \mu  }&=&2 \Delta _{0}{}   l_{[\lambda  } k_{\rho  ]}k_{[\beta  }\bar{m}_{\mu]  }+ 2 \Delta _{1}{} \bigl( l_{[\lambda  }k_{\rho ] } l_{[\beta  }k_{\mu]  }   -   m_{[\lambda  }l_{\rho ] } \bar{m}_{[\beta  } k_{\mu ] }+ l_{
[\lambda  }k_{\rho]  } \bar{m}_{[\beta  } m_{\mu ] } \bigr) -2 \Delta _{2}{} \bar{m}_{[\lambda  }l_{\rho]  } \bar{m}_{[\beta  }k_{\mu  ]}  \nonumber\\
  &&+\, 2 \Delta _{3}{} \bigl(l_{[\lambda  } k_{\rho ] } m_{[\beta  }l_{\mu]  }  -  m_{[\lambda  }l_{\rho ] }  k_{[\beta  } l_{\mu]  } - l_{[\lambda  } m_{\rho  ]}  \bar{m}_{[\beta  }m_{\mu]  }\bigr)+ 2 \Delta _{4}{} \bigl(\bar{m}_{[\lambda  }l_{\rho ] } l_{[\beta  } k_{\mu  ]}  + \bar{m}_{[\lambda  }l_{\rho]  } \bar{m}_{[\beta  }m_{\mu]  }  \bigr) \nonumber\\
  &&+\,2 \Delta _{6}{}  m_{[\lambda  }l_{\rho ] } m_{[\beta  } l_{\mu ] }    -2 \Delta _{7}{} \bar{m}_{[\lambda  }l_{\rho]  }l_{[\beta  }  m_{\mu ] }+\textrm{c.c.}=0\,.\label{case11c}
\end{eqnarray}
Clearly, these two conditions together are equivalent to Eq.~\eqref{case11a}.

%\begin{eqnarray}
 %   \Z_{\alpha\beta\lambda\mu} = (k_\lambda l_\mu -k_\mu l_\lambda) (l_\alpha v_\beta + l_\beta v_\alpha) +(l_\mu R_\lambda - l_\lambda R_\mu) (l_\alpha k_\beta +l_\beta k_\alpha) \nonumber\\
%+l_\alpha l_\beta (k_\lambda X_\mu -k_\mu X_\lambda +d_{\lambda\mu})
%+l_\mu (l_\alpha m_{\beta\lambda} +l_\beta m_{\alpha\lambda})-l_\lambda %(l_\alpha m_{\beta\mu} +l_\beta m_{\alpha\mu})\nonumber\\
%+l_\mu b_{\alpha\beta\lambda} -l_\lambda b_{\alpha\beta\mu} +l_\alpha d_{\beta\lambda\mu} +l_\beta d_{\alpha\lambda\mu} 
%\end{eqnarray}

\subsection{Contraction of $T_{\alpha\beta\lambda\mu\tau\nu}(\Z)$ with two copies of $l^\mu$: Class IIe, Class IIIb, Class IVa and Class V}

In this step, we remove four null vectors $l^\mu$ from Expression~\eqref{pnd}, which gives rise to four independent possibilities. Notice that the different contractions of the resulting expression with null vectors $l^\mu$ always lead at least to $\Delta_0=\Delta_1=\Delta_2=\Delta_3=\Delta_4=0$. Since the equations in this step become cumbersome and we have already explained in detail how the computation works, in the following we will omit explicit equations and just present the important results.

\subsubsection{ Class IIe}

The first possible contraction with two copies of $l^\mu$ is
\begin{eqnarray}
T_{\alpha\beta\lambda\mu\tau\nu}l^\alpha l^\beta=0\,,\label{case12a}
\end{eqnarray}
which gives $\Delta_0=\Delta_1=\Delta_3=0$ by contracting it with $l^\lambda l^\mu l^\tau l^\nu$, $l^\lambda l^\mu l^\tau$ and $l^\lambda l^\mu$, respectively. By taking into account these conditions and contracting Eq.~\eqref{case12a} with $l^\lambda l^\tau l^\nu$ and $l^\lambda l^\tau$, we also find $\Delta_2=\Delta_4=0$, respectively, whereas the contraction with $l^\lambda$ provides then $\Delta_6=\Delta_7=0$. Finally, by replacing all of these conditions in the equation, we find $T_{\alpha\beta\lambda\mu\tau\nu}l^\alpha l^\beta=-\,8 \Delta _{10}{} \bar{\Delta }_{10}{} l_{\lambda  } l_{\mu  } l_{\nu  } l_{\tau  }=0$, which implies $\Delta_{10}=0$. In summary, we find the $\Delta$-characterisation
\begin{equation}
    \Delta_0= \Delta_1 =\Delta_2=\Delta_3=\Delta_4=\Delta_6=\Delta_7=\Delta_{10}=0\,.\label{case12b}
\end{equation}
The maximum $bo(l)$ in this case is $1$, so that we will name this case as {\bf Class IIe}. 

One can notice that Eq.~\eqref{case12a} always depends on $l^\alpha \Z_{\alpha\beta\lambda\mu}$, in such a way that the intrinsic characterisation for this case is
\begin{eqnarray}
    l^\alpha \Z_{\alpha\beta\lambda\mu} &=& 2 \Delta _{0}{} k_{\beta  } \bar{m}_{[\lambda  } k_{\mu  ]}+ 2 \Delta _{1}{} \bigl( m_{\beta  }  k_{[\lambda  }\bar{m}_{\mu ] } + k_{\beta  } m_{[\lambda  } \bar{m}_{\mu ] }- k_{\beta  } l_{[\lambda  }  k_{\mu  ]}\bigr)  + 2 \Delta _{2}{}  \bar{m}_{\beta  } k_{[\lambda  }\bar{m}_{\mu ] } \nonumber\\
    &&+\, 2 \Delta _{3}{} \bigl(m_{\beta  }l_{[\lambda  }k_{\mu  ]}    - k_{\beta  } m_{[\lambda  }  l_{\mu  ]}+ m_{\beta  } \bar{m}_{[\lambda  } m_{\mu]  }\bigr) + 2 \Delta _{4}{} \bigl( \bar{m}_{\beta  }l_{[\lambda  } k_{\mu  ]} + l_{\beta  } \bar{m}_{[\lambda  } k_{\mu  ]} +  \bar{m}_{\beta  } \bar{m}_{[\lambda  }m_{\mu  ]}\bigr) \nonumber\\
    &&+\,2 \Delta _{6}{} m_{\beta  } m_{[\lambda  }l_{\mu  ]}  + 2 \Delta _{7}{} \bigl(l_{\beta  }k_{[\lambda  }  l_{\mu]  } +\bar{m}_{\beta  } m_{[\lambda  }l_{\mu]  }   + l_{\beta  } m_{[\lambda  } \bar{m}_{\mu ] }\bigr)+ 2 \Delta _{10}{} l_{\beta  } l_{[\lambda  } m_{\mu ] }+\textrm{c.c.}=0\,,\label{case12d}
\end{eqnarray}
that is equivalent as~\eqref{case12b}.

\subsubsection{Class IIIb}

The second possible case with two null vectors $l^\mu$ is
\begin{eqnarray}
T_{\alpha\beta\lambda\mu\tau\nu}l^\tau l^\nu=0\,,
\end{eqnarray}
which provides $\Delta_0=\Delta_1=\Delta_2=\Delta_4=0$ by contracting it with $l^\alpha l^\beta l^\lambda l^\mu,l^\alpha l^\beta l^\lambda$ and $l^\alpha l^\beta$, respectively. Moreover, if we use these conditions, further contractions with $l^\alpha l^\lambda$ and $l^\mu$ lead to $\Delta_3=\Delta_5=\Delta_7=\Delta_8=0$. By using all of these conditions, we find $T_{\alpha\beta\lambda\mu\tau\nu}l^\tau l^\nu=-\,8 \Delta _{11}{} \bar{\Delta }_{11}{} l_{\alpha  } l_{\beta  } l_{\lambda  } l_{\mu  }=0$, which means $\Delta_{11}=0$. Putting all of these conditions together, we find the $\Delta$-characterisation 
\begin{equation}
    \Delta_0= \Delta_1 =\Delta_2=\Delta_3=\Delta_4=\Delta_5=\Delta_7=\Delta_8=\Delta_{11}=0\,.\label{case13b}
\end{equation}
The maximum $bo(l)$ is zero and thus we will name this case as {\bf Class IIIb}. 

In addition, the intrinsic characterisation reads
\begin{eqnarray}
  \Z_{\alpha\beta \lambda[\mu}l^\lambda l_{\nu]}&=&\Delta _{0}{} k_{\alpha  } k_{\beta } l_{[\mu  } \bar{m}_{\nu  ]}   -2 \Delta _{1}{} k_{(\alpha  }m_{\beta  )} l_{[\mu  }  \bar{m}_{\nu]  } -2 \Delta _{2}{} \bar{m}_{(\alpha  }k_{\beta ) } l_{[\mu  }  \bar{m}_{\nu  ]}+ \Delta _{3}{}  m_{\alpha  } m_{\beta  } l_{[\mu  } \bar{m}_{\nu ] }  \nonumber\\
  &&+\, 2 \Delta _{4}{} \bigl(k_{(\alpha  } l_{\beta)  } l_{[\mu  } \bar{m}_{\nu]  } +  \bar{m}_{(\alpha  }m_{\beta)  } l_{[\mu  } \bar{m}_{\nu]  }\bigr) + \Delta _{5}{}\bar{m}_{\alpha  } \bar{m}_{\beta  }  l_{[\mu  } \bar{m}_{\nu ] } \nonumber\\
  && -\,2 \Delta _{7}{} l_{(\alpha  } m_{\beta  )} l_{[\mu  } \bar{m}_{\nu]  } -2 \Delta _{8}{} \bar{m}_{(\alpha  } l_{\beta ) } l_{[\mu  } \bar{m}_{\nu]  } + \Delta _{11}{} l_{\alpha  } l_{\beta  } l_{[\mu  } \bar{m}_{\nu]  }+\textrm{c.c.}=0\,,
\end{eqnarray}
which is equivalent to the condition~\eqref{case13b}.

\subsubsection{ Class IVa}

The third possible case with two copies of $l^\mu$ reads
\begin{eqnarray} 
T_{\alpha\beta\lambda\mu\tau\nu}l^\lambda l^\tau =0\,,
\end{eqnarray}
which gives us $\Delta_0=\Delta_1=\Delta_2=\Delta_3=\Delta_4=0$ if we contract it with $l^\alpha l^\beta l^\mu l^\nu, l^\alpha l^\beta l^\mu$ and $l^\alpha l^\beta$, respectively. Furthermore, by using these conditions and contracting the previous equation with $l^\alpha$, we find $\Delta_5=\Delta_6=\Delta_7=\Delta_8=0$, respectively. Then, using all these conditions together, the equation becomes $T_{\alpha\beta\lambda\mu\tau\nu}l^\lambda l^\tau=-\,4\bigl(\Delta _{10}{} \bar{\Delta }_{10}{} + \Delta _{11}{} \bar{\Delta }_{11}{}\bigr) l_{\alpha  } l_{\beta  } l_{\mu  } l_{\nu  }=0$, implying $\Delta_{10}=\Delta_{11}=0$. Therefore, this case gives us the $\Delta$-characterisation
\begin{equation}
    \Delta_0= \Delta_1 =\Delta_2=\Delta_3=\Delta_4=\Delta_5=\Delta_6=\Delta_7=\Delta_8=\Delta_{10}=\Delta_{11}=0\,,\label{case14b}
\end{equation}
and then, the maximum $bo(l)$ is $-1$ and we name this case as {\bf Class IVa}. 

On the other hand, the intrinsic characterisation for this case can be given by two expressions. The first one reads
\begin{eqnarray}
l_{[\gamma}\Z_{\alpha]\beta\lambda\mu}l^{\mu}&=&-\, \Delta _{0}{}  k_{[\gamma  } l_{\alpha ] } k_{\beta  }\bar{m}_{\lambda  } + \Delta _{1}{} \bigl( k_{[\gamma  } l_{\alpha ] }k_{\beta  } l_{\lambda  } -  l_{[\gamma  } k_{\alpha  ]}m_{\beta  } \bar{m}_{\lambda  } +  m_{[\gamma  }l_{\alpha  ]}  k_{\beta  }\bar{m}_{\lambda  }\bigr)+ \Delta _{2}{} \bigl( k_{[\gamma  } l_{\alpha]  } \bar{m}_{\beta  } \bar{m}_{\lambda  }-  l_{[\gamma  } \bar{m}_{\alpha  ]} k_{\beta  }\bar{m}_{\lambda  }\bigr) \nonumber\\
&&+ \,\Delta _{3}{} \bigl(l_{[\gamma  }m_{\alpha  ]}k_{\beta  }  l_{\lambda  }   - k_{[\gamma  } l_{\alpha]  }  m_{\beta  }l_{\lambda  } + l_{[\gamma  } m_{\alpha  ]} m_{\beta  } \bar{m}_{\lambda  }\bigr)- \Delta _{4}{} \bigl(k_{[\gamma  } l_{\alpha ] }  \bar{m}_{\beta  } l_{\lambda  }+  \bar{m}_{[\gamma  }l_{\alpha ] } k_{\beta  }l_{\lambda  }  + k_{[\gamma  } l_{\alpha ] } l_{\beta  } \bar{m}_{\lambda  } \nonumber\\
&&- \,l_{[\gamma  } \bar{m}_{\alpha  ]}m_{\beta  }  \bar{m}_{\lambda  }  - l_{[\gamma  } m_{\alpha]  } \bar{m}_{\beta  } \bar{m}_{\lambda  }\bigr)+ \Delta _{5}{} l_{[\gamma  }  \bar{m}_{\alpha  ]}\bar{m}_{\beta  } \bar{m}_{\lambda  }+ \Delta _{6}{} m_{[\gamma  } l_{\alpha]  } l_{\lambda  } m_{\beta  } \nonumber\\
&&+ \,\Delta _{7}{} \bigl(k_{[\gamma  } l_{\alpha]  } l_{\beta  } l_{\lambda  }  - l_{[\gamma  }\bar{m}_{\alpha  ]}  m_{\beta  }l_{\lambda  }   - l_{[\gamma  } m_{\alpha  ]}\bar{m}_{\beta  } l_{\lambda  }  + m_{[\gamma  }  l_{\alpha ] } l_{\beta  }\bar{m}_{\lambda  }\bigr)\nonumber\\
&&- \,\Delta _{8}{} \bigl(l_{[\gamma  } \bar{m}_{\alpha  ]} \bar{m}_{\beta  }l_{\lambda  }  + l_{[\gamma  } \bar{m}_{\alpha]  }l_{\beta  }  \bar{m}_{\lambda  }\bigr)-\frac{1}{2} \bigl(\Delta _{10}{} + \bar{\Delta }_{11}{}\bigr) m_{[\gamma  }l_{\alpha ] } l_{\beta  } l_{\lambda  } \nonumber\\
&&-\,\frac{1}{2} \bigl(\bar{\Delta }_{10}{} + \Delta _{11}{}\bigr)  \bar{m}_{[\gamma  }l_{\alpha ] } l_{\beta  } l_{\lambda  }   +\textrm{c.c.}    =0\,,\label{case14e}
\end{eqnarray}
whereas the second one is simply
\begin{equation}
l_{[\gamma}\Z_{\alpha]\beta[\lambda\mu}l_{\nu]}=4 \Delta _{10}{} l_{[\gamma  }m_{\alpha ] } l_{\beta  }  \bar{m}_{[\lambda  } m_{\mu  }l_{\nu  ]}   -4 \bar{\Delta }_{10}{}l_{[\gamma  } \bar{m}_{\alpha  ]} l_{\beta  }  \bar{m}_{[\lambda  } m_{\mu  }l_{\nu  ]} =0\,.\label{case14f}
\end{equation}
It is then clear that~\eqref{case14e} and~\eqref{case14f}, together, are equivalent to the condition~\eqref{case14b}. 

\subsubsection{Class V}

The last case with two copies of $l^\mu$ is
\begin{eqnarray}
T_{\alpha\beta\lambda\mu\tau\nu}l^\alpha l^\lambda  =0\,,\label{classV}
\end{eqnarray}
and gives us $\Delta_0=\Delta_1=0$ by contracting with $l^\beta l^\mu l^\tau$. Thus, by using these conditions and contracting the previous equation with $l^\beta l^\mu$ and $l^\beta l^\nu$, we find $\Delta_2=\Delta_3=\Delta_4=0$. A further contraction with $l^\beta$ and $l^\nu$ then yields  $\Delta_5=\Delta_6=\Delta_7=\Delta_8=0$. Finally, if we use all of these conditions in Eq.~\eqref{classV}, we find $T_{\alpha\beta\lambda\mu\tau\nu}l^\alpha l^\lambda  =-\,2 \bigl( \Delta_{9}{} \bar{\Delta}_{9}{}+2 \Delta_{10}{} \bar{\Delta}_{10}{} + \Delta_{11}{} \bar{\Delta}_{11}{} \bigr) l_{\beta } l_{\mu } l_{\nu } l_{\tau }=0$ and then $\Delta_{9}=\Delta_{10}=\Delta_{11}=0$. Putting it all together, the $\Delta$-characterisation is
\begin{equation}
    \Delta_0= \Delta_1 =\Delta_2=\Delta_3=\Delta_4=\Delta_5=\Delta_6=\Delta_7=\Delta_8=\Delta_{9}=\Delta_{10}=\Delta_{11}=0\,.\label{case15a}
\end{equation}
Therefore, the maximum $bo(l)$ is $-2$ and we name this case as {\bf Class V}.

The intrinsic characterisation for this case reads
\begin{eqnarray}
    l_{[\gamma  }\, ^{(1)}{}\tilde{Z}_{\alpha]}{}^{ [\beta}{}_{  \lambda  \mu  } l^{\rho ] }&=&2 \Delta _{0}{} k_{[\gamma  } l_{\alpha  ]}l^{[\beta  } \bar{m}_{[\lambda  }k_{\mu ] }   k^{\rho  ]}  -2 \Delta _{1}{} \bigl(k_{[\gamma  } l_{\alpha  ]}l^{[\beta  }k_{[\mu  }  l_{\lambda ] }k^{\rho]  }    - l_{[\gamma  } m_{\alpha ] }l^{[\beta  }  \bar{m}_{[\lambda  }k_{\mu ] } k^{\rho]  }  + k_{[\gamma  } l_{\alpha ] } l^{[\beta  } \bar{m}_{[\lambda  } m_{\mu ] }k^{\rho  ]}  \nonumber\\
    && - \, l_{[\gamma  }k_{\alpha]  }l^{[\beta  }\bar{m}_{[\lambda  } k_{\mu ] }  m^{\rho]  } \bigr)+ 2 \Delta _{2}{} \bigl(l_{[\gamma  } \bar{m}_{\alpha]  }l^{[\beta  } \bar{m}_{[\lambda  }k_{\mu ] }   k^{\rho ] }  - k_{[\gamma  } l_{\alpha  ]}l^{[\beta  } \bar{m}_{[\lambda  } k_{\mu ] }  \bar{m}^{\rho  ]}\bigr)\nonumber\\
    &&-\,2 \Delta _{3}{} \bigl(l_{[\gamma  } m_{\alpha ] } l^{[\beta  } l_{[\lambda  }k_{\mu]  } k^{\rho ] }  + l_{[\gamma  }k_{\alpha ] }  l^{[\beta  }  l_{[\lambda  } m_{\mu ] }k^{\rho  ]} + k_{[\gamma  } l_{\alpha  ]}  l^{[\beta  }k_{[\lambda  } l_{\mu]  } m^{\rho]  } + k_{[\gamma  } l_{\alpha]  }  m^{[\beta  } \bar{m}_{[\lambda  }m_{\mu  ]} l^{\rho]  } \nonumber\\
    &&-\, m_{[\gamma  } l_{\alpha ] } m^{[\beta  }k_{[\lambda  } \bar{m}_{\mu ] } l^{\rho ] }   - l_{[\gamma  } m_{\alpha]  }l^{[\beta  }  m_{[\lambda  } \bar{m}_{\mu ] }k^{\rho  ]} \bigr) +2 \Delta _{4}{} \bigl(  \bar{m}_{[\gamma  } l_{\alpha ] } l^{[\beta  } \bar{m}_{[\lambda  }m_{\mu]  } k^{\rho  ]}-  l_{[\gamma  }\bar{m}_{\alpha  ]}l^{[\beta  } l_{[\lambda  } k_{\mu ] } k^{\rho ] }   \nonumber\\
    &&-\, k_{[\gamma  }l_{\alpha]  } \bar{m}^{[\beta  }l_{[\lambda  }k_{\mu ] }   l^{\rho]  }  +l_{[\gamma  } \bar{m}_{\alpha]  }   l^{[\beta  } k_{[\lambda  }\bar{m}_{\mu ] } m^{\rho  ]}  -l_{[\gamma  } k_{\alpha  ]}\bar{m}^{[\beta  }  m_{[\lambda  }  \bar{m}_{\mu ] } l^{\rho  ]}  - m_{[\gamma  }l_{\alpha ] }l^{[\beta  } k_{[\lambda  }   \bar{m}_{\mu ] } \bar{m}^{\rho ] }\bigr)\nonumber\\
    &&+\,2 \Delta _{5}{} l_{[\gamma  } \bar{m}_{\alpha ] } \bar{m}^{[\beta  } \bar{m}_{[\lambda  }k_{\mu]  } l^{\rho]  }  -2 \Delta _{6}{} \bigl(l_{[\gamma  }m_{\alpha  ]}  m^{[\beta  }l_{[\lambda  }k_{\mu  ]}    l^{\rho  ]}+ m_{[\gamma  }l_{\alpha  ]} l^{[\beta  }  l_{[\lambda  } m_{\mu]  } k^{\rho ] }  - k_{[\gamma  } l_{\alpha]  } l^{[\beta  }  m_{[\lambda  }l_{\mu ] } m^{\rho ] } \nonumber\\
    &&-\, m_{[\gamma  }l_{\alpha  ]} l^{[\beta  }  m_{[\lambda  }\bar{m}_{\mu ] } m^{\rho ] } \bigr)-2 \Delta _{7}{} \bigl(l_{[\gamma  }\bar{m}_{\alpha  ]} m^{[\beta  } l_{[\lambda  }k_{\mu  ]}  l^{\rho]  }  +  l_{[\gamma  } m_{\alpha]  }\bar{m}^{[\beta  }l_{[\lambda  } k_{\mu  ]}l^{\rho  ]}   +   \bar{m}_{[\gamma  }l_{\alpha]  } l^{[\beta  } l_{[\lambda  } m_{\mu]  }k^{\rho]  } \nonumber\\
    &&-\,m_{[\gamma  } l_{\alpha]  }\bar{m}^{[\beta  }  \bar{m}_{[\lambda  } m_{\mu]  }   l^{\rho  ]}+  \bar{m}_{[\gamma  }l_{\alpha ] } l^{[\beta  }  \bar{m}_{[\lambda  }m_{\mu ] } m^{\rho]  } +  l_{[\gamma  }k_{\alpha ] } l^{[\beta  }  m_{[\lambda  }l_{\mu ] } \bar{m}^{\rho  ]}\bigr) \nonumber\\
    &&-\,2 \Delta _{8}{} \bigl(l_{[\gamma  } \bar{m}_{\alpha]  } \bar{m}^{[\beta  }l_{[\lambda  }k_{\mu  ]}   l^{\rho ] } +\bar{m}_{[\gamma  } l_{\alpha]  }  \bar{m}^{[\beta  }  m_{[\lambda  }\bar{m}_{\mu  ]}l^{\rho]  } \bigr)-2 \Delta _{9}{} l_{[\gamma  } m_{\alpha  ]} m^{[\beta  }m_{[\lambda  }l_{\mu ] } l^{\rho ] }\nonumber\\
   & & -\,2 \Delta _{10}{} \bigl(l_{[\gamma  } \bar{m}_{\alpha]  } m^{[\beta  } m_{[\lambda  }l_{\mu ] } l^{\rho]  }   - m_{[\gamma  }l_{\alpha]  } l^{[\beta  } l_{[\lambda  }  m_{\mu]  } \bar{m}^{\rho]  }\bigr)      -2 \Delta _{11}{} l_{[\gamma  } \bar{m}_{\alpha  ]}\bar{m}^{[\beta  }m_{[\lambda  }l_{\mu ] } l^{\rho]  }  + \textrm{c.c.}=0\,,\label{case15b}
\end{eqnarray}
which indeed is equivalent to Eq.~\eqref{case15a}.

\subsection{Contraction of $T_{\alpha\beta\lambda\mu\tau\nu}(\Z)$ with one copy of $l^\mu$: Class IVb and Class VI}

The last step is obtained by removing five null vectors $l^\mu$ in Expression~\eqref{pnd}, which allows only two independent possibilities.

\subsubsection{ Class IVb}

The first possible case with one null vector $l^\mu$ reads
\begin{eqnarray}
    T_{\alpha\beta\lambda\mu\tau\nu} l^\nu  =0\,.\label{case16a}
\end{eqnarray}
First, it is clear that this condition gives $\Delta_i=0$ with $i=1,...,5$. By applying this in Eq.~\eqref{case16a} and contracting the resulting expression with $l^\alpha l^\beta$, one finds $\Delta_7=0$. Then, by contracting with $l^\alpha l^\mu$, one also finds $\Delta_6=\Delta_8=0$. Furthermore, by replacing all of these conditions and contracting Eq.~\eqref{case16a} with $l^\alpha$, one finds $\Delta_{10}=\Delta_{11}=0$, which ends up reducing the equation itself to $\Delta_{13}=0$. Hence, the $\Delta$-characterisation is
\begin{equation}
    \Delta_0= \Delta_1 =\Delta_2=\Delta_3=\Delta_4=\Delta_5=\Delta_6=\Delta_7=\Delta_8=\Delta_{10}=\Delta_{11}=\Delta_{13}=0\,.\label{case16b}
\end{equation}
In other words, in this case only the complex scalars $\Delta_9, \Delta_{12}$ and $\Delta_{14}$ are nonvanishing. The maximum $bo(l)$ is $-1$ and, thus, we will name this case as {\bf Class IVb}. 

On the other hand, it is possible to find an intrinsic characterisation in terms of two different conditions. The first one is
\begin{eqnarray}
  l^{\nu  } \,^{(1)}{}\tilde{Z}_{\alpha  \beta  \mu  \nu  }&=&\Delta _{0}{} k_{\alpha  } k_{\beta } \bar{m}_{\mu  } - \Delta _{1}{} \bigl(k_{\alpha  } k_{\beta  } l_{\mu  } + 2 m_{(\alpha  }k_{\beta)  }  \bar{m}_{\mu  }\bigr)-2 \Delta _{2}{}  \bar{m}_{(\alpha  }k_{\beta ) } \bar{m}_{\mu  }   \nonumber\\
  &&+ \,\Delta _{3}{} \bigl(2 m_{(\alpha  }k_{\beta)  } l_{\mu  }  + m_{\alpha  } m_{\beta  } \bar{m}_{\mu  }\bigr) + 2 \Delta _{4}{} \bigl(\bar{m}_{(\alpha  }k_{\beta)  } l_{\mu  }  + k_{(\alpha  } l_{\beta ) } \bar{m}_{\mu  } + \bar{m}_{(\alpha  }m_{\beta ) }  \bar{m}_{\mu  }\bigr)\nonumber\\
  && +\, \Delta _{5}{} \bar{m}_{\alpha  } \bar{m}_{\beta  } \bar{m}_{\mu  } - \Delta _{6}{}m_{\alpha  } m_{\beta  } l_{\mu  }  -2 \Delta _{7}{} \bigl(l_{(\alpha  }k_{\beta  )}  l_{\mu  } +  \bar{m}_{(\alpha  } m_{\beta ) } l_{\mu  }+  m_{(\alpha  }l_{\beta ) } \bar{m}_{\mu  }\bigr) \nonumber\\
  &&- \,\Delta _{8}{} \bigl( \bar{m}_{\alpha  } \bar{m}_{\beta  }l_{\mu  } + 2 \bar{m}_{(\alpha  } l_{\beta ) } \bar{m}_{\mu  }\bigr)+ 2 \Delta _{10}{} l_{(\alpha  } m_{\beta)  }l_{\mu  }\nonumber\\
  &&  + \,\Delta _{11}{} \bigl(2 \bar{m}_{(\alpha  }l_{\beta)  } l_{\mu  }  + l_{\alpha  } l_{\beta  } \bar{m}_{\mu  }\bigr)- \frac{1}{2}\bigl(\Delta _{13}{} + \bar{\Delta }_{13}{}\bigr) l_{\alpha  } l_{\beta  } l_{\mu  } +\textrm{c.c.}=0\,,\label{case16c}
\end{eqnarray}
and the second one is
\begin{eqnarray}
\, ^{(1)}{}\tilde{Z}_{\alpha  \beta [ \lambda  \mu  }l_{\nu ] }  &=-\,2 \bigl( \Delta_{13}{} -\bar{\Delta}_{13}{}\bigr) l_{\alpha } l_{\beta } \bar{m}_{[\lambda }m_{\mu } l_{\nu] } +f^{(1)}_{\alpha\beta\lambda\mu\nu}=0\,,\label{16fff}
\end{eqnarray}
where for simplicity we have introduced the tensor $f^{(1)}_{\alpha\beta\lambda\mu\nu}$ that depends on $\Delta_{0}, \, \Delta_{1},\, \Delta_{2},\,\Delta_{3},\,\Delta_{4},\, \Delta_{5},\, \Delta_{6},\, \Delta_{7},\, \Delta_{8},\, \Delta_{11}$ and their conjugates. Notice that the first condition does not directly imply $\Delta_{13}=0$, but the second condition is needed to vanish it. Hence, these two expressions together provide the condition~\eqref{case16b}.

\subsubsection{Class VI}

The last possibility remaining in the classification is
\begin{eqnarray}
    T_{\alpha\beta\lambda\mu\tau\nu} l^\alpha  =0\,.\label{case17a}
\end{eqnarray}
As the previous case, clearly we first have $\Delta_i=0$, with $i=1,...,5$. By taking into account this condition in the explicit expression of Eq.~\eqref{case17a} and contracting it with $l^\tau l^\mu$, one then finds $\Delta_6=\Delta_7=\Delta_8=0$. Then, a further contraction of Eq.~\eqref{case17a} with $l^\lambda$ and $l^\tau$ leads to $\Delta_9=0$ and $\Delta_{10}=\Delta_{11}=0$. Finally, by replacing all of these conditions in Eq.~\eqref{case17a}, one straightforwardly finds $\Delta_{12}=\Delta_{13}=0$. In summary, the $\Delta$-characterisation of this case reads
\begin{equation}
    \Delta_0= \Delta_1 =\Delta_2=\Delta_3=\Delta_4=\Delta_5=\Delta_6=\Delta_7=\Delta_8=\Delta_{9}=\Delta_{10}=\Delta_{11}=\Delta_{12}=\Delta_{13}=0\,,\label{delta17case}
\end{equation}
in such a way that the only nonvanishing complex scalar is $\Delta_{14}$. The maximum $bo(l)$ is $-3$, so that this case will be named as {\bf Class VI}.

From Expression~\eqref{tensorZ1}, it is clear that the tensor $\Z_{\alpha\beta\lambda\mu}$ acquires the form
\begin{equation}
    \Z_{\alpha\beta\lambda\mu} = 2l_\alpha l_\beta \bigl(l_\lambda Y_\mu-l_\mu Y_\lambda\bigr) + f^{(2)}_{\alpha\beta\lambda\mu}\,,
\end{equation}
 where $Y_\mu$ is orthogonal to $l^\mu$ and $k^\mu$, while $f^{(2)}_{\alpha\beta\lambda\mu}$ depends on all of the complex scalars and their conjugates, except on $\Delta_{14}$ and $\bar{\Delta}_{14}$. In our tetrad, we have $Y_\mu=(1/2)(\Delta_{14}m_\mu+\bar{\Delta}_{14}\bar{m}_\mu)$.

Thereby, the intrinsic characterisation for this case reads
\begin{eqnarray}
    l_{[\gamma}\Z_{\alpha]\beta\lambda\mu}&=& -\,2 \Delta _{0}{}  k_{[\gamma  }l_{\alpha ] }k_{\beta  } \bar{m}_{[\lambda  } k_{\mu]  }  + 2 \Delta _{1}{}\bigl(k_{[\gamma  }l_{\alpha  ]}k_{\beta  }   l_{[\lambda  }k_{\mu  ]}   - l_{[\gamma  } m_{\alpha ] }k_{\beta  } \bar{m}_{[\lambda  }k_{\mu ] }   + k_{[\gamma  } l_{\alpha  ]} m_{\beta  } \bar{m}_{[\lambda  } k_{\mu]  } +  k_{[\gamma  } l_{\alpha ] }k_{\beta  }\bar{m}_{[\lambda  } m_{\mu ] }\bigr)\nonumber\\
    &&-\,2 \Delta _{2}{} \bigl(l_{[\gamma  } \bar{m}_{\alpha  ]}k_{\beta  } \bar{m}_{[\lambda  }k_{\mu ] }   +  l_{[\gamma  }k_{\alpha ] } \bar{m}_{\beta  } \bar{m}_{[\lambda  } k_{\mu  ]}\bigr)+ 2 \Delta _{3}{} \bigl(l_{[\gamma  } m_{\alpha ] }k_{\beta  } l_{[\lambda  }  k_{\mu]  }  - k_{[\gamma  } l_{\alpha  ]} m_{\beta  }l_{[\lambda  }k_{\mu  ]}   + k_{[\gamma  } l_{\alpha ] }k_{\beta  }   m_{[\lambda  }l_{\mu  ]}\nonumber\\
    &&+\,l_{[\gamma  } m_{\alpha ] }  m_{\beta  } \bar{m}_{[\lambda  } k_{\mu]  } + l_{[\gamma  } m_{\alpha ] }k_{\beta  } \bar{m}_{[\lambda  } m_{\mu]  }   - k_{[\gamma  } l_{\alpha]  } m_{\beta  } \bar{m}_{[\lambda  }m_{\mu  ]} \bigr)-2 \Delta _{4}{} \bigl( l_{[\gamma  } \bar{m}_{\alpha ] }k_{\beta  } k_{[\lambda  }l_{\mu]  }   - k_{[\gamma  } l_{\alpha ] } \bar{m}_{\beta  }k_{[\lambda  } l_{\mu ] } \nonumber\\
    &&- \,l_{[\gamma  } \bar{m}_{\alpha  ]}m_{\beta  }  \bar{m}_{[\lambda  } k_{\mu  ]}  -  l_{[\gamma  }\bar{m}_{\alpha  ]}  k_{\beta  }\bar{m}_{[\lambda  }  m_{\mu  ]}+   m_{[\gamma  } l_{\alpha  ]}\bar{m}_{\beta  } \bar{m}_{[\lambda  }k_{\mu ] }  -l_{[\gamma  } k_{\alpha  ]}  \bar{m}_{\beta  } \bar{m}_{[\lambda  } m_{\mu ] }  - k_{[\gamma  } l_{\alpha]  } l_{\beta  }k_{[\lambda  }  \bar{m}_{\mu ] }\bigr)\nonumber\\
    &&+ \,2 \Delta _{5}{}  \bar{m}_{[\gamma  }l_{\alpha  ]} \bar{m}_{\beta  }  k_{[\lambda  }\bar{m}_{\mu ] } -2 \Delta _{6}{} \bigl(l_{[\gamma  }m_{\alpha ] }m_{\beta  }l_{[\lambda  }k_{\mu  ]}     -  l_{[\gamma  } k_{\alpha ] } m_{\beta  }m_{[\lambda  }  l_{\mu  ]} - m_{[\gamma  } l_{\alpha ] }  k_{\beta  }l_{[\lambda  } m_{\mu]  } + l_{[\gamma  } m_{\alpha]  } m_{\beta  }\bar{m}_{[\lambda  } m_{\mu  ]} \bigr)\nonumber\\ 
     && + \,2 \Delta _{7}{} \bigl(k_{[\gamma  } l_{\alpha ] } l_{\beta  } l_{[\lambda  }k_{\mu  ]}   -  l_{[\gamma  } \bar{m}_{\alpha ] } m_{\beta  } l_{[\lambda  } k_{\mu  } - m_{[\gamma  }l_{\alpha ] } \bar{m}_{\beta  }k_{[\lambda  }l_{\mu]  }    +  l_{[\gamma  }k_{\alpha ] } \bar{m}_{\beta  } m_{[\lambda  } l_{\mu  ]} - \bar{m}_{[\gamma  }l_{\alpha]  } k_{\beta  }  m_{[\lambda  } l_{\mu  ]}\nonumber\\
     &&+ \,k_{[\gamma  } l_{\alpha ] } l_{\beta  }\bar{m}_{[\lambda  } m_{\mu]  }   - l_{[\gamma  }  \bar{m}_{\alpha ] }m_{\beta  } \bar{m}_{[\lambda  } m_{\mu ] }  - l_{[\gamma  } m_{\alpha]  }  \bar{m}_{\beta  } \bar{m}_{[\lambda  } m_{\mu ] }- m_{[\gamma  } l_{\alpha]  } l_{\beta  } k_{[\lambda  } \bar{m}_{\mu  ]}\bigr) \nonumber \\
    &&  -\,2 \Delta _{8}{} \bigl(l_{[\gamma  }\bar{m}_{\alpha  ]}\bar{m}_{\beta  }  l_{[\lambda  }  k_{\mu  ]} + l_{[\gamma  } \bar{m}_{\alpha]  } l_{\beta  }\bar{m}_{[\lambda  } k_{\mu  ]}   - l_{[\gamma  } \bar{m}_{\alpha]  } \bar{m}_{\beta  } m_{[\lambda  } \bar{m}_{\mu  ]}\bigr)-2 \Delta _{9}{}m_{[\gamma  } l_{\alpha]  }  m_{\beta  } l_{[\lambda  } m_{\mu ] } \nonumber\\
    &&  +\, 2 \Delta _{10}{} \bigl(l_{[\gamma  } m_{\alpha ] }l_{\beta  }  l_{[\lambda  }k_{\mu  ]}  + k_{[\gamma  } l_{\alpha]  } l_{\beta  }  m_{[\lambda  } l_{\mu]  } - l_{[\gamma  }  \bar{m}_{\alpha  ]} m_{\beta  } m_{[\lambda  } l_{\mu ] } - l_{[\gamma  }m_{\alpha ] }  \bar{m}_{\beta  }m_{[\lambda  }l_{\mu  ]}   +  l_{[\gamma  } m_{\alpha ] } l_{\beta  }\bar{m}_{[\lambda  }m_{\mu]  } \bigr)  \nonumber\\
    &&-\,2 \Delta _{11}{}\bigl(l_{[\gamma  } \bar{m}_{\alpha ] }\bar{m}_{\beta  } m_{[\lambda  } l_{\mu ] }  + \bar{m}_{[\gamma  }l_{\alpha ] } l_{\beta  } l_{[\lambda  }  k_{\mu  ]}  - l_{[\gamma  }\bar{m}_{\alpha ] }  l_{\beta  }  \bar{m}_{[\lambda  }m_{\mu ] }\bigr) \nonumber\\
    &&+\,2 \Delta _{12}{} l_{[\gamma  }m_{\alpha ] }l_{\beta  }  m_{[\lambda  }l_{\mu]  }  + 2 \Delta _{13}{} l_{[\gamma  }\bar{m}_{\alpha ] }l_{\beta  }  m_{[\lambda  } l_{\mu]  } +\textrm{c.c.}=0\,,\label{case17d}
\end{eqnarray}
which is indeed equivalent to the condition~\eqref{delta17case}.

\section{Bernstein's theorem applied to $\Delta_0'=0$}\label{App:Bernstein}

Consider the main equation for the rotated complex scalar $\Delta_{0}'$:
\begin{align}
    \Delta_{0}'=&\,\Delta _0+4 \epsilon  \Delta _1+2 \bar{\epsilon } \Delta _2+6 \epsilon ^2 \Delta _3+8 \epsilon  \bar{\epsilon }
   \Delta _4+\bar{\epsilon }^2 \Delta _5+4 \epsilon ^3 \Delta _6+12 \epsilon ^2 \bar{\epsilon } \Delta _7+4 \epsilon 
   \bar{\epsilon }^2 \Delta _8+\epsilon ^4 \Delta _9+8 \epsilon ^3 \bar{\epsilon } \Delta _{10}+6 \epsilon ^2
   \bar{\epsilon }^2 \Delta _{11}\nonumber\\
   &+2 \epsilon ^4 \bar{\epsilon } \Delta _{12}+4 \epsilon ^3 \bar{\epsilon }^2 \Delta
   _{13}+\epsilon ^4 \bar{\epsilon }^2 \Delta _{14}=0\,,
\end{align}
where we want to solve for \(\epsilon\) and \(\bar{\epsilon}\). In the following, we will assume that both quantities are independent of each other.

To determine the maximum number of solutions, we can use the Bernstein's theorem, which involves calculating the areas of certain polytopes associated with the equation~\cite{bernshtein1975number}.

%\subsection*{Figure 1: Primary Area}

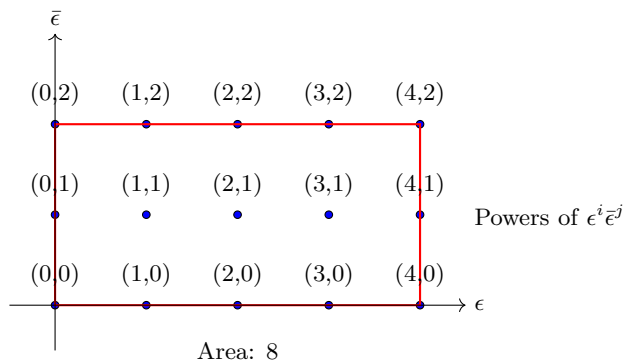
\begin{figure}[H]
\centering
\begin{tikzpicture}[scale=1.2]
    % Dibuja los puntos del área de 8
    \foreach \x/\y in {0/0, 1/0, 0/1, 2/0, 1/1, 0/2, 3/0, 2/1, 1/2, 4/0, 2/2, 3/1, 4/1, 3/2, 4/2} {
        \node[draw, circle, fill=blue, inner sep=1pt] at (\x, \y) {};
    }

    % Dibuja la envolvente convexa del área de 8
    \draw[thick, red] (0,0) -- (4,0) -- (4,2) -- (0,2) -- cycle;

    % Etiquetas de los puntos
    \foreach \x/\y in {0/0, 1/0, 0/1, 2/0, 1/1, 0/2, 3/0, 2/1, 1/2, 4/0, 2/2, 3/1, 4/1, 3/2, 4/2} {
        \node[anchor=south] at (\x,\y+0.1) {(\x,\y)};
    }

    % Ejes
    \draw[->] (-0.5, 0) -- (4.5, 0) node[right] {$\epsilon$};
    \draw[->] (0, -0.5) -- (0, 3) node[above] {$\bar{\epsilon}$};

    % Etiqueta del área
    \node at (2, -0.5) {Area: 8};

    % Etiquetas de las potencias
    \node[anchor=west] at (4.5, 1) {Powers of $\epsilon^i \bar{\epsilon}^j$};
\end{tikzpicture}
\caption{Polygon generated from the powers of $\epsilon$ and $\bar{\epsilon}$ }
\label{fig:primary-area}
\end{figure}

Figure~\ref{fig:primary-area} shows the points corresponding to the terms of the equation where \(\epsilon\) and \(\bar{\epsilon}\) appear with different powers. These points are:
\[
(0,0), (1,0), (0,1), (2,0), (1,1), (0,2), (3,0), (2,1), (1,2), (4,0), (2,2), (3,1), (4,1), (3,2), (4,2)
\]
and they form a polygon with an area of 8 where the polygon is drawn with red lines in Figure~\ref{fig:primary-area}.

%\subsection*{Figure 2: Conjugate Area}

\begin{figure}[H]
\centering
\begin{tikzpicture}[scale=1.2]
    % Dibuja los puntos del área conjugada de 8
    \foreach \x/\y in {0/0, 1/0, 0/1, 2/0, 1/1, 0/2, 3/0, 2/1, 1/2, 4/0, 2/2, 3/1, 4/1, 3/2, 4/2} {
        \node[draw, circle, fill=blue, inner sep=1pt] at (\y, \x) {}; % Invertir x e y para el conjugado
    }

    % Dibuja la envolvente convexa del área conjugada de 8
    \draw[thick, red] (0,0) -- (2,0) -- (2,4) -- (0,4) -- cycle;

    % Etiquetas de los puntos
    \foreach \x/\y in {0/0, 1/0, 0/1, 2/0, 1/1, 0/2, 3/0, 2/1, 1/2, 4/0, 2/2, 3/1, 4/1, 3/2, 4/2} {
        \node[anchor=south] at (\y,\x+0.1) {(\x,\y)};
    }

    % Ejes
    \draw[->] (-0.5, 0) -- (3, 0) node[right] {$\epsilon$};
    \draw[->] (0, -0.5) -- (0, 5) node[above] {$\bar{\epsilon}$};

    % Etiqueta del área
    \node at (1, -0.5) {Area: 8};

    % Etiquetas de las potencias
    \node[anchor=west] at (3, 2) {Powers of $\epsilon^i \bar{\epsilon}^j$};
\end{tikzpicture}
\caption{Polygon generated from the conjugate of powers of $\epsilon$ and $\bar{\epsilon}$. }
\label{fig:primary-area2}
\end{figure}
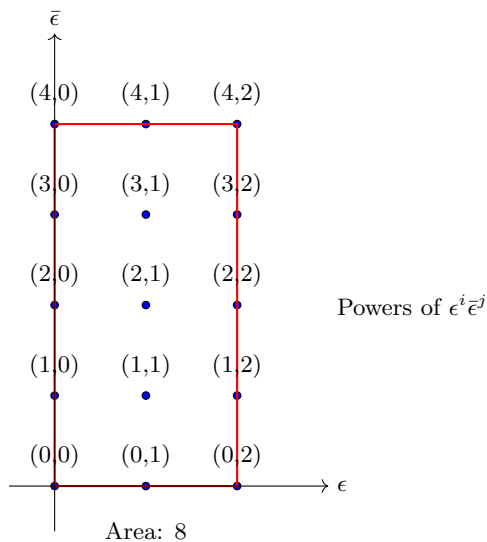

Figure~\ref{fig:primary-area2} represents the conjugate area of the same points. For the conjugate terms, the powers of \(\epsilon\) and \(\bar{\epsilon}\) are swapped. The same points form a different polygon with an area of 8, also with red lines.

%\subsection*{Figure 3: Total Area}

\begin{figure}[H]
\centering
\begin{tikzpicture}[scale=1.2]
    % Segundo gráfico: Área total de 36
    \begin{scope}[shift={(0,0)}]
        % Dibuja los puntos del área total de 36
        \foreach \x/\y in {0/0, 1/0, 2/0, 3/0, 4/0, 5/0, 6/0, 0/1, 1/1, 2/1, 3/1, 4/1, 5/1, 6/1, 0/2, 1/2, 2/2, 3/2, 4/2, 5/2, 6/2, 0/3, 1/3, 2/3, 3/3, 4/3, 5/3, 6/3, 0/4, 1/4, 2/4, 3/4, 4/4, 5/4, 6/4, 0/5, 1/5, 2/5, 3/5, 4/5, 5/5, 6/5, 0/6, 1/6, 2/6, 3/6, 4/6, 5/6, 6/6} {
            \node[draw, circle, fill=blue, inner sep=1pt] at (\x, \y) {};
        }

        % Dibuja la envolvente convexa del área total de 36
        \draw[thick, green] (0,0) -- (6,0) -- (6,6) -- (0,6) -- cycle;

      % Etiquetas de los puntos del área total
        \foreach \x/\y in {0/0, 1/0, 2/0, 3/0, 4/0, 5/0, 6/0, 0/1, 1/1, 2/1, 3/1, 4/1, 5/1, 6/1, 0/2, 1/2, 2/2, 3/2, 4/2, 5/2, 6/2, 0/3, 1/3, 2/3, 3/3, 4/3, 5/3, 6/3, 0/4, 1/4, 2/4, 3/4, 4/4, 5/4, 6/4, 0/5, 1/5, 2/5, 3/5, 4/5, 5/5, 6/5, 0/6, 1/6, 2/6, 3/6, 4/6, 5/6, 6/6} {
            \node[anchor=south] at (\x,\y+0.1) {(\x,\y)};
        }

        % Ejes
        \draw[->] (-0.5, 0) -- (7, 0) node[right] {$\epsilon$};
        \draw[->] (0, -0.5) -- (0, 7) node[above] {$\bar{\epsilon}$};

        % Etiqueta del área
        \node at (3, -0.5) {Total Area: 36};

        % Etiquetas de las potencias
        \node[anchor=west] at (6.5, 3) {Sum of all powers};
    \end{scope}

\end{tikzpicture}
\caption{Total polygon generated from the possible existing powers of $\epsilon$ and $\bar{\epsilon}$. }
\label{fig:primary-area3}
\end{figure}
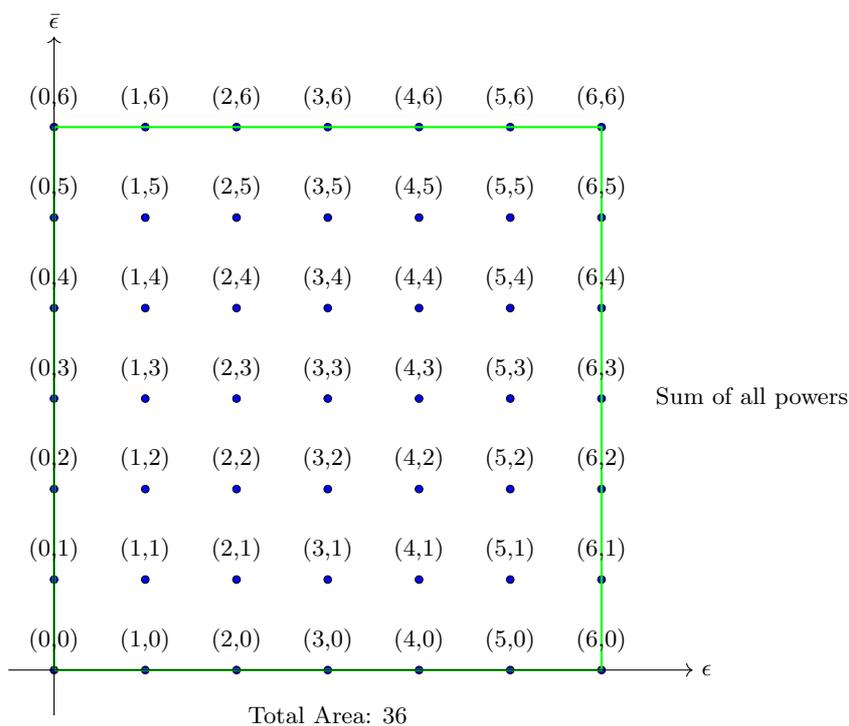

Finally, Figure~\ref{fig:primary-area3} shows the overall polytope which includes all combinations of \(\epsilon\) and \(\bar{\epsilon}\).  To obtain such a figure, all the points from the first drawing are added to all the points from the second drawing (see Figs.~\ref{fig:primary-area} and~\ref{fig:primary-area2}). This means each \((i,j)\) from the first drawing is added to each \((i',j')\) from the second drawing, where \(i,j,i',j'\) are the powers of \(\epsilon\) and \(\bar{\epsilon}\). The sum is performed as \((i,j) + (i',j') = (i+i',j+j')\).

When summing, many points will appear multiple times. However, this repetition is not important. What matters is the resulting polytope and the convex polygon that encompasses it.

The points that form the large square are:
\[
\begin{aligned}
&(0,0), (1,0), (2,0), (3,0), (4,0), (5,0), (6,0), \\
&(0,1), (1,1), (2,1), (3,1), (4,1), (5,1), (6,1), \\
&(0,2), (1,2), (2,2), (3,2), (4,2), (5,2), (6,2), \\
&(0,3), (1,3), (2,3), (3,3), (4,3), (5,3), (6,3), \\
&(0,4), (1,4), (2,4), (3,4), (4,4), (5,4), (6,4), \\
&(0,5), (1,5), (2,5), (3,5), (4,5), (5,5), (6,5), \\
&(0,6), (1,6), (2,6), (3,6), (4,6), (5,6), (6,6).
\end{aligned}
\]

Each point represents the sum of the corresponding powers of \(\epsilon\) and \(\bar{\epsilon}\) from the original polygons. The resulting polytope is the convex hull that includes all these points, forming a square with an area of 36.

Using the Bernstein's theorem, the maximum number of solutions is given by the total area minus the areas of the individual polygons.
In this case, the calculation is $36 - 8 - 8 = 20$. Therefore, the maximum number of solutions for the equation, when considering $\epsilon$ and $\bar\epsilon$ as two independent complex variables, is $20$. This result comes from subtracting the areas of the primary and conjugate polygons from the total area, effectively accounting for the overlap and ensuring the count of unique solutions. 

In order to analyse the nongeneric cases within the context of the Bernstein's theorem, we need to consider the (sub)polynomials associated with the edges that form the final polygon and determine when these, considered jointly, have solutions. The edges of the first polygon correspond to combinations of powers of \(\epsilon\) and \(\bar{\epsilon}\) from the terms in the original equation. Similarly, the edges of the second polygon correspond to combinations of powers of \(\epsilon\) and \(\bar{\epsilon}\) from the conjugate terms. When summing the edges of the two original polygons, we obtain the edges of the final polygon. This is done by summing the corresponding points from the edges of each polygon:
\[
(i,j) + (i',j') = (i+i', j+j')\,.
\]
Each edge in the original polygons is associated with a subpolynomial. The nongeneric cases arise in general when the sum of these subpolynomials, considered jointly, has solutions.

Then, for our equation, there are only two independent combinations. The first set would be
\begin{align}
    0=&\,\Delta _0+2 \bar{\epsilon } \Delta _2+\bar{\epsilon }^2 \Delta _5\,,\label{app1}\\
    0=&\,\bar{\Delta }_0+4 \bar{\epsilon } \bar{\Delta }_1+6 \bar{\epsilon }^2 \bar{\Delta }_3+4 \bar{\epsilon }^3 \bar{\Delta
   }_6+\bar{\epsilon }^4 \bar{\Delta }_9\,,\label{app2}
\end{align}
while the second set related to particular cases is
\begin{align}
    0=&\,\Delta _5+4 \epsilon  \Delta _8+6 \epsilon ^2 \Delta _{11}+4 \epsilon ^3 \Delta _{13}+\epsilon ^4 \Delta _{14}\,,\label{app3}\\
    0=&\,\bar{\Delta }_9+2 \epsilon  \bar{\Delta }_{12}+\epsilon ^2 \bar{\Delta }_{14}\,.\label{app4}
\end{align}

Let us start by solving the system~\eqref{app1}-\eqref{app2}. 
We can first proceed isolating $\bar{\epsilon}$ in Eq.~\eqref{app1} and, then, replacing it into Eq.~\eqref{app2}. If $\Delta_{5}\neq 0,$ this leads to the following constraint:
\begin{align}
    \bar{\Delta }_0=&\,\frac{4 \Delta _2 \bar{\Delta }_1}{\Delta _5}\pm \frac{4\bar{\Delta }_1 \sqrt{\Delta _2^2-\Delta _0 \Delta _5}}{\Delta _5}-\frac{12 \Delta _2^2 \bar{\Delta }_3}{\Delta _5^2}+\frac{6 \Delta _0
   \bar{\Delta }_3}{\Delta _5}\mp \frac{12 \Delta _2 \bar{\Delta }_3\sqrt{\Delta _2^2-\Delta _0 \Delta _5}}{\Delta
   _5^2}+\frac{16 \Delta _2^3 \bar{\Delta }_6}{\Delta _5^3}\nonumber\\
   &-\frac{12 \Delta _0 \Delta _2
   \bar{\Delta }_6}{\Delta _5^2}\mp \frac{16 \Delta _2^2 \bar{\Delta
   }_6\sqrt{\Delta _2^2-\Delta _0 \Delta _5}}{\Delta _5^3}\pm \frac{4 \Delta _0 \bar{\Delta }_6\sqrt{\Delta _2^2-\Delta _0 \Delta _5}}{\Delta
   _5^2}-\frac{8 \Delta _2^4 \bar{\Delta }_9}{\Delta _5^4}+\frac{8 \Delta _0 \Delta _2^2 \bar{\Delta
   }_9}{\Delta _5^3}\nonumber\\
   &-\frac{\Delta _0^2 \bar{\Delta }_9}{\Delta _5^2}\pm \frac{8 \Delta _2^3 \bar{\Delta }_9\sqrt{\Delta
   _2^2-\Delta _0 \Delta _5}}{\Delta _5^4}\mp \frac{4 \Delta _0 \Delta _2 \bar{\Delta }_9\sqrt{\Delta
   _2^2-\Delta _0 \Delta _5}}{\Delta _5^3}\,,
\end{align}
whereas, if $\Delta_5=0$ and $\Delta_2\neq0$, we find:
\begin{equation}
    \bar{\Delta }_0=\frac{2 \Delta _0 \bar{\Delta }_1}{\Delta _2}-\frac{3 \Delta _0^2 \bar{\Delta }_3}{2 \Delta
   _2^2}+\frac{\Delta _0^3 \bar{\Delta }_6}{2 \Delta _2^3}-\frac{\Delta _0^4 \bar{\Delta }_9}{16 \Delta _2^4}\,.
\end{equation}
Finally, if $\Delta_5=\Delta_2=0$, one has:
\begin{equation}
  4 \bar{\epsilon } \bar{\Delta }_1+6 \bar{\epsilon }^2 \bar{\Delta }_3+4 \bar{\epsilon }^3 \bar{\Delta
   }_6+\bar{\epsilon }^4 \bar{\Delta }_9 = 0  \,,\quad \Delta_0=0\,.
\end{equation}

Now, let us solve the second system composed by~\eqref{app3}-\eqref{app4}. Similarly, if $\Delta_{14}\neq 0$, we find the following constraint:
\begin{align}
   \Delta _5=&\,-\frac{8 \Delta _{14} \bar{\Delta }_{12}^4}{\bar{\Delta }_{14}^4}+\frac{8 \Delta _{14} \bar{\Delta }_9 \bar{\Delta
   }_{12}^2}{\bar{\Delta }_{14}^3}+\frac{16 \Delta _{13} \bar{\Delta }_{12}^3}{\bar{\Delta }_{14}^3}-\frac{\Delta
   _{14} \bar{\Delta }_9^2}{\bar{\Delta }_{14}^2}-\frac{12 \Delta _{13} \bar{\Delta }_9 \bar{\Delta
   }_{12}}{\bar{\Delta }_{14}^2}-\frac{12 \Delta _{11} \bar{\Delta }_{12}^2}{\bar{\Delta }_{14}^2}+\frac{6 \Delta
   _{11} \bar{\Delta }_9}{\bar{\Delta }_{14}}+\frac{4 \Delta _8 \bar{\Delta }_{12}}{\bar{\Delta }_{14}}\nonumber\\
   &\mp \frac{8 \Delta _{14} \bar{\Delta }_{12}^3 \sqrt{\bar{\Delta }_{12}^2-\bar{\Delta }_9 \bar{\Delta
   }_{14}}}{\bar{\Delta }_{14}^4}\pm \frac{4 \Delta _{14} \bar{\Delta }_9 \bar{\Delta }_{12} \sqrt{\bar{\Delta
   }_{12}^2-\bar{\Delta }_9 \bar{\Delta }_{14}}}{\bar{\Delta }_{14}^3}\pm \frac{16 \Delta _{13} \bar{\Delta }_{12}^2 \sqrt{\bar{\Delta }_{12}^2-\bar{\Delta }_9 \bar{\Delta
   }_{14}}}{\bar{\Delta }_{14}^3}\nonumber\\
   &\mp \frac{4 \Delta _{13} \bar{\Delta }_9 \sqrt{\bar{\Delta }_{12}^2-\bar{\Delta }_9
   \bar{\Delta }_{14}}}{\bar{\Delta }_{14}^2}\mp \frac{12 \Delta _{11} \bar{\Delta }_{12} \sqrt{\bar{\Delta }_{12}^2-\bar{\Delta }_9 \bar{\Delta
   }_{14}}}{\bar{\Delta }_{14}^2}\pm \frac{4 \Delta _8 \sqrt{\bar{\Delta }_{12}^2-\bar{\Delta }_9 \bar{\Delta
   }_{14}}}{\bar{\Delta }_{14}}\,,
\end{align}
and, if $\Delta_{14}=0$, but $\Delta_{12}\neq 0$:
\begin{equation}
  \Delta _5=\frac{\Delta _{13} \bar{\Delta }_9^3}{2 \bar{\Delta }_{12}^3}-\frac{3 \Delta _{11} \bar{\Delta }_9^2}{2
   \bar{\Delta }_{12}^2}+\frac{2 \Delta _8 \bar{\Delta }_9}{\bar{\Delta }_{12}}\,.
\end{equation}
The last possibility is given by $\Delta_{14}=\Delta_{12}=0$, which from~\eqref{app3}-\eqref{app4} gives rise to:
\begin{equation}
    4 \epsilon ^3 \Delta _{13}+6\epsilon ^2 \Delta _{11} +4\epsilon \Delta _8  -\Delta _5 = 0\,,\quad \Delta_9=0\,.
\end{equation}

Thereby, any particular case satisfying the aforementioned constraints constitutes a nongeneric case, and is excluded from the application of the Bernstein's theorem.

\bibliographystyle{utphys}
\bibliography{references}

\end{document}